\documentclass[11pt,a4paper]{article}
\usepackage[T1]{fontenc}
\usepackage[utf8]{inputenc}
\usepackage{lmodern,microtype}
\usepackage[margin=27mm]{geometry}
\usepackage{amsmath,amssymb,amsthm,mathtools,mathrsfs}
\usepackage{authblk,booktabs,array,enumitem,aliascnt}
\usepackage{graphicx}
\usepackage{tikz}
\usetikzlibrary{arrows.meta,calc,positioning}
\usepackage[section]{placeins}
\usepackage[numbers,sort&compress]{natbib}
\usepackage{xurl}
\usepackage[hidelinks,unicode,pdfdisplaydoctitle]{hyperref}
\usepackage[nameinlink,noabbrev]{cleveref}
\providecommand{\doi}[1]{\href{https://doi.org/#1}{\nolinkurl{https://doi.org/#1}}}
\hypersetup{pdftitle={Quantum Trajectories: Ergodicity, Perturbative Dynamics, and Stationary Response},pdfauthor={Guocheng Zhen, Xuanrong Yang, Chengkai Zhu, Ranyiliu Chen, Xin Wang},pdfsubject={Stationary quantum trajectories, tangent limits, rare excursions, and Clifford resource response},pdfkeywords={quantum trajectories, stationary tangent limits, singular perturbations, rare excursions, Clifford circuits, stabilizer magic}}
\numberwithin{equation}{section}
\newtheorem{theorem}{Theorem}[section]
\newaliascnt{proposition}{theorem}
\newtheorem{proposition}[proposition]{Proposition}
\aliascntresetthe{proposition}
\newaliascnt{lemma}{theorem}
\newtheorem{lemma}[lemma]{Lemma}
\aliascntresetthe{lemma}
\newaliascnt{corollary}{theorem}
\newtheorem{corollary}[corollary]{Corollary}
\aliascntresetthe{corollary}
\crefname{theorem}{Theorem}{Theorems}
\crefname{proposition}{Proposition}{Propositions}
\crefname{lemma}{Lemma}{Lemmas}
\crefname{corollary}{Corollary}{Corollaries}
\theoremstyle{definition}
\newaliascnt{definition}{theorem}
\newtheorem{definition}[definition]{Definition}
\aliascntresetthe{definition}
\crefname{definition}{Definition}{Definitions}
\theoremstyle{remark}

\newcommand{\R}{\mathbb R}
\newcommand{\C}{\mathbb C}
\newcommand{\F}{\mathbb F}
\newcommand{\E}{\mathbb E}
\newcommand{\PP}{\mathbb P}
\newcommand{\HH}{\mathcal H}
\newcommand{\X}{\mathsf X}
\newcommand{\TX}{\widehat{\mathsf X}}
\newcommand{\Cl}{\mathcal C_{d,N}}
\newcommand{\St}{\mathcal S_{d,N}}
\newcommand{\Tr}{\operatorname{Tr}}
\newcommand{\rank}{\operatorname{rank}}

\newcommand{\diag}{\operatorname{diag}}
\newcommand{\Herm}{\operatorname{Herm}}
\newcommand{\Lip}{\operatorname{Lip}}
\newcommand{\TV}{\mathrm{TV}}
\newcommand{\op}{\mathrm{op}}
\newcommand{\HS}{\mathrm{HS}}
\newcommand{\dd}{\,\mathrm d}
\newcommand{\one}{\mathbf 1}
\newcommand{\Ph}{\widehat P}
\newcommand{\Pt}{\widetilde P}
\newcommand{\pih}{\widehat\pi}
\newcommand{\floor}[1]{\lfloor #1\rfloor}
\newcommand{\ceil}[1]{\lceil #1\rceil}
\newcommand{\condK}{\hyperref[cond:kraus]{\textup{(K)}}}
\newcommand{\condS}{\hyperref[cond:reference]{\textup{(S)}}}
\newcommand{\condQ}{\hyperref[cond:quadratic]{\textup{(Q)}}}
\newcommand{\condH}{\hyperref[cond:coherent]{\textup{(H)}}}
\newcommand{\condR}{\hyperref[cond:reset]{\textup{(R)}}}
\newcommand{\condO}{\hyperref[cond:overlap]{\textup{(O)}}}
\DeclareMathOperator{\Flip}{Flip}
\setlist[enumerate]{label=\textup{(\roman*)},leftmargin=*,itemsep=3pt,topsep=5pt}
\allowdisplaybreaks[1]
\title{Quantum Trajectories: Ergodicity,\\
Perturbative Dynamics, and Stationary Response}
\author[1]{Guocheng Zhen}
\author[1]{Xuanrong Yang}
\author[2]{Chengkai Zhu}
\author[3]{Ranyiliu Chen\thanks{Corresponding author: Ranyiliu Chen. email: chenranyiliu@quantumsc.cn.}}
\author[1]{Xin Wang\thanks{Corresponding author: Xin Wang. email: felixxinwang@hkust-gz.edu.cn.}}
\affil[1]{\small Thrust of Artificial Intelligence, Information Hub, The Hong Kong University of Science and Technology (Guangzhou), Guangdong 511453, China}
\affil[2]{\small QudeLeap Research, Shanghai 200030, China}
\affil[3]{\small Quantum Science Center of Guangdong--Hong Kong--Macao Greater Bay Area, Shenzhen 518045, China}
\date{}
\begin{document}
\maketitle

\begin{abstract}
Quantum trajectories capture measurement-conditioned statistics beyond ensemble averages. For finite-dimensional Kraus dynamics, complete-reset and overlap conditions yield a unique invariant law with explicit geometric total-variation contraction. Under finite-reference mixing and quantitative transverse contraction, $C^2$ Kraus perturbations yield two complementary stationary limits: an affine tangent law for local fluctuations and an occupation measure for rare excursions. These limits give leading observable-response formulas, including an excursion contribution at quadratic order; both limits and coefficients depend only on reference Kraus operators and their first derivatives.

For Clifford-driven monitored trajectories, group structure verifies the general hypotheses and yields geometric Wasserstein mixing uniformly over intermediate unitaries at fixed dimensions, an equivalence between uniform total-variation ergodicity and a finite nonzero rank-one Kraus product for each fixed unitary, and exact stationary tangent energies. Under partial measurement, tangent dynamics have almost surely finite memory with a sharp exponential tail; every $C^2$ state observable has vanishing first-order stationary response. For weak Hamiltonian perturbations, odd-prime Gross--Wigner mana responds linearly in the perturbation strength $\theta$, whereas prime-dimensional stabilizer R\'enyi entropies scale as $\theta^{2\alpha}$, $\theta^2\ln(1/\theta)$, and $\theta^2$ for fixed $0<\alpha<1$, $\alpha=1$, and $1<\alpha<\infty$, respectively. Under partial measurement, the leading coefficients are strictly positive precisely for Hamiltonians outside the real span of the measurement projectors; on this quotient, the mana coefficient defines a norm and the order-one logarithmic coefficient an explicit positive-definite quadratic form.
\end{abstract}

\tableofcontents

\section{Introduction}\label{sec:intro}

Quantum trajectories describe the evolution of quantum systems conditioned
on measurement outcomes, rather than only the ensemble-averaged dynamics.
They provide a wave-function approach to open-system simulations
\cite{DalibardCastinMolmer1992} and a description of individual quantum
states accessible through continuous monitoring \cite{Murch2013}.
Their stationary laws retain fluctuations and nonlinear state statistics
that cannot, in general, be recovered from the stationary density matrix
alone. Understanding these laws is therefore important for characterizing
the long-time behavior of monitored quantum systems and physical
observables. This raises two related
questions: under what conditions does the conditioned evolution forget
its initial state, and how do its stationary statistics respond to weak
perturbations of the dynamics?

The mathematical theory of quantum trajectories has developed
substantially. Rigorous results include pathwise ergodic properties
and purification \cite{KummererMaassen2004,MaassenKummerer2006},
uniqueness of invariant probability measures and geometric Wasserstein
convergence under suitable conditions \cite{Benoist2019}, and further
spectral and limit theorems \cite{BenoistHautecoeur2025}.
Related perturbative investigations have addressed linear response
for Lindbladian dynamics \cite{AlbertEtAl2016}, weak-measurement
perturbation theory and homogenization limits of quantum trajectories
\cite{BallesterosEtAl2019,BenoistEtAlHomogenization2021}, and response
bounds for trajectory observables \cite{Vu2025}. These developments
provide important mathematical foundations for studying the long-time
behavior and perturbations of measurement-conditioned dynamics.

A further challenge is to characterize how weak perturbations affect
the long-time dynamics of quantum trajectories, particularly when
the unperturbed stationary law is supported on a finite invariant
set $S$. Weak perturbations produce local transverse
fluctuations but can also activate measurement outcomes whose reference
probabilities vanish. Such outcomes may induce excursions away from $S$
with probabilities of order $\theta^2$, where $\theta$ is the
perturbation strength, and contribute to stationary observables at the
same order as local quadratic fluctuations. Resolving this interaction
requires a perturbation theory that captures both effects. We develop
such a theory for finite-dimensional quantum trajectories. First,
complete-reset and overlap conditions yield explicit geometric
total-variation contraction. For normalized $C^2$ Kraus families with
finite-reference mixing and quantitative transverse contraction, we
establish two complementary stationary limits: an affine tangent law
describing local fluctuations, with Wasserstein convergence for
$1\le p<2$ and convergence of stationary finite-time distributions,
and an occupation measure describing rare excursions. These limits
yield leading-order response formulas for three classes of observables,
distinguished by their local behavior: subquadratic, logarithmically
enhanced quadratic, and ordinary quadratic. The first two classes are governed by the tangent law.
At ordinary quadratic order, however, local
linearization alone is insufficient: rare departures
generate an additional occupation term through the
reference dynamics killed upon return to $S$.
This separation of typical fluctuations from rare
excursions is central to the response theory developed
here. Technically, both limiting objects and their
response coefficients depend only on the reference
Kraus operators and their first derivatives.
The conclusions hold for any choice of stationary laws
of the perturbed dynamics, without assuming their uniqueness.

Clifford-driven quantum trajectories provide a particularly structured
setting in which these general results can be realized and strengthened.
Clifford operations and Pauli measurements are fundamental to stabilizer
quantum error correction and efficient classical simulation
\cite{Gottesman1997,AaronsonGottesman2004}, while Clifford ensembles
play an important role in unitary designs and randomized measurements
\cite{GrossAudenaertEisert2007,HuangKuengPreskill2020}.
Their monitored dynamics also connect scrambling and measurement
to quantum error correction and purification \cite{Choi2020,Gullans2020}.
We consider an architecture combining fresh uniform global Clifford
scrambling, an intermediate unitary, projective measurement on a nonempty
subset of qudits, and inverse Clifford evolution. Beyond verifying the
hypotheses of the general perturbation theory, the distinctive
algebraic structure of Clifford dynamics yields several stronger
conclusions. First, we establish geometric Wasserstein mixing uniformly
over intermediate unitaries at fixed dimensions, and characterize
uniform total-variation ergodicity for each fixed unitary by the
existence of a finite nonzero rank-one Kraus product. Second, Clifford symmetry makes the stationary law
Clifford-invariant and determines its first two state
moments---the stationary density matrix and its second
tensor moment---independently of the intermediate unitary,
even though nonlinear stationary resource statistics may
vary with it. We also establish exact Wasserstein
transport identities and quantitative stability under
changes of the intermediate unitary. Third, for weak
Hamiltonian perturbations, the Clifford and stabilizer structures yield
an exact stationary tangent-energy formula and identify the effective
perturbation directions. Under partial measurement, the tangent
dynamics have almost surely finite memory with a sharp exponential
tail, leading to exponential integrability and quantitative stability
of their stationary laws. Finally, these structures yield three
stationary Wasserstein response regimes separated by moment order two
and a quadratic stationary response formula for arbitrary $C^2$
state observables, whose first-order response vanishes.

To exhibit the resource-theoretic consequences of this theory, we study
quantum magic, a resource for computation beyond the stabilizer
subtheory \cite{BravyiKitaev2005,Veitch2014}, and its interplay with
measurement in Clifford-based dynamics
\cite{BejanMcLauchlanBeri2024,Niroula2024MagicTransition,Scocco2026RiseFall}.
We consider Gross--Wigner mana in odd prime local dimension
\cite{Gross2006} and stabilizer R\'enyi entropies in prime local
dimension \cite{Leone2022,WangLi2023}. Their distinct local geometries
give a linear-in-strength stationary mana response and three R\'enyi
response regimes: $\theta^{2\alpha}$ for $0<\alpha<1$,
$\theta^2\ln(1/\theta)$ for $\alpha=1$, and $\theta^2$ for
$\alpha>1$, at every fixed finite order $\alpha>0$. The coefficients
above order one also contain the contribution of rare excursions. Under partial measurement,
these leading response coefficients are strictly positive precisely
for Hamiltonians outside the real span of the measurement projectors.
The mana coefficient defines a norm on Hamiltonians modulo this span,
while the logarithmic order-one coefficient is an explicit
positive-definite quadratic form on the same quotient. Thus the
response laws reveal both the local geometry of quantum resources
and the geometry of effective Hamiltonian perturbations.

Beyond the asymptotic scaling laws, these results provide analytical tools for monitored quantum-information models: finite Kraus-word certificates for uniform total-variation mixing, Hamiltonian-dependent susceptibilities that identify resource-active perturbations, and finite-memory and excursion representations with quantitative truncation bounds. The applications below illustrate how these tools can place existing analyses of monitored quantum dynamics on a rigorous mathematical footing, while yielding further structural and quantitative insights. Together, these results connect structural properties of quantum trajectories with verifiable assumptions and quantitative predictions for stationary quantum resources.

\section{Guide to the main results}
\label{sec:main-results-guide}

Figure~\ref{fig:main-results-map} organizes the results into general
quantum-trajectory theory, Clifford-specific structure, and stationary
resource response. The Clifford layer both realizes the general
perturbative hypotheses and supplies additional exact conclusions.
Here TV denotes total variation and $W_p$ denotes Wasserstein distance
of order $p$. In the Clifford model, $N$ is the number of qudits and
$1\le m\le N$ is the number measured.

\begin{figure}[!ht]
\centering
\newcommand{\GuideResultCard}[4]{%
  \begin{minipage}[c][#1][s]{\linewidth}
    \centering
    \textbf{#2}\par
    \vfill
    #3\par
    \vfill
    {\scriptsize #4\par}
  \end{minipage}%
}
\begin{tikzpicture}[
  x=\linewidth, y=1cm, font=\footnotesize,
  panel/.style={draw=black!35,fill=black!2,rounded corners=3pt,
                line width=.5pt},
  heading/.style={anchor=west,font=\sffamily\bfseries\small},
  result/.style={draw=black!55,fill=white,rounded corners=2pt,
                 line width=.55pt,align=center,
                 inner xsep=4pt,inner ysep=5pt},
  general/.style={result,text width=.27\linewidth},
  cliffcard/.style={result,text width=.27\linewidth},
  resource/.style={result,text width=.42\linewidth},
  flow/.style={-{Latex[length=2mm]},draw=black!65,line width=.7pt},
  every node/.append style={outer sep=0pt}
]
\draw[panel] (.005,0) rectangle (.995,-4.35);
\node[heading] at (.035,-.30)
  {I. General quantum-trajectory theory};
\node[general] at (.17,-1.96) {
  \GuideResultCard{2.50cm}
    {Complete resets\\and overlap}
    {Unique invariant law;\\explicit geometric\\TV contraction}
    {\cref{prop:resets}}
};
\node[general] at (.50,-1.96) {
  \GuideResultCard{2.50cm}
    {Stationary\\tangent dynamics}
    {Concentration and process limit;\\
     $W_p$ convergence, $1\le p<2$}
    {\cref{prop:concentration,prop:affine};\\
     \cref{thm:tangent}}
};
\node[general] at (.83,-1.96) {
  \GuideResultCard{2.50cm}
    {Local and\\excursion response}
    {Three observable classes;\\quadratic coefficient:\\
     tangent + occupation}
    {\cref{cor:subquadratic};\\\cref{thm:log,thm:quadratic}}
};
\node[align=center,text width=.93\linewidth,font=\scriptsize]
  at (.50,-3.90)
  {Reference Kraus data and first derivatives fix both limits and coefficients\\
   (\cref{cor:jets}).};

\draw[flow] (.50,-4.35) -- (.50,-4.75);
\draw[panel] (.005,-4.75) rectangle (.995,-9.25);
\node[heading] at (.035,-5.05)
  {II. Clifford-specific structure};
\node[cliffcard] at (.17,-6.80) {
  \GuideResultCard{2.80cm}
    {Global ergodicity}
    {Uniform $W_1$ over $R$;\\for fixed $R$, uniform TV\\
     iff a rank-one word exists}
    {\cref{thm:global};\\
     \cref{prop:clifford-TV-characterization}}
};
\node[cliffcard] at (.50,-6.80) {
  \GuideResultCard{2.80cm}
    {Tangent dynamics}
    {TV mixing; exact energy;\\for $m<N$: finite memory\\
     and exponential moments}
    {\cref{prop:verification,thm:energy};\\
     \cref{thm:tangent-exponential-moments}}
};
\node[cliffcard] at (.83,-6.80) {
  \GuideResultCard{2.80cm}
    {Transport and\\smooth response}
    {Three $W_p$ regimes;\\second-order expansion\\
     for every $C^2$ observable}
    {\cref{cor:Clifford-Wasserstein-response};\\
     \cref{cor:clifford-response}}
};
\node[align=center,text width=.93\linewidth,font=\scriptsize]
  at (.50,-8.83)
  {Stationary symmetry, moments, exact transport, and gate stability
   (\cref{prop:stationary-structure});\\
   finite excursion occupation and exponential truncation for $m<N$
   (\cref{prop:finite-Clifford-occupation}).};

\draw[flow] (.50,-9.25) -- (.50,-9.65);
\draw[panel] (.005,-9.65) rectangle (.995,-13.28);
\node[heading] at (.035,-9.95)
  {III. Stationary magic response};
\node[resource] at (.26,-11.37) {
  \GuideResultCard{1.95cm}
    {Odd-prime mana}
    {Linear-in-strength response;\\coefficient is a norm on\\
     effective Hamiltonian directions}
    {\cref{thm:mana}}
};
\node[resource] at (.74,-11.37) {
  \GuideResultCard{1.95cm}
    {Stabilizer R\'enyi entropies}
    {$\theta^{2\alpha}$, $\theta^2\ln(1/\theta)$, $\theta^2$\\
     for fixed finite orders\\below, at, and above one}
    {\cref{thm:sre}}
};
\node[align=center,text width=.93\linewidth,font=\scriptsize]
  at (.50,-12.95)
  {Under partial measurement: positivity precisely outside sector phases;\\
   norms and quadratic forms on the Hamiltonian quotient.};
\end{tikzpicture}
\caption{Overview of the main conclusions. The three layers distinguish
  general dynamical and response theorems, Clifford-specific refinements,
  and resource-theoretic consequences. Vertical arrows indicate
  specialization and application.}
\label{fig:main-results-map}
\end{figure}

\par\medskip
\noindent\textbf{I. General ergodicity and stationary dynamics.}
Conditions~\condR\ and~\condO\ give a complete-reset criterion for a
unique invariant law and explicit geometric TV contraction
(\cref{prop:resets}). The perturbative results instead use
Conditions~\condK, \condS, and~\condQ: a normalized finite $C^2$ Kraus
family, a finite invariant reference set $S$ with primitive dynamics,
and quantitative transverse contraction. They hold for every choice
of invariant laws $\pi_\theta$ of the perturbed kernels. Writing
$r_S(x)=d(x,S)$, we obtain $\pi_\theta(r_S^2)=O(\theta^2)$ and a unique
affine tangent invariant law $\pih$ with a finite moment of some order
$p_*>2$ (\cref{prop:concentration,prop:affine}). If $B_\theta$ rescales
transverse coordinates by $\theta^{-1}$, then, as $\theta\downarrow0$,
\[
  (B_\theta)_\#\pi_\theta\Rightarrow\pih,
  \qquad
  \theta^{-2}\pi_\theta|_{\X\setminus S}
       \xrightarrow{\mathrm{vague}}\mu^{\mathrm{occ}}.
\]
The first limit holds in tangent-space Wasserstein distance for
$1\le p<2$ and extends to stationary joint distributions on every fixed
finite time window (\cref{thm:tangent}). The second identifies a Radon
occupation measure $\mu^{\mathrm{occ}}$ on $\X\setminus S$, generated by rare departures
and the reference evolution killed upon reaching $S$
(\cref{thm:quadratic}).

These two limits determine the responses of the three observable
classes in Section~\ref{sec:response}. Subquadratic and logarithmically
enhanced quadratic responses admit expansions at scales $\theta^k$,
$0<k<2$, and
$\theta^2[\ln(1/\theta)]^\zeta$, $\zeta>0$, respectively; their leading
coefficients are tangent averages
(\cref{cor:subquadratic,thm:log}). For an ordinary quadratic observable
$F$, with quadratic local term $q_F$, the coefficient also records
excursions:
\[
  \pi_\theta(F)
   =\theta^2\bigl[\pih(q_F)+\mu^{\mathrm{occ}}(F)\bigr]+o(\theta^2).
\]
The reference Kraus operators and their first derivatives determine
both limits and all these coefficients. Moreover, within any
finite-dimensional space of admissible first-order Kraus directions,
the stationary tangent variables admit a common coupling that is
real-linear in the direction (\cref{cor:jets}).

\par\medskip
\noindent\textbf{II. Clifford-specific dynamics and stationary structure.}
Fix prime local dimension $d$, set $D=d^N$, and let $S=\St$, with
uniform law $u_S$. For every intermediate unitary $R\in U(D)$, the
Clifford-driven kernel has a unique invariant law $\pi_R$ and geometric
$W_1$ mixing uniformly in $R$ at fixed $d,N,m$ (\cref{thm:global}).
For each fixed $R$, geometric TV convergence uniformly over initial
laws is equivalent to the existence of a finite nonzero rank-one Kraus product
(\cref{prop:clifford-TV-characterization}). Unitaries supported on the
measured subsystem, product unitaries, and computational-basis diagonal
unitaries admit explicit uniform TV bounds
(\cref{cor:clifford-reset-families}). The laws $\pi_R$ are Clifford
invariant, share the same first and second state moments, and satisfy
\[
  W_p(\pi_R,u_S)^p=\pi_R(r_S^p),\qquad 1\le p<\infty,
  \qquad
  W_1(\pi_R,\pi_{R'})\le C_{\mathrm{gate}}\|R-R'\|_{\op},
\]
where $C_{\mathrm{gate}}$ depends only on the fixed dimensions
(\cref{prop:stationary-structure}).

For a path $R_\theta=I_D-i\theta H+O(\theta^2)$ satisfying
Condition~\condH, write $\pi_\theta=\pi_{R_\theta}$.
The general perturbative hypotheses hold, and the affine tangent chain
is itself uniformly geometrically ergodic in TV, with constants
depending only on the reference model (\cref{prop:verification}).
Denote its invariant law by $\pih_H$ and the occupation measure by
$\mu_H^{\mathrm{occ}}$. For $p>0$, set
\[
  M_p(H):=\int_{\TX}\|v\|^p\,\pih_H(\mathrm d(s,v)),
  \qquad
  \mathcal A_M:=\left\{\sum_a h_a\Pi_a:h_a\in\R\right\},
\]
where $\Pi_a$ are the measurement projectors.
The exact energy recursion and stationary formula of \cref{thm:energy}
resolve the within-sector traceless and inter-sector components of $H$.
For $m<N$, $M_2$ is an explicit positive-definite quadratic form on
$\Herm(D)/\mathcal A_M$, with sharp Hilbert--Schmidt bounds.
In this partial-measurement regime, the stationary tangent recursion
has almost surely finite memory
with a sharp exponential tail, giving exponential integrability
(\cref{thm:tangent-exponential-moments}). Consequently, for every
$1\le p<\infty$, $M_p^{1/p}$ is a norm on the same quotient and
$W_p^{\widehat d}(\pih_H,\pih_G)\le M_p(H-G)^{1/p}$
(\cref{cor:tangent-Hamiltonian-stability}). Uniform exponential return
to $S$ also makes $\mu_H^{\mathrm{occ}}$ finite and yields exponential truncation
bounds for its occupation series
(\cref{prop:finite-Clifford-occupation}).

The exact transport identity converts the local--excursion decomposition
into three regimes. For each fixed $1\le p<\infty$, as
$\theta\downarrow0$,
\[
 W_p(\pi_\theta,u_S)^p=
 \begin{cases}
  \theta^pM_p(H)+o(\theta^p),&1\le p<2,\\[2pt]
  \theta^2\bigl[M_2(H)+\mu_H^{\mathrm{occ}}(r_S^2)\bigr]+o(\theta^2),&p=2,\\[2pt]
  \theta^2\mu_H^{\mathrm{occ}}(r_S^p)+o(\theta^2),&2<p<\infty.
 \end{cases}
\]
The displayed coefficients are respectively tangent, mixed, and purely
excursion contributions
(\cref{cor:Clifford-Wasserstein-response}). Clifford averaging also gives,
for every $f\in C^2(\X;\R)$,
\[
  \pi_\theta(f)=u_S(f)+\theta^2\mathcal L_H(f)+o(\theta^2),
\]
with an explicit local--excursion coefficient $\mathcal L_H(f)$ that is real-linear in
$f$; in particular, the first-order response vanishes
(\cref{cor:clifford-response}).

\par\medskip
\noindent\textbf{III. Stationary magic and effective perturbations.}
Gross--Wigner mana in odd prime local dimension has a linear-in-strength
stationary expansion whose coefficient is the tangent average of its
local norm (\cref{thm:mana}). In prime local dimension, stabilizer
R\'enyi entropies have expansions at scales $\theta^{2\alpha}$,
$\theta^2\ln(1/\theta)$, and $\theta^2$ for fixed finite orders
$0<\alpha<1$, $\alpha=1$, and $\alpha>1$, respectively
(\cref{thm:sre}). In particular,
\[
  \kappa_{1,\log}(H)=\frac4{\ln2}M_2(H),
  \qquad
  \kappa_\alpha^{(2)}(H)
   =\frac{2\alpha}{(\alpha-1)\ln2}M_2(H)+\mu_H^{\mathrm{occ}}(S_\alpha),
  \quad \alpha>1.
\]
For $m<N$, all these resource coefficients are strictly positive
exactly when $H\notin\mathcal A_M$. The mana coefficient defines a
norm on $\Herm(D)/\mathcal A_M$, and $\kappa_{1,\log}$ is an explicit
positive-definite quadratic form on this quotient. More generally,
the same nondegeneracy criterion applies to nonnegative observables
in the three classes whose leading local term is positive off the
zero section (\cref{cor:clifford-response}). For full measurement,
$m=N$, one has $\pi_R=u_S$ for every $R$, and all these stationary
magic averages vanish.

\section{Quantum trajectories and finite reference dynamics}\label{sec:framework}

\subsection{Kraus dynamics and stationary measures}
Let $\HH=\C^D$, $D\ge2$, and let $\X=\mathbb P(\HH)$ be its projective space. For $x\ne0$, write $[x]$ for its ray. A unit representative defines the pure-state density matrix $\rho_x=|x\rangle\langle x|$. We equip $\X$ with
\begin{equation}\label{eq:metric}
 d([x],[y])=\frac12\|\rho_x-\rho_y\|_1
          =\sqrt{1-|\langle x,y\rangle|^2},
 \qquad \|x\|=\|y\|=1.
\end{equation}
Inner products are conjugate-linear in the first argument. We write $\|\cdot\|_{\op}$ for the operator norm and $\|A\|_{\HS}=(\Tr A^\dagger A)^{1/2}$ for the Hilbert--Schmidt norm. The Wasserstein distance associated with $d$ is denoted by $W_p$. Our total-variation convention is
\[
 \|\mu-\nu\|_{\TV}=\sup_{B\in\mathcal B(\X)}|\mu(B)-\nu(B)|,
 \qquad \mu,\nu\in\mathcal P(\X).
\]
Fix $\theta_0>0$ and a finite index set $J$ labeling the possible one-step Kraus branches. We impose the following condition on the Kraus family.

\phantomsection\label{cond:kraus}
\noindent\textbf{(K)} \textit{Kraus regularity and normalization.}
For each $j\in J$, the map
$K_j:(-\theta_0,\theta_0)\to M_D(\C)$ is of class $C^2$,
and the Kraus family satisfies the normalization condition
\begin{equation}\label{eq:complete}
 \sum_{j\in J}K_j(\theta)^\dagger K_j(\theta)=I_D,
 \qquad \theta\in(-\theta_0,\theta_0).
\end{equation}
Although the Kraus family is defined on a two-sided neighborhood of zero to ensure differentiability, all perturbative limits below are taken as $\theta \downarrow 0$. For fixed $\theta$, the quantum trajectory $(X_n^\theta)_{n\ge0}$ is defined by
\[
 \begin{split}
 \PP(J_{n+1}=j\mid X_n^\theta=[x])&=\|K_j(\theta)x\|^2,\\
 X_{n+1}^\theta&=[K_{J_{n+1}}(\theta)x],
 \qquad \|x\|=1.
 \end{split}
\]
Each branch label $j$ encodes the measurement outcomes and any recorded random gate choices in a single step, and is associated with one Kraus operator. The transition kernel and averaged quantum channel are
\begin{align}
 P_\theta([x],B)
   &=\sum_{j\in J}\|K_j(\theta)x\|^2
                    \one_B([K_j(\theta)x]),\label{eq:trajectory-kernel}\\
 \Phi_\theta(A)&=\sum_{j\in J}K_j(\theta)A K_j(\theta)^\dagger.
 \label{eq:averaged-channel}
\end{align}
A term with $K_j(\theta)x=0$ is assigned the value zero. We use
\[
 (P_\theta f)(x)=\int f(y)P_\theta(x,\mathrm dy),
 \qquad
 (\mu P_\theta)(B)=\int P_\theta(x,B)\,\mu(\mathrm dx).
\]
For each fixed $\theta$, the kernel $P_\theta$ is a finite-support instance of the quantum-trajectory framework of \cite{Benoist2019}, associated with the matrix-space measure $\nu_\theta:=\sum_{j\in J}\delta_{K_j(\theta)}$.

A probability $\pi_\theta$ is stationary if $\pi_\theta P_\theta=\pi_\theta$. Such a probability exists for every $\theta$. Indeed, for $f\in C(\X)$ the map
\[
 z\longmapsto
 \begin{cases}\|z\|^2f([z]),&z\ne0,\\0,&z=0,\end{cases}
\]
is continuous. Hence $P_\theta f\in C(\X)$. On the compact space $\X$, the probabilities
\[
 \mu_n=\frac1n\sum_{k=0}^{n-1}\delta_xP_\theta^k
\]
have a weakly convergent subsequence. If $\mu_{n_l}\Rightarrow\pi_\theta$, then
\[
 \mu_n(P_\theta f-f)=\frac{P_\theta^nf(x)-f(x)}n\longrightarrow0
 \quad\Longrightarrow\quad
 \pi_\theta(P_\theta f)=\pi_\theta(f).
\]
In Sections~\ref{sec:tangent}--\ref{sec:response}, $\pi_\theta$ denotes any stationary probability of $P_\theta$.

\subsection{A complete-reset criterion for total-variation ergodicity}

The preceding Kraus formulation also admits a quantitative
criterion for total-variation ergodicity. In this subsection,
fix a normalized finite Kraus family $\{K_j\}_{j\in J}$,
write $P$ for its transition kernel, and set
$K_\omega=K_{j_\ell}\cdots K_{j_1}$ for a word
$\omega=(j_1,\ldots,j_\ell)\in J^\ell$.

For a Markov kernel $B$, its Dobrushin coefficient is
\[
 \delta(B):=\sup_{x,y\in\X}\|B(x,\cdot)-B(y,\cdot)\|_{\TV}.
\]
Consider the following conditions on a fixed normalized Kraus family.

\phantomsection\label{cond:reset}
\noindent\textbf{(R)} \textit{Complete reset family.}
There exist $\ell\ge1$, $p\in(0,1]$, and $\mathcal E\subseteq J^\ell$ such that
\begin{equation}\label{eq:complete-reset}
 \rank K_\omega\le1\quad(\omega\in\mathcal E),
 \qquad
 \sum_{\omega\in\mathcal E}K_\omega^\dagger K_\omega=pI_D.
\end{equation}
The event $\{\omega\in\mathcal E\}$ occurs with probability $p$ independently of the initial state. Conditioning on this event defines the reset kernel
\[
 B^*([x],\cdot)
 :=p^{-1}\sum_{\omega\in\mathcal E}
       \|K_\omega x\|^2\delta_{[K_\omega x]},
 \qquad \|x\|=1,
\]
where zero outputs contribute zero. Each nonzero rank-one word has an output ray independent of the initial state, although its probability may depend on that state.

\phantomsection\label{cond:overlap}
\noindent\textbf{(O)} \textit{Overlap of reset laws.}
There exists $a\in[0,1)$ such that $\delta(B^*)\le a$.
\begin{proposition}[Total-variation contraction from complete resets]
\label{prop:resets}
Let $P$ be a quantum-trajectory kernel with a finite normalized Kraus family satisfying Conditions~\condR\ and~\condO. Set
\(
 \beta:=1-p(1-a)\in[0,1).
\)
Then $P$ has a unique invariant probability measure $\pi$, and
\begin{equation}\label{eq:TV-contraction}
 \|\mu P^n-\nu P^n\|_{\TV}
 \le\beta^{\floor{n/\ell}}\|\mu-\nu\|_{\TV},
 \qquad n\ge0,\quad\mu,\nu\in\mathcal P(\X).
\end{equation}
Here $\beta^0=1$, also when $\beta=0$. In particular,
\[
 \|\mu P^n-\pi\|_{\TV}
 \le\beta^{\floor{n/\ell}}.
\]
\end{proposition}

\begin{proof}
By Kraus completeness and Condition~\condR,
\[
 P^\ell=pB^*+(1-p)\mathcal R,
\]
where, for $p<1$, the residual Markov kernel is
\[
 \mathcal R([x],B)
 :=\frac1{1-p}\sum_{\omega\notin\mathcal E}
       \|K_\omega x\|^2\mathbf1_B([K_\omega x]).
\]
The residual term is absent when $p=1$. Hence
Condition~\condO\ gives
\[
 \delta(P^\ell)
 \le p\delta(B^*)+(1-p)\delta(\mathcal R)
 \le pa+1-p=\beta.
\]

For any two probability measures $\mu,\nu$, their difference admits a Jordan decomposition $\mu-\nu=b(\alpha-\gamma)$, where $\alpha,\gamma$ are probability measures and $b=\|\mu-\nu\|_{\TV}$. Consequently,
\[
 \begin{split}
 \|\mu P^\ell-\nu P^\ell\|_{\TV}
 &\le b\int
    \|P^\ell(x,\cdot)-P^\ell(y,\cdot)\|_{\TV}
       \,\alpha(\mathrm dx)\gamma(\mathrm dy)\\
 &\le\beta\|\mu-\nu\|_{\TV}.
 \end{split}
\]
The case $\mu=\nu$ is immediate. Iteration and
nonexpansiveness of Markov kernels in total variation yield \eqref{eq:TV-contraction}. To establish existence, fix $\mu_0\in\mathcal P(\X)$. For every $k\ge0$,
\[
 \|\mu_0P^{(k+1)\ell}-\mu_0P^{k\ell}\|_{\TV}
 \le\beta^k.
\]
Thus $(\mu_0P^{k\ell})_{k\ge0}$ is Cauchy in total variation and converges to a probability measure $\pi$ satisfying $\pi P^\ell=\pi$. The strict contraction above makes this $P^\ell$-invariant probability unique. Since $\pi P$ is also $P^\ell$-invariant, it follows that $\pi P=\pi$. 
\end{proof}

\subsection{Finite reference dynamics and mean-square contraction}\label{subsec:Q}

To analyze stationary laws under weak perturbations, we introduce a finite reference set for the unperturbed kernel $P_0$, which serves as the base for the concentration and tangent-limit analysis below. We impose the following invariance and mixing condition.

\phantomsection\label{cond:reference}
\noindent\textbf{(S)} \textit{Finite invariant reference set.}
There is a nonempty finite set $S\subset\X$ such that
\[
 s\in S,\ j\in J
 \quad\Longrightarrow\quad
 K_j(0)s=0\ \text{or}\ [K_j(0)s]\in S,
\]
and the restricted kernel $Q=P_0|_S$ is primitive:
\[
 \exists n_Q\ge1\quad Q^{n_Q}(s,t)>0
 \qquad(s,t\in S).
\]

We choose a unit representative for each $s\in S$ and use the same symbol for the ray and its representative. The invariant probability of $Q$ is denoted by $\nu$; thus $\nu(s)>0$ for all $s\in S$.

Define
\begin{equation}\label{eq:distance-cone}
 E_s=I-|s\rangle\langle s|,
 \qquad r_S(x)=\min_{s\in S}d(x,s),
 \qquad \delta_S(z)=\min_{s\in S}\|E_sz\|
\end{equation}
where $x\in\X$ and $z\in\HH$. The function $\delta_S$ is the distance to $\bigcup_{s\in S}\C s$, and
\[
 |\delta_S(z)-\delta_S(w)|\le\|z-w\|,
 \qquad
 \delta_S(z)=\|z\|r_S([z])\quad(z\ne0).
\]
For a word $\omega=(j_1,\ldots,j_n)\in J^n$, set
\[
 K_{\omega,\theta}=K_{j_n}(\theta)\cdots K_{j_1}(\theta).
\]
Successive use of \eqref{eq:complete} gives
\begin{equation}\label{eq:word-completeness}
 \sum_{\omega\in J^n}K_{\omega,\theta}^\dagger K_{\omega,\theta}=I_D.
\end{equation}
Condition~\condS\ describes the reference dynamics on $S$, but does not control trajectories starting outside this set. We therefore impose a quantitative condition ensuring finite-step contraction toward $S$ under the unperturbed dynamics.

\phantomsection\label{cond:quadratic}
\noindent\textbf{(Q)} \textit{Quadratic contraction toward the reference set.}
There exist $L\ge1$ and $\eta\in[0,1)$ such that, for each
$s\in S$, one can choose $\tau(\omega,s)\in S$,
$\omega\in J^L$, with
\begin{equation}\label{eq:Q}
 \begin{split}
 E_{\tau(\omega,s)}K_{\omega,0}s&=0,\\
 \sum_{\omega\in J^L}K_{\omega,0}^\dagger
 E_{\tau(\omega,s)}K_{\omega,0}&\preceq\eta E_s.
 \end{split}
\end{equation}
The first identity specifies the target
$\tau(\omega,s)$ as the output ray whenever
$K_{\omega,0}s\ne0$; for zero-output branches,
the target is constrained by the second inequality which is equivalent to
\[
 \sum_{\omega\in J^L}
 \|E_{\tau(\omega,s)}K_{\omega,0}z\|^2
 \le \eta\|E_sz\|^2,
 \qquad z\in\HH.
\]
The left-hand side measures the total squared
transverse deviation of the unnormalized Kraus
outputs from their designated reference rays,
while the right-hand side bounds it by a strict
fraction of the squared transverse deviation
of the input from $s$.

For any $x=[z]\in\X$ with $\|z\|=1$, choose
$s\in S$ minimizing $\|E_sz\|$. By the Born rule
and the definition of $\delta_S$, we obtain
\[
 \begin{aligned}
 (P_0^L r_S^2)(x)
 &=\sum_{\omega\in J^L}
   \delta_S(K_{\omega,0}z)^2\\
 &\le\sum_{\omega\in J^L}
   \|E_{\tau(\omega,s)}K_{\omega,0}z\|^2\\
 &\le\eta\|E_sz\|^2
 =\eta r_S(x)^2.
 \end{aligned}
\]
Thus Condition~\condQ\ ensures that, after
$L$ unperturbed steps, the squared distance
to $S$ contracts uniformly in expectation,
without requiring contraction along every
individual branch. 

\subsection{Homogeneous and affine tangent kernels}\label{subsec:tangent-data}

We identify $T_s\X$ with $s^\perp$, regarded as a real Hilbert space. On $\{[x]:\langle s,x\rangle\ne0\}$, use the chart
\begin{equation}\label{eq:charts}
 \kappa_s([x])=\frac{E_sx}{\langle s,x\rangle},
 \qquad \kappa_s^{-1}(v)=[s+v].
\end{equation}
Write $K_j=K_j(0)$ and $\dot K_j=\partial_\theta K_j(0)$. Define
\[
 p_j(s)=\|K_js\|^2,
 \qquad J_+(s)=\{j\in J:p_j(s)>0\},
 \qquad J_0(s)=J\setminus J_+(s).
\]
For $j\in J_+(s)$, let $s_j=[K_js]$ and choose
$\lambda_{j,s}\ne0$ so that $K_js=\lambda_{j,s}s_j$.
Then $p_j(s)=|\lambda_{j,s}|^2$ and we define
\begin{equation}\label{eq:A-b}
 A_{j,s}v:=\lambda_{j,s}^{-1}E_{s_j}K_jv,
 \qquad b_{j,s}:=\lambda_{j,s}^{-1}E_{s_j}\dot K_js.
\end{equation}
Here $A_{j,s}:s^\perp\to s_j^\perp$ describes the linearized reference dynamics, while $b_{j,s}\in s_j^\perp$ represents the first-order transverse forcing induced by the perturbation of the Kraus family.
Both terms belong to $s_j^\perp$. On
\[
 \TX=\bigsqcup_{s\in S}(\{s\}\times s^\perp),
 \qquad V(s,v)=\|v\|,
\]
we define
\begin{align}
 \Pt((s,v),\cdot)
  &:=\sum_{j\in J_+(s)}p_j(s)\delta_{(s_j,A_{j,s}v)},\label{eq:tangent-kernels}\\
 \Ph((s,v),\cdot)
  &:=\sum_{j\in J_+(s)}p_j(s)\delta_{(s_j,A_{j,s}v+b_{j,s})}.
 \label{eq:affine-kernel}
\end{align}
The first kernel describes the linearized reference motion, while the second also incorporates the first-order forcing
induced by the perturbation. Let
$\mathrm{pr}_S:\TX\to S$ be the natural projection $\mathrm{pr}_S(s,v)=s$. Then
\[
\begin{aligned}
(\mathrm{pr}_S)_\#\Pt((s,v),\cdot)
 &=\sum_{j\in J_+(s)}p_j(s)\delta_{s_j}
 =Q(s,\cdot),\\
(\mathrm{pr}_S)_\#\Ph((s,v),\cdot)
 &=\sum_{j\in J_+(s)}p_j(s)\delta_{s_j}
 =Q(s,\cdot).
\end{aligned}
\]
Both tangent kernels induce the same reference chain $Q$ on $S$, differing in their fiber dynamics.

\section{Tangent limits of stationary quantum trajectories}\label{sec:tangent}

In this section, we establish a stationary tangent-limit theorem near the finite invariant set $S$. We first prove quantitative concentration of stationary laws around $S$ and moment contraction of the homogeneous tangent dynamics. We then construct the unique invariant law of the affine tangent kernel and prove convergence of the rescaled stationary laws to this invariant law, together with finite-dimensional distributional convergence of the rescaled stationary trajectories to the stationary affine chain as $\theta\downarrow0$.

\subsection{Stationary concentration and tangent contraction}

\begin{proposition}[Quadratic concentration and tangent contraction]\label{prop:concentration}
Under Conditions~\condK, \condS, and~\condQ, the following assertions hold.
\begin{enumerate}
\item \textit{Reference and tangent contraction.} For every $n\ge0$,
\begin{equation}\label{eq:reference-drift}
 P_0^n r_S^2\le\eta^{\floor{n/L}}r_S^2,
 \qquad
 \Pt^n V^2(s,v)\le\eta^{\floor{n/L}}\|v\|^2.
\end{equation}
\item \textit{Stationary concentration.} There exist $C<\infty$ and $\theta_*>0$ such that
\begin{equation}\label{eq:stationary-concentration}
 \pi_\theta(r_S^2)\le C\theta^2
 \qquad
 (|\theta|\le\theta_*,\ \pi_\theta P_\theta=\pi_\theta).
\end{equation}
The reference kernel $P_0$ has the unique stationary probability
\[
 \pi_0=\sum_{s\in S}\nu(s)\delta_s.
\]
\item \textit{Higher tangent moments.} There exist $p_*>2$, $C_*<\infty$, and $\rho_*\in(0,1)$ such that
\begin{equation}\label{eq:higher-tangent}
 \Pt^n V^{p_*}(s,v)\le C_*\rho_*^n\|v\|^{p_*}
 \qquad(n\ge0,\ s\in S,\ v\in s^\perp).
\end{equation}
\end{enumerate}
\end{proposition}

\begin{proof}
\emph{(i) Reference and tangent contraction.}
Fix $z\in\HH$ and choose $s\in S$ with $\delta_S(z)=\|E_sz\|$. For each branch choose a reference target $t_j$ satisfying $E_{t_j}K_js=0$. Then
\begin{align*}
 \sum_{j\in J}\delta_S(K_jz)^2
 &\le\sum_{j\in J}\|E_{t_j}K_jz\|^2\\
 &=\sum_{j\in J}\|E_{t_j}K_jE_sz\|^2\\
 &\le\sum_{j\in J}\|K_jE_sz\|^2
 =\|E_sz\|^2=\delta_S(z)^2.
\end{align*}
For words of length $L$, Condition~\condQ gives the sharper bound
\begin{align*}
 \sum_{\omega\in J^L}\delta_S(K_{\omega,0}z)^2
 &\le\sum_{\omega\in J^L}\|E_{\tau(\omega,s)}K_{\omega,0}E_sz\|^2\\
 &\le\eta\|E_sz\|^2=\eta\delta_S(z)^2.
\end{align*}
For $z\ne0$,
\begin{equation}\label{eq:lift-distance-identity}
 \|z\|^2(P_0^nr_S^2)([z])
   =\sum_{\omega\in J^n}\delta_S(K_{\omega,0}z)^2.
\end{equation}
Writing $n=kL+l$, $0\le l<L$, and iterating the two bounds proves the first inequality. 

We next consider the tangent chain. If $K_{\omega,0}s\ne0$, every successive reference branch is positive. With
\[
 K_{\omega,0}s=\lambda_{\omega,s}s_\omega,
 \qquad p_\omega(s)=|\lambda_{\omega,s}|^2,
\]
the chain rule for \eqref{eq:charts} gives
\[
 A_{\omega,s}v
  =\lambda_{\omega,s}^{-1}E_{s_\omega}K_{\omega,0}v.
\]
Thus
\begin{align*}
 (\Pt^LV^2)(s,v)
 &=\sum_{\omega:p_\omega(s)>0}
       p_\omega(s)\|A_{\omega,s}v\|^2\\
 &=\sum_{\omega:p_\omega(s)>0}
       \|E_{s_\omega}K_{\omega,0}v\|^2\\
 &\le\sum_{\omega\in J^L}
       \|E_{\tau(\omega,s)}K_{\omega,0}v\|^2
 \le\eta\|v\|^2.
\end{align*}
For any remaining length $l<L$, completeness gives
$\Pt^lV^2\le V^2$. Iteration proves the second bound in \eqref{eq:reference-drift}.

\emph{(ii) Stationary concentration.}
Choose $0<\theta_*<\theta_0$. By Condition~\condK, each word matrix $K_{\omega,\theta}$ is continuously differentiable in $\theta$. Since $J^L$ is finite, there exists a constant $C<\infty$, independent of $\theta$ and $z$, such that
\[
 \sum_{\omega\in J^L}
   \|(K_{\omega,\theta}-K_{\omega,0})z\|^2
 \le C^2\theta^2\|z\|^2,
 \qquad |\theta|\le\theta_*,\quad z\in\HH.
\]
The Lipschitz property of $\delta_S$ and Minkowski's inequality therefore give, for $\|z\|=1$,
\begin{align}
 (P_\theta^Lr_S^2([z]))^{1/2}
 &\le
 \left(\sum_{\omega\in J^L}\delta_S(K_{\omega,0}z)^2\right)^{1/2}
 +\left(\sum_{\omega\in J^L}
          \|(K_{\omega,\theta}-K_{\omega,0})z\|^2\right)^{1/2}
 \notag\\
 &\le\sqrt\eta\,r_S([z])+C|\theta|.
 \label{eq:perturbed-root-drift}
\end{align}
Set $a_\theta=(\pi_\theta(r_S^2))^{1/2}$. Stationarity and the $L^2(\pi_\theta)$ triangle inequality yield
\[
 a_\theta
 =\bigl(\pi_\theta(P_\theta^Lr_S^2)\bigr)^{1/2}
 \le\sqrt\eta\,a_\theta+C|\theta|,
 \qquad
 a_\theta\le\frac{C}{1-\sqrt\eta}|\theta|.
\]
This proves \eqref{eq:stationary-concentration}. At $\theta=0$, every invariant probability is supported on $S$. Its restriction is $Q$-invariant and hence equals $\nu$.

\emph{(iii) Higher tangent moments.}
Suppose first that $\eta>0$ and set
\[
 M_A=\max\left(1,
       \max_{s\in S,\,\omega\in J^L:\,p_\omega(s)>0}
                    \|A_{\omega,s}\|_{\op}\right).
\]
For $\varepsilon>0$,
\begin{align*}
 (\Pt^LV^{2+\varepsilon})(s,v)
 &\le M_A^\varepsilon\|v\|^\varepsilon
       \sum_{\omega:p_\omega(s)>0}
           p_\omega(s)\|A_{\omega,s}v\|^2\\
 &\le M_A^\varepsilon\eta\|v\|^{2+\varepsilon}.
\end{align*}
Choose $\varepsilon>0$ so that $\beta_*=M_A^\varepsilon\eta<1$ and put $p_*=2+\varepsilon$. The finitely many lengths $0\le l<L$ satisfy
$\Pt^lV^{p_*}\le C_0V^{p_*}$. Therefore
\[
 \Pt^nV^{p_*}\le C_0\beta_*^{\floor{n/L}}V^{p_*}
                \le C_*\rho_*^nV^{p_*}
\]
for suitable $\rho_*\in(0,1)$. If $\eta=0$, all positive $L$-word derivatives vanish, so the same conclusion holds for any fixed $p_*>2$.
\end{proof}

\subsection{Invariant law of the affine tangent chain}

\begin{proposition}[The affine stationary recursion]\label{prop:affine}
Under Conditions~\condK, \condS, and~\condQ, the kernel $\Ph$ has a unique invariant probability $\pih$. Its marginal on $S$ is $\nu$, and
\[
 \int_{\TX}\|v\|^{p_*}\,\pih(\mathrm d(s,v))<\infty
\]
for some $p_*>2$ as in \cref{prop:concentration}.
\end{proposition}
\begin{proof}
Let $(S_n,J_n)_{n\in\mathbb Z}$ be a two-sided stationary reference chain with
\[
 \PP(S_n=s)=\nu(s),\qquad
 \PP(J_n=j,S_{n+1}=t\mid S_n=s)
       =p_j(s)\one_{\{t=s_j\}}.
\]
Conditional on $S_n$, the future records are independent of the past. Set
\[
 A_n=A_{J_n,S_n},\qquad b_n=b_{J_n,S_n},\qquad
 B_0=\max_{s,j\in J_+(s)}\|b_{j,s}\|.
\]
In the fiber over $S_0$, consider
\begin{equation}\label{eq:backward}
 V_0=\sum_{k=1}^\infty A_{-1}\cdots A_{-k+1}b_{-k}.
\end{equation}

The product is the identity when $k=1$.
Conditional on the history up to time $-k+1$,
the vector $b_{-k}\in S_{-k+1}^{\perp}$ is fixed,
and the subsequent records evolve according to
the reference Markov chain. Hence, by
\eqref{eq:higher-tangent},
\[
\begin{aligned}
&\E\!\left[
 \|A_{-1}\cdots A_{-k+1}b_{-k}\|^{p_*}
 \,\middle|\,\mathcal F_{-k+1}^{-}
 \right]\\
&\quad=
 (\Pt^{k-1}V^{p_*})(S_{-k+1},b_{-k})\\
&\quad\le
 C_*\rho_*^{k-1}\|b_{-k}\|^{p_*}.
\end{aligned}
\]
Here $\mathcal F_{-k+1}^{-}$ denotes the history
through the transition into $S_{-k+1}$.
Taking expectations and using
$\|b_{-k}\|\le B_0$ gives
\[
 \E\|A_{-1}\cdots A_{-k+1}b_{-k}\|^{p_*}
 \le C_*\rho_*^{k-1}B_0^{p_*}.
\]
Consequently,
\[
 \sum_{k=1}^\infty
   \|A_{-1}\cdots A_{-k+1}b_{-k}\|_{L^{p_*}}
 \le C_*^{1/p_*}B_0\sum_{k=1}^\infty\rho_*^{(k-1)/p_*}<\infty.
\]
The series converges in $L^{p_*}$. The sum of the expected norms is finite as well, so it converges absolutely almost surely. Define its time shifts by
\[
 V_n=\sum_{k=1}^\infty A_{n-1}\cdots A_{n-k+1}b_{n-k}.
\]
By stationarity of the two-sided reference environment, the finite truncations of the backward series have time-independent joint laws with $S_n$. Passing to the $L^{p_*}$ limit gives
\[
 \mathcal L(S_n,V_n)=\mathcal L(S_0,V_0).
\]
Moreover, the series satisfies
$V_{n+1}=A_nV_n+b_n$. Since $V_n$ depends only
on the past, the conditional distribution of $J_n$
given $(S_n,V_n)$ is $p_j(S_n)$. Consequently,
for every bounded Borel function $f$ on $\TX$,
\[
 \E[f(S_{n+1},V_{n+1})\mid S_n,V_n]
 =(\Ph f)(S_n,V_n).
\]
Thus $(S_n,V_n)$ is a stationary Markov chain
with transition kernel $\Ph$, and its common law
$\pih:=\mathcal L(S_0,V_0)$ is $\Ph$-invariant.
The $L^{p_*}$ convergence also gives the stated
moment bound.

For uniqueness, let $\xi$ and $\xi'$ be two invariant probabilities of $\Ph$. Their base marginals are both $\nu$. Couple $(S_0,V_0)$ and $(S_0,V'_0)$ with a common base and use the same future records. Writing $D_n=V_n-V'_n$, we obtain
\[
 D_n=A_{n-1}\cdots A_0D_0.
\]
For $R,\varepsilon>0$, conditional mean-square contraction gives
\begin{align*}
 \PP(\|D_n\|>\varepsilon)
 &\le\PP(\|D_0\|>R)
      +\varepsilon^{-2}
          \E\bigl[\|D_n\|^2\one_{\{\|D_0\|\le R\}}\bigr]\\
 &\le\PP(\|D_0\|>R)
      +\varepsilon^{-2}R^2\eta^{\floor{n/L}}.
\end{align*}
Letting $n\to\infty$ and then $R\to\infty$ proves $D_n\to0$ in probability. For any bounded Lipschitz function $f$ on $\TX$,
\[
 |\xi(f)-\xi'(f)|
 \le\E\min\{2\|f\|_\infty,\Lip(f)\|D_n\|\}
 \longrightarrow0.
\]
Thus $\xi=\xi'$.
\end{proof}

\subsection{Stationary blow-up and tangent convergence}

Since $S$ is finite, its distinct reference rays have a positive minimum pairwise distance. We may therefore choose $\epsilon\in(0,1)$ sufficiently small that the balls
\[
 U_s=\{x:d(x,s)<\epsilon\},\qquad
 U=\bigcup_{s\in S}U_s,
\]
are pairwise disjoint, and every $x\in U_s$ has
$s$ as its unique nearest reference ray.
Adjoin a point $\dagger$ to $\TX$ and write $\TX_\dagger=\TX\sqcup\{\dagger\}$. For $\theta>0$, set
\begin{equation}\label{eq:blowup}
 B_\theta(x)=
 \begin{cases}
 (s,\theta^{-1}\kappa_s(x)),&x\in U_s,\\
 \dagger,&x\notin U,
 \end{cases}
 \qquad \pih_\theta=(B_\theta)_\#\pi_\theta.
\end{equation}
We use the complete proper metric
\[
 \begin{split}
 \widehat d((s,v),(s,w))&=\|v-w\|,\\
 \widehat d((s,v),(t,w))&=1+\|v\|+\|w\|\quad(s\ne t),\\
 \widehat d(\dagger,(s,v))&=1+\|v\|.
 \end{split}
\]
Extend $V$ by $V(\dagger)=0$ and regard $\pih$ as a probability on $\TX_\dagger$ with zero mass at $\dagger$. For $p\ge1$, we write $W_p^{\widehat d}$ for the $p$-Wasserstein distance on $(\TX_\dagger,\widehat d)$.

\begin{theorem}[Stationary tangent limit]\label{thm:tangent}
Suppose Conditions~\condK, \condS, and~\condQ\ hold. For each $0<\theta<\theta_0$, let $\pi_\theta$ be any invariant probability measure of $P_\theta$, and set $\pih_\theta=(B_\theta)_\#\pi_\theta$.
\begin{enumerate}
\item \textit{Concentration after rescaling.} There are $C<\infty$ and $\theta_*>0$ such that
\[
 \pih_\theta(\{\dagger\})\le C\theta^2,
 \qquad \pih_\theta(V^2)\le C
 \qquad(0<\theta\le\theta_*).
\]
\item \textit{Stationary distributional limit.}
The rescaled stationary laws converge to the unique invariant law of the affine tangent kernel, with Wasserstein convergence at every order below two:
\begin{equation}\label{eq:tangent-limit}
 \pih_\theta\Rightarrow\pih,
 \qquad
 W_p^{\widehat d}(\pih_\theta,\pih)\longrightarrow0
 \quad(1\le p<2),\qquad \theta\downarrow0.
\end{equation}
\item \textit{Stationary process limit.} 
For $T\in\mathbb N_0$, let $(X_n^\theta)_{n\ge0}$ be a stationary Markov chain with transition kernel $P_\theta$ and marginal law $\pi_\theta$, and let $(S_n,V_n)_{n\ge0}$ be the nonnegative-time restriction of the stationary affine chain constructed in Proposition~\ref{prop:affine}. Then
\[
 \mathcal L\bigl(B_\theta X_0^\theta,\ldots,B_\theta X_T^\theta\bigr)
 \Rightarrow
 \mathcal L\bigl((S_0,V_0),\ldots,(S_T,V_T)\bigr).
\]
\end{enumerate}
\end{theorem}
\begin{proof}
\emph{(i) Concentration and tightness.}
For $x\in U_s$, write $w=\kappa_s(x)$. Then
\[
 r_S(x)=\frac{\|w\|}{\sqrt{1+\|w\|^2}},
 \qquad
 \|w\|^2\le\frac{r_S(x)^2}{1-\epsilon^2}.
\]
Since $r_S\ge\epsilon$ on $U^c$, \cref{prop:concentration} yields
\begin{align*}
 \pih_\theta(\{\dagger\})
 &=\pi_\theta(U^c)
 \le\epsilon^{-2}\pi_\theta(r_S^2)\le C\theta^2,\\
 \pih_\theta(V^2)
 &=\theta^{-2}\sum_s\int_{U_s}\|\kappa_s(x)\|^2\,\pi_\theta(\mathrm dx)
 \le C\theta^{-2}\pi_\theta(r_S^2)\le C.
\end{align*}
Thus, for $\mathcal K_R=\{(s,v)\in\TX:\|v\|\le R\}$,
\begin{equation}\label{eq:tangent-tail}
 \pih_\theta(\TX_\dagger\setminus\mathcal K_R)
       \le CR^{-2}+C\theta^2.
\end{equation}
Since $S$ is finite and each tangent fiber is
finite-dimensional, $\mathcal K_R\cup\{\dagger\}$ is compact. Moreover,
\[
 \pih_\theta\bigl(
 \TX_\dagger\setminus(\mathcal K_R\cup\{\dagger\})
 \bigr)\le CR^{-2},
\]
so the family $\{\pih_\theta:0<\theta\le\theta_*\}$
is tight. Since $\dagger$ is isolated and
$\pih_\theta(\{\dagger\})\to0$, every weak limit
is supported on $\TX$.

\emph{(ii) Stationary distributional limit.}
For $\|v\|\le R$,
\[
 K_j(\theta)(s+\theta v)
 =K_js+\theta(K_jv+\dot K_js)+O_R(\theta^2).
\]
If $j\in J_+(s)$, division by the chart denominator gives
\begin{equation}\label{eq:branch-limit}
 \begin{split}
 \frac{\|K_j(\theta)(s+\theta v)\|^2}{1+\theta^2\|v\|^2}
   &=p_j(s)+O_R(\theta),\\
 \theta^{-1}\kappa_{s_j}([K_j(\theta)(s+\theta v)])
   &=A_{j,s}v+b_{j,s}+O_R(\theta).
 \end{split}
\end{equation}
The remainders are uniform over $s$ and $j$ because
$\min_{s,j\in J_+(s)}|\lambda_{j,s}|>0$. If $j\in J_0(s)$, the branch probability is $O_R(\theta^2)$. Hence, for $f\in C_b(\TX_\dagger)$,
\begin{equation}\label{eq:rescaled-transition-limit}
 \sup_{(s,v)\in\mathcal K_R}
 \left|P_\theta(f\circ B_\theta)([s+\theta v])
                  -(\Ph f)(s,v)\right|\longrightarrow0.
\end{equation}
The positive-branch outputs lie in a fixed compact set, on which $f$ is uniformly continuous; the zero-branch sum is bounded by $O_R(\theta^2)\|f\|_\infty$. Suppose $\pih_{\theta_l}\Rightarrow\xi$. For $f\in C_c(\TX)$, extend $f$ and $\Ph f$ by zero at $\dagger$. Stationarity gives
\[
 \pih_\theta(f)
 =\int P_\theta(f\circ B_\theta)(x)\,\pi_\theta(\mathrm dx).
\]
Using \eqref{eq:rescaled-transition-limit} on $\mathcal K_R$ and \eqref{eq:tangent-tail} on its complement, we obtain
\[
 \limsup_{l\to\infty}
 |\pih_{\theta_l}(f)-\pih_{\theta_l}(\Ph f)|
       \le 2C\|f\|_\infty R^{-2}.
\]
The function $\Ph f$ is bounded and continuous. Letting $l\to\infty$ and $R\to\infty$ gives
\[
 \xi(f)=\xi(\Ph f),\qquad f\in C_c(\TX).
\]
These tests determine Radon measures, so $\xi$ is invariant. By \cref{prop:affine}, $\xi=\pih$. Every convergent subsequence has this limit, proving weak convergence.

For $1\le p<2$,
\[
 \int_{V>R}V^p\,\mathrm d\pih_\theta
       \le R^{p-2}\pih_\theta(V^2)\le CR^{p-2}.
\]
The same estimate holds for $\pih$. Let $T_R$ truncate each tangent vector radially to the ball of radius $R$ and leave $\dagger$ fixed. Then
\[
 W_p^{\widehat d}(\pih_\theta,(T_R)_\#\pih_\theta)^p
       \le CR^{p-2}.
\]
On the compact image of $T_R$, weak convergence implies Wasserstein convergence. The triangle inequality, followed by $R\to\infty$, proves the second assertion of \eqref{eq:tangent-limit}.

\emph{(iii) Stationary process limit.}
Stationarity and the union bound give
\[
 \PP\left(\exists n\le T:\ B_\theta X_n^\theta\notin\mathcal K_R\right)
       \le (T+1)(CR^{-2}+C\theta^2).
\]
Fix $T\in\mathbb N_0$ and
$F\in C_b(\TX_\dagger^{T+1})$.
For $(s,v)\in\TX$, write
\[
\begin{aligned}
G_{\theta,T}(s,v)
&:=\E_{[s+\theta v]}
 F(B_\theta X_0^\theta,\ldots,
   B_\theta X_T^\theta),\\
G_T(s,v)
&:=\E_{(s,v)}
 F(Y_0,\ldots,Y_T),
\end{aligned}
\]
where $(Y_n)_{n\ge0}$ is the Markov chain with
transition kernel $\Ph$.

For fixed $R$ and $T$, the positive-reference-probability
branch expansions in \eqref{eq:branch-limit} can be
iterated uniformly over $\mathcal K_R$: their
tangent coordinates remain bounded over the
finite time window, while paths containing a
zero-reference-probability branch have total
probability $O_{R,T}(\theta^2)$. Since the branch
set is finite, it follows that
\[
 \sup_{(s,v)\in\mathcal K_R}
 |G_{\theta,T}(s,v)-G_T(s,v)|
 \longrightarrow0.
\]
Moreover, $G_T$ is bounded and continuous.
Combining this local uniform convergence with
$\pih_\theta\Rightarrow\pih$, using compact
cutoffs and the uniform tail bound
\eqref{eq:tangent-tail}, gives
\[
 \E F(B_\theta X_0^\theta,\ldots,
       B_\theta X_T^\theta)
 \longrightarrow
 \E F((S_0,V_0),\ldots,(S_T,V_T)).
\]
This proves the finite-dimensional distributional convergence.
\end{proof}

\section{Stationary response of observables}\label{sec:response}

For observables vanishing on $S$, the leading response depends on how their local asymptotic scale compares with the $\theta^2$ scale of rare departures. We distinguish subquadratic, logarithmically enhanced quadratic, and ordinary quadratic behavior.

\subsection{Observable classes and subquadratic response}

For any bounded Borel function $F:\X\to\R$, define
\begin{equation}\label{eq:lift}
 \Gamma_F(z)=
 \begin{cases}\|z\|^2F([z]),&z\ne0,\\0,&z=0.\end{cases}
\end{equation}
This homogeneous extension incorporates the Born probabilities:
\begin{equation}\label{eq:lift-intertwining}
 \Gamma_{P_\theta^nF}(z)
       =\sum_{\omega\in J^n}\Gamma_F(K_{\omega,\theta}z).
\end{equation}
Throughout the following definitions, let $F:\X\to\R$ and $\mathcal K_R=\{(s,v)\in\TX:\|v\|\le R\}$.

\begin{definition}[Subquadratic observables]\label{def:subquadratic}
Let $0<k<2$. A bounded Borel function $F$ is a subquadratic observable of order $k$ if $F|_S=0$ and there are $f_k\in C(\TX)$ and $C_F<\infty$ such that
\begin{equation}\label{eq:subquadratic-class}
 \theta^{-k}F([s+\theta v])\longrightarrow f_k(s,v),
 \qquad |F(x)|\le C_Fr_S(x)^k\quad(x\in\X),
\end{equation}
with convergence uniform on every $\mathcal K_R$. We call $f_k$ its leading local term.
\end{definition}

The local limit and the global bound in
\eqref{eq:subquadratic-class} imply that $f_k$
is positively homogeneous of degree $k$ and satisfies
\[
 f_k(s,tv)=t^kf_k(s,v)\quad(t>0),
 \qquad |f_k(s,v)|\le C_F\|v\|^k.
\]

\begin{definition}[Ordinary quadratic observables]\label{def:quadratic}
An ordinary quadratic observable is a globally Lipschitz function $F$ that is $C^2$ on a neighborhood of $S$ and satisfies
\[
 F(s)=0,\qquad \mathrm dF_s=0\qquad(s\in S).
\]
We define its quadratic local term by
\begin{equation}\label{eq:quadratic-germ}
 q_F(s,v)
 :=\frac12D^2(F\circ\kappa_s^{-1})(0)[v,v].
\end{equation}

\end{definition}

By Taylor's theorem and the vanishing of the zeroth- and first-order terms,
\[
 F([s+\theta v])
 =\theta^2q_F(s,v)+o_R(\theta^2),
\]
uniformly on every $\mathcal K_R$ as $\theta\downarrow0$.

\begin{definition}[Logarithmic quadratic observables]\label{def:logarithmic}
Let $\zeta>0$. A logarithmic quadratic observable of exponent $\zeta$ is a function $F\in C^1(\X)$ with $F|_S=0$ satisfying both of the following properties.
\begin{enumerate}
\item There is $q_{\log}\in C(\TX)$ such that
\begin{equation}\label{eq:log-class}
 \frac{F([s+\theta v])}{\theta^2[\ln(1/\theta)]^\zeta}
       \longrightarrow q_{\log}(s,v)
\end{equation}
uniformly on every $\mathcal K_R$.

\item There exists a constant $C_F>0$ such that
\begin{equation}\label{eq:log-gradient}
 \|\nabla\Gamma_F(z)\|
 \le C_F\delta_S(z)
       \left[\ln\frac{e\|z\|}{\delta_S(z)}\right]^\zeta,
 \qquad z\in\HH\setminus\{0\}.
\end{equation}
Here $\nabla$ denotes the gradient with respect to the real Euclidean structure on $\HH\simeq\R^{2D}$. When $\delta_S(z)=0$, the right-hand side is understood by continuous extension and hence equals zero.
\end{enumerate}
\end{definition}

Again, a change of scale in \eqref{eq:log-class} gives
\[
 q_{\log}(s,tv)=t^2q_{\log}(s,v)\quad(t>0),
 \qquad |q_{\log}(s,v)|\le C\|v\|^2.
\]

\begin{proposition}[Subquadratic response]\label{cor:subquadratic}
Suppose Conditions~\condK, \condS, and~\condQ\ hold, and let $F$ be a subquadratic observable of order $0<k<2$ in \cref{def:subquadratic}. For each sufficiently small $\theta>0$, let $\pi_\theta$ be any invariant probability measure of $P_\theta$. Then, as $\theta\downarrow0$,
\begin{equation}\label{eq:subquadratic-response}
 \pi_\theta(F)=\theta^k\pih(f_k)+o(\theta^k).
\end{equation}
\end{proposition}

\begin{proof}
For $\theta>0$, define $F_\theta$ on $\TX_\dagger$ by
\[
 F_\theta(s,v)=\theta^{-k}F([s+\theta v]),
 \qquad F_\theta(\dagger)=0,
\]
and extend $f_k$ by $f_k(\dagger)=0$.
By the definition of $B_\theta$,
\begin{equation}\label{eq:subquadratic-decomposition}
 \theta^{-k}\pi_\theta(F)
 =\int_{\TX_\dagger}F_\theta\,\mathrm d\pih_\theta
  +\theta^{-k}\int_{U^c}F\,\mathrm d\pi_\theta.
\end{equation}
The global bound in \eqref{eq:subquadratic-class}
and $r_S([s+\theta v])\le d([s+\theta v],s)$ give
\[
 |F_\theta(s,v)|
 \le C_F\frac{\|v\|^k}
 {(1+\theta^2\|v\|^2)^{k/2}}
 \le C_F\|v\|^k.
\]
The same growth bound holds for $f_k$.

Choose a continuous function
$\chi:[0,\infty)\to[0,1]$ such that
$\chi=1$ on $[0,1]$ and $\chi=0$ on $[2,\infty)$,
and set
\[
 \chi_R(s,v)=\chi(\|v\|/R),
 \qquad \chi_R(\dagger)=0.
\]
For each fixed $R>0$, the local uniform convergence
in \eqref{eq:subquadratic-class} implies
\[
 \sup_{\TX_\dagger}
 |\chi_R(F_\theta-f_k)|\longrightarrow0.
\]
Since $\chi_Rf_k\in C_c(\TX_\dagger)$,
the weak convergence in \cref{thm:tangent} yields
\[
 \int\chi_RF_\theta\,\mathrm d\pih_\theta
 \longrightarrow
 \int\chi_Rf_k\,\mathrm d\pih.
\]

To remove the cutoff, the uniform second-moment
bound in \cref{thm:tangent} gives
\[
 \begin{aligned}
 \int_{V>R}|F_\theta|\,\mathrm d\pih_\theta
 &\le C_F\int_{V>R}V^k\,\mathrm d\pih_\theta\\
 &\le C_FR^{k-2}\pih_\theta(V^2)
 \le CR^{k-2}.
 \end{aligned}
\]
The same estimate holds for $f_k$ under $\pih$.
Moreover, the cemetery estimate gives
\[
 \theta^{-k}
 \left|\int_{U^c}F\,\mathrm d\pi_\theta\right|
 \le \|F\|_\infty\theta^{-k}\pi_\theta(U^c)
 \le C\theta^{2-k}.
\]
Since $0<k<2$, combining these estimates with
\eqref{eq:subquadratic-decomposition} yields
\[
 \limsup_{\theta\downarrow0}
 \left|
 \theta^{-k}\pi_\theta(F)-\pih(f_k)
 \right|
 \le CR^{k-2}.
\]
Letting $R\to\infty$ completes the proof.
\end{proof}

\subsection{Poisson estimates and logarithmic quadratic response}

Having established the subquadratic response using the stationary tangent limit, we now turn to the remaining two classes of observables. Their response analysis relies on the reference Poisson potential $u_F=\sum_{n\ge0}P_0^nF$. The following lemma establishes the convergence and regularity of this series, identifies its local rescaled limits, and provides estimates for the perturbation increment $(P_\theta-P_0)u_F$.

For the estimates below, we write
\[
 \operatorname{lip}G(z)
 :=\limsup_{w\to z,\,w\ne z}
   \frac{|G(w)-G(z)|}{\|w-z\|}
\]
for a locally Lipschitz function $G$ on $\HH$, and set $\ell(t):=1+\ln^+(1/t)$ for $t>0$, where $\ln^+(x):=\max\{\ln x,0\}$.

\begin{lemma}[Reference potentials]\label{lem:poisson}
Suppose Conditions~\condK, \condS, and~\condQ\ hold. Let $F$ be either an ordinary quadratic observable in \cref{def:quadratic} or a logarithmic quadratic observable of exponent $\zeta>0$ in \cref{def:logarithmic}. In the ordinary quadratic case, set $\zeta=0$ for notational uniformity. Then the following assertions hold.

\begin{enumerate}
\item \textit{Poisson potential and decay estimates.}
The series
\begin{equation}\label{eq:poisson}
 u_F=\sum_{n=0}^\infty P_0^nF
\end{equation}
converges uniformly on $\X$ to a Lipschitz function satisfying
\[
 u_F|_S=0,\qquad (I-P_0)u_F=F.
\]
There exist $C_1>0$ and $\gamma\in(0,1)$ such that,
for every $n\ge0$ and $z\in\HH\setminus\{0\}$,
\begin{align}
 |\Gamma_{P_0^nF}(z)|
 &\le C_1\gamma^n\delta_S(z)^2
          \ell\left(\frac{\delta_S(z)}{\|z\|}\right)^\zeta,
          \label{eq:potential-value}\\
 \operatorname{lip}\Gamma_{P_0^nF}(z)
 &\le C_1\gamma^n\delta_S(z)
          \ell\left(\frac{\delta_S(z)}{\|z\|}\right)^\zeta.
          \label{eq:potential-slope}
\end{align}
Moreover, there exists $C_2>0$ such that
\[
 \begin{aligned}
 |\Gamma_{u_F}(z)|
 &\le C_2\delta_S(z)^2
          \ell\left(\frac{\delta_S(z)}{\|z\|}\right)^\zeta,\\
 \operatorname{lip}\Gamma_{u_F}(z)
 &\le C_2\delta_S(z)
          \ell\left(\frac{\delta_S(z)}{\|z\|}\right)^\zeta.
 \end{aligned}
\]
In these inequalities, products involving
$\delta_S(z)^m\ell(\delta_S(z)/\|z\|)^\zeta$,
$m\in\{1,2\}$, are understood by continuous extension
and equal zero when $\delta_S(z)=0$.

\item \textit{Local limits of the potential.}
There exist a continuous function $h_F:\TX\to\R$,
positively homogeneous of degree two in the tangent variable,
and a constant $C_3>0$ with the following properties.
For an ordinary quadratic observable,
\begin{equation}\label{eq:h-quadratic}
 \theta^{-2}u_F([s+\theta v])\longrightarrow h_F(s,v),
 \qquad |h_F(s,v)|\le C_3\|v\|^2.
\end{equation}
Define the zero-reference-branch source term by
\[
 (Z^0u_F)(s,v)
 :=\sum_{j\in J_0(s)}\Gamma_{u_F}(K_jv).
\]
Then $h_F$ satisfies
\begin{equation}\label{eq:h-poisson}
 h_F=q_F+\Pt h_F+Z^0u_F.
\end{equation}
For a logarithmic quadratic observable,
\begin{equation}\label{eq:h-log}
 \begin{split}
 \frac{u_F([s+\theta v])}
 {\theta^2[\ln(1/\theta)]^\zeta}
 &\longrightarrow
 h_F(s,v)=\sum_{n=0}^\infty\Pt^nq_{\log}(s,v),\\
 (I-\Pt)h_F&=q_{\log},
 \qquad |h_F(s,v)|\le C_3\|v\|^2.
 \end{split}
\end{equation}
Both limits are uniform on every $\mathcal K_R$
as $\theta\downarrow0$.

\item \textit{Perturbation increment.}
There exist $0<\theta_*<\min\{\theta_0,e^{-1}\}$ and
constants $C_4,C_5>0$ such that, for all
$0<\theta\le\theta_*$ and $x\in\X$,
\begin{equation}\label{eq:increment-bound}
 |(P_\theta-P_0)u_F(x)|
 \le C_4\theta\bigl(r_S(x)+\theta\bigr)
       \ell\bigl(r_S(x)+\theta\bigr)^\zeta.
\end{equation}
Set $a_\theta:=\theta^2[\ln(1/\theta)]^\zeta$.
For every choice of invariant probability measures
$\pi_\theta$ of $P_\theta$,
\begin{equation}\label{eq:rescaled-increment-bound}
 \begin{split}
 a_\theta^{-1}|(P_\theta-P_0)u_F([s+\theta v])|
 &\le C_5(1+\|v\|),
 \qquad (s,v)\in B_\theta(U),\\
 a_\theta^{-1}\int_{U^c}|(P_\theta-P_0)u_F|
 \,\mathrm d\pi_\theta
 &\longrightarrow0
 \qquad(\theta\downarrow0).
 \end{split}
\end{equation}
\end{enumerate}
\end{lemma}

\begin{proof}
\emph{(i) Poisson potential and decay estimates.}
We first establish the bounds for $\Gamma_F$.
For ordinary quadratic $F$, Taylor's theorem and the vanishing
of $F$ and its first derivative on $S$ give
$|F(x)|\le Cr_S(x)^2$ and a local $d$-Lipschitz slope bound
of order $Cr_S(x)$ near $S$. Since $F$ is globally
Lipschitz and $\X$ is compact, both estimates extend to all
of $\X$ after increasing $C$. Applying these bounds to the
homogeneous lift \eqref{eq:lift} yields
\[
 |\Gamma_F(z)|\le C\delta_S(z)^2,
 \qquad \operatorname{lip}\Gamma_F(z)\le C\delta_S(z).
\]

For logarithmic quadratic $F$, choose $s\in S$ with
$\delta_S(z)=\|E_sz\|$ and set
$z_t=(I-E_s)z+tE_sz$ for $0\le t\le1$.
Then $\Gamma_F(z_0)=0$,
$\delta_S(z_t)\le t\delta_S(z)$, and
$\|z_t\|\le\|z\|$. For $c>\zeta+1$, the function
$a\mapsto a[c+\ln(\|z\|/a)]^\zeta$ is increasing on
$(0,\|z\|]$. Thus \eqref{eq:log-gradient}, with a harmless
shift of the logarithm, and integration along $z_t$ give
\[
\begin{aligned}
 |\Gamma_F(z)|
 &\le C\delta_S(z)^2\int_0^1
 t\left[c+\ln\frac{\|z\|}{t\delta_S(z)}\right]^\zeta\,\mathrm dt\\
 &\le C\delta_S(z)^2
 \ell\left(\frac{\delta_S(z)}{\|z\|}\right)^\zeta.
\end{aligned}
\]
The corresponding slope bound is \eqref{eq:log-gradient}.
At $z=0$ both types of $\Gamma_F$ have local Lipschitz
slope zero, since $|\Gamma_F(z)|\le\|F\|_\infty\|z\|^2$.
Hence in either case, for $z\ne0$,
\[
\begin{aligned}
 |\Gamma_F(z)|
 &\le C\delta_S(z)^2
    \ell\left(\frac{\delta_S(z)}{\|z\|}\right)^\zeta,\\
 \operatorname{lip}\Gamma_F(z)
 &\le C\delta_S(z)
    \ell\left(\frac{\delta_S(z)}{\|z\|}\right)^\zeta.
\end{aligned}
\]
All expressions at $\delta_S(z)=0$ are understood by continuous
extension.

We use the following elementary logarithmic summation estimate.
For finite arrays $0\le a_i\le b_i$, put
$A^2=\sum_i a_i^2$ and $B^2=\sum_i b_i^2$.
For every $t\ge0$,
\begin{equation}\label{eq:log-array}
 \sum_i a_i^2\left[\ln\frac{e b_i}{a_i}\right]^t
 \le C_t A^2\left[\ln\frac{eB}{A}\right]^t,
\end{equation}
with zero terms interpreted by continuity.
Indeed, for $c>t+1$, the function
$\psi(x)=x(c-\frac12\ln x)^t$, extended by $\psi(0)=0$,
is increasing and concave on $[0,1]$.
Jensen's inequality with weights $b_i^2/B^2$ yields
$\sum_i b_i^2\psi(a_i^2/b_i^2)
 \le B^2\psi(A^2/B^2)$; comparison of the shifted
logarithms proves \eqref{eq:log-array}.
Cauchy--Schwarz and \eqref{eq:log-array} with $t=2\zeta$
further give, for $c_i\ge0$,
\begin{equation}\label{eq:log-array-cross}
 \sum_i a_i c_i\left[\ln\frac{e b_i}{a_i}\right]^\zeta
 \le C_\zeta A\left[\ln\frac{eB}{A}\right]^\zeta
       \left(\sum_i c_i^2\right)^{1/2}.
\end{equation}
The cases $A=0$ or $B=0$ follow by continuity.

For $z,w\in\HH$ and words $\omega\in J^n$, set
$a_\omega=\delta_S(K_{\omega,0}z)$,
$b_\omega=\|K_{\omega,0}z\|$, and
$c_\omega=\|K_{\omega,0}w\|$.
Kraus completeness and \eqref{eq:reference-drift} imply
\begin{equation}\label{eq:reference-array-bounds}
 \sum_\omega b_\omega^2=\|z\|^2,
 \qquad \sum_\omega c_\omega^2=\|w\|^2,
 \qquad \sum_\omega a_\omega^2
       \le\eta^{\floor{n/L}}\delta_S(z)^2.
\end{equation}
Apply these initial bounds to
\eqref{eq:lift-intertwining}, and then
\eqref{eq:log-array}--\eqref{eq:log-array-cross}.
Writing $A_n=(\sum_\omega a_\omega^2)^{1/2}$,
we obtain, for $z\ne0$,
\[
\begin{aligned}
 |\Gamma_{P_0^nF}(z)|
 &\le C A_n^2\left[\ln\frac{e\|z\|}{A_n}\right]^\zeta,\\
 \operatorname{lip}\Gamma_{P_0^nF}(z)
 &\le C A_n\left[\ln\frac{e\|z\|}{A_n}\right]^\zeta.
\end{aligned}
\]
For the slope estimate, the finite Kraus sum and the local Lipschitz property give
\[
\begin{aligned}
 \operatorname{lip}\Gamma_{P_0^nF}(z)
 &\le \sup_{\|w\|=1}
 \sum_{\omega\in J^n}
 \operatorname{lip}\Gamma_F(K_{\omega,0}z)
 \|K_{\omega,0}w\|\\
 &\le C A_n
 \left[\ln\frac{e\|z\|}{A_n}\right]^\zeta,
\end{aligned}
\]
where the last inequality follows from \eqref{eq:log-array-cross} and \eqref{eq:reference-array-bounds}.
 If $0<\eta<1$, the last bound in
\eqref{eq:reference-array-bounds} and the boundedness of
$x^m(1+\ln(1/x))^\zeta$ on $(0,1]$ ($m=1,2$) yield
\[
\begin{aligned}
 |\Gamma_{P_0^nF}(z)|
 &\le C\eta^{\floor{n/L}}\delta_S(z)^2
   \left[\ell\left(\frac{\delta_S(z)}{\|z\|}\right)+C(1+n)\right]^\zeta,\\
 \operatorname{lip}\Gamma_{P_0^nF}(z)
 &\le C\eta^{\floor{n/L}/2}\delta_S(z)
   \left[\ell\left(\frac{\delta_S(z)}{\|z\|}\right)+C(1+n)\right]^\zeta.
\end{aligned}
\]
Since $\ell\ge1$, the polynomial factor can be absorbed into a
geometric rate $\gamma\in(0,1)$, proving
\eqref{eq:potential-value}--\eqref{eq:potential-slope}
with a common constant $C_1$. If $\eta=0$,
\eqref{eq:reference-array-bounds} makes
$\Gamma_{P_0^nF}$ identically zero for $n\ge L$;
the same conclusion follows from the finitely many remaining terms.

On the unit sphere, both right-hand sides are bounded by
$C\gamma^n$. Integrating the local slope bound along line
segments joining suitably phase-aligned unit representatives
(which stay uniformly away from zero) gives
$\|P_0^nF\|_\infty+\operatorname{Lip}(P_0^nF)
 \le C\gamma^n$.
Consequently \eqref{eq:poisson} converges in the Lipschitz norm.
Because $P_0$ preserves $S$, all summands vanish on $S$,
and the identity
\[
 (I-P_0)\sum_{n=0}^{N}P_0^nF=F-P_0^{N+1}F
\]
proves $(I-P_0)u_F=F$.
Summing \eqref{eq:potential-value}--\eqref{eq:potential-slope}
proves the stated estimates for $\Gamma_{u_F}$ with $C_2$.

\emph{(ii) Local limits of the potential.}
First let $F$ be ordinary quadratic. For each fixed $n\ge0$,
\eqref{eq:lift-intertwining} and the local Taylor expansion of
$F$ give, uniformly on each $\mathcal K_R$,
\begin{equation}\label{eq:finite-profile}
\begin{split}
 h_n(s,v)
 &:=\lim_{\theta\downarrow0}
       \theta^{-2}P_0^nF([s+\theta v])\\
 &=\sum_{\omega:K_{\omega,0}s\ne0}
       p_\omega(s)q_F(s_\omega,A_{\omega,s}v)
   +\sum_{\omega:K_{\omega,0}s=0}
       \Gamma_F(K_{\omega,0}v).
\end{split}
\end{equation}
Indeed, the positive-word terms follow by expanding in the
output chart; for a zero word, the degree-two homogeneity of
$\Gamma_F$ yields
$\Gamma_F(K_{\omega,0}(s+\theta v))
 =\theta^2\Gamma_F(K_{\omega,0}v)$.
The global bound $|F|\le Cr_S^2$ and
\eqref{eq:reference-drift} imply
\[
 \theta^{-2}|P_0^nF([s+\theta v])|
 \le C\eta^{\floor{n/L}}
       \frac{\|v\|^2}{1+\theta^2\|v\|^2}
 \le C\eta^{\floor{n/L}}\|v\|^2.
\]
Thus the finite-word limits may be summed uniformly on every
$\mathcal K_R$, defining
$h_F=\sum_{n\ge0}h_n$.
Each $h_n$ is continuous and positively homogeneous of degree
two; hence so is $h_F$, and $|h_F(s,v)|\le C_3\|v\|^2$.
Taking the rescaled limit of $u_F=F+P_0u_F$ now gives
\eqref{eq:h-poisson}: positive reference branches contribute
$\Pt h_F$, while the branches $K_js=0$ contribute
$\Gamma_{u_F}(K_jv)$, namely $Z^0u_F$.

Next let $F$ be logarithmic quadratic. For fixed $n$, the
positive reference words in \eqref{eq:lift-intertwining} have
rescaled limit $\Pt^nq_{\log}$, by \eqref{eq:log-class}.
For a word with $K_{\omega,0}s=0$, the rescaled term equals
\[
 \frac{\Gamma_F(K_{\omega,0}v)}
 {(1+\theta^2\|v\|^2)[\ln(1/\theta)]^\zeta}
 \longrightarrow0.
\]
Moreover, \eqref{eq:potential-value} implies, for every fixed
$R$ and all sufficiently small $\theta>0$,
\[
 \sup_{(s,v)\in\mathcal K_R}
 \frac{|P_0^nF([s+\theta v])|}
      {\theta^2[\ln(1/\theta)]^\zeta}
 \le C_R\gamma^n.
\]
Indeed, for each fixed $R>0$, one has $\delta_S(s+\theta v)=\theta\|v\|$ on $\mathcal K_R$ for all sufficiently small $\theta>0$. Moreover,
\[
 \sup_{\substack{0<\theta\le\theta_R\\0<t\le R}}
 t^2\left[
 \frac{\ell(\theta t)}{\ln(1/\theta)}
 \right]^\zeta<\infty
\]
for sufficiently small $\theta_R>0$, with the
expression at $t=0$ understood by continuous extension as zero. Thus \eqref{eq:potential-value} provides a geometric majorant independent of $\theta$, allowing
the locally uniform limits to be summed:
\[
 h_F=\sum_{n\ge0}\Pt^nq_{\log}.
\]
The degree-two homogeneity and bound
$|q_{\log}(s,v)|\le C\|v\|^2$, together with
\eqref{eq:reference-drift}, give absolute locally uniform
convergence, $|h_F(s,v)|\le C_3\|v\|^2$, and
$(I-\Pt)h_F=q_{\log}$. This proves \eqref{eq:h-log}.

\emph{(iii) Perturbation increment.}
Choose $\theta_*>0$ sufficiently small, with
$\theta_*<\min\{\theta_0,e^{-1}\}$.
For a unit representative $x$ of a ray in $\X$, put
\[
 \Delta_j=(K_j(\theta)-K_j)x,
 \qquad z_j(t)=K_jx+t\Delta_j,
 \qquad 0\le t\le1.
\]
By \condK\ and finiteness of $J$, there is $C_K<\infty$
such that
$\sum_j\|\Delta_j\|^2\le C_K^2\theta^2$.
Kraus completeness, the triangle inequality, and the one-step
case of \eqref{eq:reference-drift} give, uniformly in $t$,
\[
\begin{aligned}
 \frac12\le\left(\sum_j\|z_j(t)\|^2\right)^{1/2}&\le2,\\
 \left(\sum_j\delta_S(z_j(t))^2\right)^{1/2}
 &\le r_S([x])+C_K\theta.
\end{aligned}
\]
The function $\Gamma_{u_F}$ is locally Lipschitz, including
at the origin. Integrating its local slope along each segment
$z_j(t)$ and applying \eqref{eq:log-array-cross} to the
$C_2$-bound in part (i), we obtain
\[
\begin{aligned}
 |(P_\theta-P_0)u_F([x])|
 &\le\sum_j\int_0^1
 \operatorname{lip}\Gamma_{u_F}(z_j(t))\|\Delta_j\|\,\mathrm dt\\
 &\le C\theta\int_0^1
 A(t)\left[\ln\frac{eB(t)}{A(t)}\right]^\zeta\,\mathrm dt,
\end{aligned}
\]
where
$A(t)^2=\sum_j\delta_S(z_j(t))^2$ and
$B(t)^2=\sum_j\|z_j(t)\|^2$, with the integrand
interpreted as zero if $A(t)=0$.
Since $A(t)\le r_S([x])+C_K\theta$ and
$1/2\le B(t)\le2$, the same elementary monotonicity
estimate used in part (i) yields
\[
 |(P_\theta-P_0)u_F([x])|
 \le C_4\theta\bigl(r_S([x])+\theta\bigr)
          \ell\bigl(r_S([x])+\theta\bigr)^\zeta,
\]
which is \eqref{eq:increment-bound}. For $0<\theta\le\theta_*$, reducing $\theta_*$ if necessary,
$\ell(r_S(x)+\theta)\le\ell(\theta)
 \le2\ln(1/\theta)$.
Thus, with $a_\theta=\theta^2[\ln(1/\theta)]^\zeta$,
\[
 a_\theta^{-1}|(P_\theta-P_0)u_F(x)|
 \le C_5\left(1+\frac{r_S(x)}{\theta}\right).
\]
If $(s,v)\in B_\theta(U)$, then
$r_S([s+\theta v])\le\theta\|v\|$, proving the
first bound in \eqref{eq:rescaled-increment-bound}.
On $U^c$, use $r_S\le1$ and the uniform cemetery estimate
$\pi_\theta(U^c)\le C\theta^2$ from
\cref{thm:tangent} to obtain
\[
 a_\theta^{-1}\int_{U^c}|(P_\theta-P_0)u_F|
 \mathrm d\pi_\theta
 \le C_5(1+\theta^{-1})\pi_\theta(U^c)
 =O(\theta).
\]
The constants are independent of the choice of $\pi_\theta$, so this convergence is uniform.
\end{proof}

We first apply the preceding Poisson estimates to establish the stationary response formula for logarithmic quadratic observables. The ordinary quadratic case requires additional
analysis and will be treated separately below.

\begin{theorem}[Logarithmically enhanced quadratic response]
\label{thm:log}
Suppose Conditions~\condK, \condS, and~\condQ\ hold. Let $F$ be a logarithmic quadratic observable of exponent $\zeta>0$ in \cref{def:logarithmic}. Let $\pi_\theta$ be any invariant probability measure of $P_\theta$. Then, as $\theta\downarrow0$,
\begin{equation}\label{eq:log-response}
 \pi_\theta(F)
 =\theta^2[\ln(1/\theta)]^\zeta\pih(q_{\log})
   +o\bigl(\theta^2[\ln(1/\theta)]^\zeta\bigr).
\end{equation}
\end{theorem}

\begin{proof}
Write $u=u_F$, $h=h_F$, and $a_\theta=\theta^2[\ln(1/\theta)]^\zeta$. The Poisson equation and stationarity give
\begin{equation}\label{eq:stationary-poisson}
 \pi_\theta(F)
 =\pi_\theta((I-P_0)u)
 =\pi_\theta((P_\theta-P_0)u).
\end{equation}
For $j\in J_+(s)$, \eqref{eq:branch-limit} and \eqref{eq:h-log} imply that the corresponding rescaled difference converges to
\[
 p_j(s)\bigl[h(s_j,A_{j,s}v+b_{j,s})
                 -h(s_j,A_{j,s}v)\bigr].
\]
For $j\in J_0(s)$,
\[
 K_j(\theta)(s+\theta v)
 =\theta(K_jv+\dot K_js)+O_R(\theta^2).
\]
The corresponding contribution is
\[
 \frac{\Gamma_u(K_jv+\dot K_js+O_R(\theta))}
 {(1+\theta^2\|v\|^2)[\ln(1/\theta)]^\zeta}
       \longrightarrow0.
\]
The same applies to the reference zero branch. Hence, locally uniformly,
\begin{equation}\label{eq:log-increment-limit}
 G_\theta(s,v):=a_\theta^{-1}(P_\theta-P_0)u([s+\theta v])
       \longrightarrow G(s,v):=(\Ph-\Pt)h(s,v).
\end{equation}

By \eqref{eq:rescaled-increment-bound},
$|G_\theta(s,v)|\le C(1+\|v\|)$ on the charts.
The local uniform convergence in
\eqref{eq:log-increment-limit} gives the same
bound for $G$. For $R\ge1$, the uniform second-moment estimate yields
\[
 \int_{V>R}(1+V)\,\mathrm d\pih_\theta
 \le (R^{-2}+R^{-1})\pih_\theta(V^2)
 \le C/R.
\]
The same tail bound holds under $\pih$.
Choose continuous cutoffs $\chi_R$ supported
on $\mathcal K_{2R}$ and equal to one on
$\mathcal K_R$, with $\chi_R(\dagger)=0$.
By local uniform convergence and
$\pih_\theta\Rightarrow\pih$,
\[
 \int\chi_R G_\theta\,\mathrm d\pih_\theta
 \longrightarrow
 \int\chi_R G\,\mathrm d\pih.
\]
The tail estimates therefore imply
\[
 \int_{\TX}G_\theta\,\mathrm d\pih_\theta
 \longrightarrow\pih(G).
\]
By \eqref{eq:stationary-poisson} and the definition
of the blow-up map,
\[
 \begin{aligned}
 a_\theta^{-1}\pi_\theta(F)
 =\int_{\TX}G_\theta\,\mathrm d\pih_\theta +
 a_\theta^{-1}\int_{U^c}
 (P_\theta-P_0)u\,\mathrm d\pi_\theta.
 \end{aligned}
\]
The second term tends to zero by
\eqref{eq:rescaled-increment-bound}. Hence
\[
 \lim_{\theta\downarrow0}
 a_\theta^{-1}\pi_\theta(F)=\pih(G).
\]
Finally, $|h|\le CV^2$, $|\Pt h|\le CV^2$, and $|\Ph h|\le C(1+V^2)$. All three functions are $\pih$-integrable. Invariance and \eqref{eq:h-log} yield
\[
 \pih(G)=\pih(\Ph h-\Pt h)
        =\pih(h-\Pt h)=\pih(q_{\log}).
\]
This proves \eqref{eq:log-response}.
\end{proof}

\subsection{Rare departures and ordinary quadratic response}

We now turn to ordinary quadratic observables. At this scale, transitions with zero probability at the unperturbed reference states may contribute to the leading stationary response. We describe these departures by a measure $\Lambda$ and their subsequent evolution under $P_0$ before returning to $S$ by an occupation measure $\mu^{\mathrm{occ}}$.

For $(s,v)\in\TX$ and $j\in J_0(s)$, define
\begin{equation}\label{eq:rare-amplitude}
 z_j^+(s,v):=K_jv+\dot K_js,
 \qquad z_j^0(s,v):=K_jv.
\end{equation}
These are the first-order output amplitudes for the perturbed and reference Kraus operators, respectively. Since $K_js=0$, Taylor expansion gives
\[
 \begin{aligned}
 K_j(\theta)(s+\theta v)
 &=\theta z_j^+(s,v)+O_R(\theta^2),\\
 K_j(s+\theta v)
 &=\theta z_j^0(s,v),
 \end{aligned}
\]
uniformly on every $\mathcal K_R$. Consequently, the Born probability of record $j$ satisfies
\[
 \frac{\|K_j(\theta)(s+\theta v)\|^2}
 {1+\theta^2\|v\|^2}
 =\theta^2\|z_j^+(s,v)\|^2+O_R(\theta^3).
\]
Whenever $z_j^+(s,v)\ne0$, the corresponding output ray converges to $[z_j^+(s,v)]$ as $\theta\downarrow0$. Thus $\|z_j^+(s,v)\|^2$ determines the leading departure intensity, while $[z_j^+(s,v)]$ identifies its limiting destination. We define a finite Borel measure $\Lambda$ on $\X\setminus S$ by
\begin{equation}\label{eq:Lambda}
 \Lambda(B):=
 \int_{\TX}
 \sum_{\substack{j\in J_0(s)\\z_j^+(s,v)\ne0}}
 \|z_j^+(s,v)\|^2
 \one_B([z_j^+(s,v)])
 \,\pih(\mathrm d(s,v)),
 \qquad B\in\mathcal B(\X\setminus S).
\end{equation}
The measure $\Lambda$ thus encodes the leading-order intensity and limiting destinations of rare departures from the reference set $S$, before their subsequent evolution under $P_0$.

To account for the subsequent evolution of these departures, let $P_0^\circ$ denote the sub-Markov kernel on $\X\setminus S$ obtained by killing the reference dynamics upon entering $S$:
\[
 P_0^\circ(x,B):=P_0(x,B),
 \qquad x\in\X\setminus S,\quad
 B\in\mathcal B(\X\setminus S).
\]
We then define the corresponding occupation measure by
\begin{equation}\label{eq:Xi}
 \mu^{\mathrm{occ}}:=\sum_{n=0}^\infty\Lambda(P_0^\circ)^n.
\end{equation}
Thus $\mu^{\mathrm{occ}}$ records the cumulative occupation of $\X\setminus S$, starting from the departure intensity $\Lambda$ and including all subsequent reference steps before the first return to $S$. The following theorem characterizes the occupation measure $\mu^{\mathrm{occ}}$ and shows how it determines both the additional contribution to ordinary quadratic response and the rescaled stationary mass away from $S$.

\begin{theorem}[Occupation measure and quadratic stationary response]
\label{thm:quadratic}
Suppose Conditions~\condK, \condS, and~\condQ\ hold, and let $\Lambda$ and $\mu^{\mathrm{occ}}$ be defined by \eqref{eq:Lambda}--\eqref{eq:Xi}. For each $\theta\in(0,\theta_0)$, let $\pi_\theta$ be any invariant probability measure of $P_\theta$. Then the following statements hold.
\begin{enumerate}
\item \textit{Occupation measure.}
The measure $\Lambda$ is finite, and $\mu^{\mathrm{occ}}$ is a Radon measure on $\X\setminus S$. Moreover, $\mu^{\mathrm{occ}}$ is the unique nonnegative Borel measure $\mu$ on $\X\setminus S$ satisfying
\begin{equation}\label{eq:potential-equation}
 \mu=\Lambda+\mu P_0^\circ,
 \qquad
 \int_{\X\setminus S}r_S(x)^2\,\mu(\mathrm dx)<\infty.
\end{equation}

\item \textit{Quadratic response.}
For every ordinary quadratic observable $F$, as $\theta\downarrow0$,
\begin{equation}\label{eq:quadratic-response}
 \pi_\theta(F) =\theta^2\bigl[\pih(q_F)+\mu^{\mathrm{occ}}(F)\bigr]+o(\theta^2).
\end{equation}
If $F\ge0$, both terms in the coefficient are nonnegative.

\item \textit{Limit away from the reference set.}
As $\theta\downarrow0$,
\begin{equation}\label{eq:outer-limit}
 \theta^{-2}\pi_\theta|_{\X\setminus S}
 \xrightarrow{\mathrm{vague}}\mu^{\mathrm{occ}}.
\end{equation}
\end{enumerate}
\end{theorem}

\begin{proof}
\emph{(i) Occupation measure.}
By Kraus completeness and the inequality
$\|a+b\|^2\le2\|a\|^2+2\|b\|^2$, we have
\[
 \sum_{j\in J_0(s)}\|K_jv+\dot K_js\|^2
 \le 2\|v\|^2
     +2\sum_{j\in J}\|\dot K_j\|_{\mathrm{op}}^2.
\]
Consequently, since $J$ is finite and
$\pih(V^2)<\infty$,
\[
 \begin{aligned}
 \Lambda(\X\setminus S)
 &\le\int_{\TX}\sum_{j\in J_0(s)}
       \|K_jv+\dot K_js\|^2\,
       \pih(\mathrm d(s,v))\\
 &\le 2\pih(V^2)
       +2\sum_{j\in J}\|\dot K_j\|_{\mathrm{op}}^2
 <\infty.
 \end{aligned}
\]
Since $S$ is $P_0$-invariant and $r_S|_S=0$,
for $x\notin S$,
\[
 (P_0^\circ)^n r_S^2(x)=P_0^n r_S^2(x)
 \le\eta^{\floor{n/L}}r_S(x)^2
 \qquad(n\ge0).
\]
Consequently, by monotone convergence,
\[
 \begin{aligned}
 \mu^{\mathrm{occ}}(r_S^2)
 &=\sum_{n\ge0}\Lambda((P_0^\circ)^n r_S^2)\\
 &\le\frac{L}{1-\eta}\Lambda(r_S^2)<\infty.
 \end{aligned}
\]
For every compact $K\subset\X\setminus S$,
the quantity $\inf_Kr_S$ is positive, so
$\mu^{\mathrm{occ}}(K)<\infty$. Thus $\mu^{\mathrm{occ}}$ is Radon on
$\X\setminus S$. The defining series and monotone
convergence also give $\mu^{\mathrm{occ}}=\Lambda+\mu^{\mathrm{occ}} P_0^\circ$.

To prove uniqueness, let $\mu$ be a nonnegative
Borel measure satisfying the equation in
\eqref{eq:potential-equation} with
$\mu(r_S^2)<\infty$. Iteration yields
\[
 \mu=\sum_{k=0}^{n-1}\Lambda(P_0^\circ)^k
       +\mu(P_0^\circ)^n.
\]
For $f\in C_c(\X\setminus S)$ there is
$C_f<\infty$ such that $|f|\le C_fr_S^2$.
Hence
\[
 |\mu((P_0^\circ)^nf)|
 \le C_f\eta^{\floor{n/L}}\mu(r_S^2)
 \longrightarrow0.
\]
Taking $n\to\infty$ gives $\mu(f)=\mu^{\mathrm{occ}}(f)$, so $\mu=\mu^{\mathrm{occ}}$.

\emph{(ii) Quadratic response.}
Fix an ordinary quadratic observable $F$, and
write $u=u_F$ and $h=h_F$ as in
\cref{lem:poisson}. By the Poisson equation
and stationarity,
\[
 \pi_\theta(F)=\pi_\theta((P_\theta-P_0)u).
\]
For $j\in J_+(s)$, the branch expansion
\eqref{eq:branch-limit} and the locally uniform
limit \eqref{eq:h-quadratic} show that the
contribution of record $j$ to the rescaled
difference converges, uniformly on each
$\mathcal K_R$, to
\[
 p_j(s)\bigl[
 h(s_j,A_{j,s}v+b_{j,s})
 -h(s_j,A_{j,s}v)\bigr].
\]
For $j\in J_0(s)$, Condition~\condK\ and
the definitions \eqref{eq:rare-amplitude} give
\[
 \begin{aligned}
 K_j(\theta)(s+\theta v)
   &=\theta\bigl(z_j^+(s,v)+O_R(\theta)\bigr),\\
 K_j(s+\theta v)&=\theta z_j^0(s,v).
 \end{aligned}
\]
Homogeneity and continuity of $\Gamma_u$,
including at zero, therefore imply the
locally uniform limits
\[
 \begin{aligned}
 \frac{\Gamma_u(K_j(\theta)(s+\theta v))}
       {\theta^2(1+\theta^2\|v\|^2)}
     &\longrightarrow\Gamma_u(z_j^+(s,v)),\\
 \frac{\Gamma_u(K_j(s+\theta v))}
       {\theta^2(1+\theta^2\|v\|^2)}
     &\longrightarrow\Gamma_u(z_j^0(s,v)).
 \end{aligned}
\]
Set
\[
 Z^+u(s,v):=
 \sum_{j\in J_0(s)}\Gamma_u(z_j^+(s,v));
\]
recall that $Z^0u$ is defined in
\eqref{eq:h-poisson}. Summing the finitely
many branches yields, uniformly on
$\mathcal K_R$ for every $R<\infty$,
\begin{equation}\label{eq:Gquadratic}
 \begin{split}
 G_\theta(s,v)
 &:=\theta^{-2}(P_\theta-P_0)u([s+\theta v])\\
 &\longrightarrow G(s,v):=(\Ph-\Pt)h(s,v)
                      +Z^+u(s,v)-Z^0u(s,v).
 \end{split}
\end{equation}
The increment estimates
\eqref{eq:rescaled-increment-bound} with
$\zeta=0$ give, on $B_\theta(U)$,
\[
 |G_\theta(s,v)|\le C(1+V(s,v)),
 \qquad
 \theta^{-2}\int_{U^c}|(P_\theta-P_0)u|\,
                \mathrm d\pi_\theta\longrightarrow0.
\]
The local uniform convergence implies
$|G(s,v)|\le C(1+V(s,v))$.
Using continuous cutoffs $\chi_R$ as in
the proof of \cref{cor:subquadratic},
extended by zero at $\dagger$, local
uniform convergence and \cref{thm:tangent}
give
\[
 \int\chi_RG_\theta\,\mathrm d\pih_\theta
 \longrightarrow\int\chi_RG\,\mathrm d\pih.
\]
Moreover, for $R\ge1$,
\[
 \int_{V>R}(1+V)\,\mathrm d\pih_\theta
 \le (R^{-2}+R^{-1})\pih_\theta(V^2)\le C/R,
\]
and the same bound holds under $\pih$.
Thus the cutoffs may be removed, giving
\[
 \int_{\TX}G_\theta\,\mathrm d\pih_\theta
 \longrightarrow\pih(G).
\]
The stationary Poisson identity and the
definition of $B_\theta$ give
\[
 \begin{aligned}
 \theta^{-2}\pi_\theta(F)
 =\int_{\TX}G_\theta\,\mathrm d\pih_\theta +\theta^{-2}\int_{U^c}
 (P_\theta-P_0)u\,\mathrm d\pi_\theta.
 \end{aligned}
\]
The last term tends to zero by
\eqref{eq:rescaled-increment-bound}; hence
$\theta^{-2}\pi_\theta(F)\to\pih(G)$.

By \cref{lem:poisson},
$|h|+|\Pt h|\le CV^2$; the affine form of
$\Ph$ and the boundedness of $u$ give
\[
 |\Ph h|+|Z^+u|+|Z^0u|
 \le C(1+V^2).
\]
These functions are $\pih$-integrable.
Thus $\pih\Ph=\pih$ and \eqref{eq:h-poisson}
imply
\begin{align}\label{eq:coefficient-cancel}
 \pih(G)
 &=\pih(h-\Pt h+Z^+u-Z^0u)\notag\\
 &=\pih(q_F)+\pih(Z^+u).
\end{align}
Since $u|_S=0$ and $\Gamma_u(0)=0$,
the definition of $\Lambda$ gives
$\pih(Z^+u)=\Lambda(u)$.
Moreover, $|F|\le Cr_S^2$, so
\eqref{eq:reference-drift} implies
\[
 \sum_{n\ge0}\Lambda(|P_0^nF|)
 \le C\sum_{n\ge0}
 \eta^{\floor{n/L}}\Lambda(r_S^2)<\infty.
\]
As $F|_S=0$ and $S$ is $P_0$-invariant,
$(P_0^\circ)^nF=P_0^nF$ on
$\X\setminus S$. Consequently,
\[
 \Lambda(u)
 =\sum_{n\ge0}\Lambda(P_0^nF)
 =\sum_{n\ge0}\Lambda(P_0^\circ)^n(F)
 =\mu^{\mathrm{occ}}(F),
\]
which proves \eqref{eq:quadratic-response}.
If $F\ge0$, then each $s\in S$ is a
local minimum, so $q_F(s,v)\ge0$;
positivity of $\mu^{\mathrm{occ}}$ also gives
$\mu^{\mathrm{occ}}(F)\ge0$.

\emph{(iii) Limit away from $S$.}
For $f\in C_c^\infty(\X\setminus S)$,
extend $f$ by zero to $\X$.
This extension is an ordinary quadratic
observable with $q_f=0$.
Part (ii) therefore gives
\[
 \theta^{-2}\pi_\theta(f)\longrightarrow\mu^{\mathrm{occ}}(f).
\]
For any compact $K\subset\X\setminus S$,
stationary concentration
\eqref{eq:stationary-concentration} yields
\[
 \sup_{0<\theta\le\theta_*}
 \theta^{-2}\pi_\theta(K)
 \le\frac{C}{\inf_Kr_S^2}<\infty.
\]
Since $\mu^{\mathrm{occ}}(K)<\infty$, uniform
approximation of any
$f\in C_c(\X\setminus S)$ by smooth
functions supported in a common compact
subset of $\X\setminus S$ extends the
convergence to all such $f$.
\end{proof}

\begin{corollary}\label{cor:jets}
Let $\{K_j^{(1)}(\theta)\}_{j\in J}$ and $\{K_j^{(2)}(\theta)\}_{j\in J}$ be two Kraus families satisfying Conditions~\condK, \condS, and~\condQ\ with the same reference set $S$.
\begin{enumerate}
\item \textit{First-order determination.}
If, for every $j\in J$,
\[
 K_j^{(1)}(0)=K_j^{(2)}(0),\qquad
 \partial_\theta K_j^{(1)}(0)
   =\partial_\theta K_j^{(2)}(0),
\]
then their stationary tangent laws $\pih^{(1)},\pih^{(2)}$, departure measures $\Lambda^{(1)},\Lambda^{(2)}$, and occupation measures $\mu_{(1)}^{\mathrm{occ}},\mu_{(2)}^{\mathrm{occ}}$ coincide. Consequently, the leading coefficients in \cref{cor:subquadratic,thm:log,thm:quadratic} agree for each fixed observable in the respective class.

\item \textit{Dependence on perturbation directions.}
Fix a normalized reference Kraus family $\{K_j\}_{j\in J}$ satisfying Conditions~\condS\ and~\condQ. Let $\mathscr D\subseteq\bigoplus_{j\in J}M_D(\C)$ be a finite-dimensional real vector subspace such that, for every $h=(h_j)_{j\in J}\in\mathscr D$, there exists a $C^2$ normalized Kraus path $\{K_j^{(h)}(\theta)\}_{j\in J}$ near $\theta=0$ with
\[
 K_j^{(h)}(\theta)
   =K_j+\theta h_j+O_h(\theta^2),\qquad j\in J.
\]
Write $\pih_h$, $\Lambda_h$, and $\mu_h^{\mathrm{occ}}$ for its stationary tangent law, departure measure, and occupation measure, respectively. There exist a random state $S_0\sim\nu$ and vectors $V_h\in S_0^\perp$, $h\in\mathscr D$, on a common probability space such that
\[
 \mathcal L(S_0,V_h)=\pih_h,\qquad
 V_{\alpha h+\beta g}=\alpha V_h+\beta V_g
 \quad(h,g\in\mathscr D,\ \alpha,\beta\in\R)
\]
almost surely. The occupation measures satisfy
\begin{equation}\label{eq:source-homogeneity}
 \mu_{th}^{\mathrm{occ}}=t^2\mu_h^{\mathrm{occ}}\qquad(h\in\mathscr D,\ t\in\R).
\end{equation}
Hence for every ordinary quadratic observable $F$, the stationary response coefficient
\[
 \kappa_F(h):=\pih_h(q_F)+\mu_h^{\mathrm{occ}}(F)
\]
satisfies $\kappa_F(th)=t^2\kappa_F(h)$ for all $t\in\R$.
\end{enumerate}
\end{corollary}

\begin{proof}
\emph{(i) First-order determination.}
Use the same unit representatives for the common reference
set $S$. The assumed equalities give, for $s\in S$ and
$j\in J_+(s)$,
\[
 p_j^{(1)}(s)=p_j^{(2)}(s),\qquad
 A_{j,s}^{(1)}=A_{j,s}^{(2)},\qquad
 b_{j,s}^{(1)}=b_{j,s}^{(2)}.
\]
Thus the affine tangent kernels coincide, and their
unique invariant laws satisfy $\pih^{(1)}=\pih^{(2)}$
by \cref{prop:affine}. The amplitudes in
\eqref{eq:rare-amplitude} are also identical, so
\eqref{eq:Lambda} gives $\Lambda^{(1)}=\Lambda^{(2)}$.
The common reference operators give the same killed kernel
$P_0^\circ$; hence \eqref{eq:Xi} yields
$\mu_{(1)}^{\mathrm{occ}}=\mu_{(2)}^{\mathrm{occ}}$. The response formulas then imply
equality of their leading coefficients.

\emph{(ii) Dependence on perturbation directions.}
The reference probabilities $p_j(s)$ and tangent maps
$A_{j,s}$ are independent of $h$, whereas the affine
forcing is real-linear in $h$:
\[
 b_{j,s}(h)=\lambda_{j,s}^{-1}E_{s_j}h_js,
 \qquad j\in J_+(s).
\]
Use the common two-sided stationary reference chain
$(S_n,J_n)_{n\in\mathbb Z}$ from \cref{prop:affine},
and put $A_n=A_{J_n,S_n}$ and
$b_n(h)=b_{J_n,S_n}(h)$. Its backward series is
\[
 V_h=\sum_{k=1}^{\infty}
 A_{-1}\cdots A_{-k+1}b_{-k}(h).
\]
Choose a real basis of $\mathscr D$. For each basis vector, the series converges absolutely almost surely by \cref{prop:affine}. On the intersection of these finitely many full-probability events, it converges for every $h\in\mathscr D$ and is real-linear in $h$.
Moreover, $(S_0,V_h)$ has law $\pih_h$, proving the claimed coupling. For $j\in J_0(s)$, the departure amplitude is
\[
 z_{j,h}^+(s,v)=K_jv+h_js.
\]
The coupling gives $V_{th}=tV_h$ and hence
\[
 z_{j,th}^+(S_0,V_{th})
   =t\,z_{j,h}^+(S_0,V_h).
\]
For $t\ne0$, multiplication by $t$ preserves every
nonzero output ray and multiplies its squared weight
by $t^2$. For $t=0$, all these weights vanish.
Thus $\Lambda_{th}=t^2\Lambda_h$ as measures on
$\X\setminus S$. Since $P_0^\circ$ is independent
of $h$, \eqref{eq:Xi} gives
$\mu_{th}^{\mathrm{occ}}=t^2\mu_h^{\mathrm{occ}}$.

Finally, $\pih_{th}=(T_t)_\#\pih_h$, where
$T_t(s,v)=(s,tv)$. Since
$q_F(s,tv)=t^2q_F(s,v)$ for every real $t$,
\[
 \kappa_F(th)
 =\pih_{th}(q_F)+\mu_{th}^{\mathrm{occ}}(F)
 =t^2\bigl[\pih_h(q_F)+\mu_h^{\mathrm{occ}}(F)\bigr].
\]
\end{proof}

\section{Clifford-driven quantum trajectories}\label{sec:clifford}

We now apply the abstract framework to a class of quantum trajectories combining Clifford scrambling, unitary injection and projective measurements. Beyond verifying the relevant structural conditions, we exploit the algebraic and design properties of the Clifford group to obtain sharper ergodic estimates and a more detailed characterization of the stationary laws.

\subsection{Model and Kraus representation}
Fix a prime $d$, an integer $N\ge1$, and a nonempty measured subset $M\subseteq\{1,\ldots,N\}$. Set $m=|M|$ and
\begin{equation}\label{eq:dimensions}
 \begin{gathered}
 \HH=(\C^d)^{\otimes N},\qquad
 D=d^N,\qquad q=d^m,\qquad r=d^{N-m},\qquad D=qr,\\
 \Pi_a=|a\rangle\langle a|_M\otimes I_{M^c},
 \qquad a\in\F_d^m.
 \end{gathered}
\end{equation}
Thus $\rank\Pi_a=r$, $\Pi_a\Pi_b=\one_{\{a=b\}}\Pi_a$, and $\sum_a\Pi_a=I_D$. Tensor factors are ordered according to $M\sqcup M^c$ whenever this is convenient.

Write $\omega_d=e^{2\pi i/d}$. For $p,b,x\in\F_d^N$, define
\[
 X(b)|x\rangle=|x+b\rangle,
 \qquad
 Z(p)|x\rangle=\omega_d^{p\cdot x}|x\rangle.
\]
The projective Pauli group is
\[
 \mathcal P_{d,N}
 :=\{[X(b)Z(p)]:b,p\in\F_d^N\}
 \subseteq\mathrm{PU}(D),
\]
where $[V]$ denotes the class of $V\in U(D)$ modulo scalar phases. Its normalizer in $\mathrm{PU}(D)$ is the projective Clifford group $\Cl$. We write $g=|\Cl|$ and use arbitrary unitary representatives of its elements. The conjugation-based expressions and Kraus operators below are independent of
these phase choices.

Let $S=\St$ denote the finite set of pure stabilizer rays, namely the common eigenrays of maximal commuting families of Pauli operators. We recall the following standard facts \cite{Gottesman1997,Gross2006}:
\begin{enumerate}
\item \emph{Symplectic structure.}
Modulo scalar phases, Pauli operators are labeled by the symplectic space $\F_d^{2N}$, with commutation relations determined by its standard symplectic form.

\item \emph{Clifford action.}
Clifford conjugation induces symplectic transformations of Pauli labels, and $\Cl$ acts transitively on $S$.

\item \emph{Stabilizer preservation.}
Every nonzero branch of a projective Pauli measurement maps a stabilizer ray to a stabilizer ray.
\end{enumerate}

The uniform Clifford ensemble is also a unitary two-design \cite{GrossAudenaertEisert2007,Dankert2009}, so its two-copy conjugation averages agree with Haar averages. In particular, for every pure density matrix $\rho$,
\begin{equation}\label{eq:two-design}
 \frac1g\sum_{U\in\Cl}(U\rho U^\dagger)^{\otimes2}
 =\frac{I+\Flip}{D(D+1)},
 \qquad \Flip(x\otimes y)=y\otimes x.
\end{equation}

Fix a unitary $R\in U(D)$. Each quantum-trajectory step draws $U$ uniformly from $\Cl$, applies $U$ followed by $R$, performs the projective measurement $(\Pi_a)_a$, and finally applies $U^\dagger$. Its branch index set and Kraus operators are
\begin{equation}\label{eq:clifford-Kraus}
 J=\Cl\times\F_d^m,\qquad
 K^R_{U,a}=g^{-1/2}U^\dagger\Pi_aRU.
\end{equation}
For a unit input vector $x$, the branch law and update are
\[
 \PP(U,a\mid[x])=g^{-1}\|\Pi_aRUx\|^2,
 \qquad [x]\longmapsto[U^\dagger\Pi_aRUx].
\]
As usual, zero outputs have zero probability. The identity
\[
 \sum_{U,a}(K^R_{U,a})^\dagger K^R_{U,a}
 =\frac1g\sum_U U^\dagger R^\dagger
       \Bigl(\sum_a\Pi_a\Bigr)RU=I_D
\]
verifies normalization. Denote the resulting pure-state transition kernel by $P_R$. The associated averaged quantum channel describes the nonselective evolution obtained by averaging over the Clifford draw and all measurement outcomes:
\[
 \Phi_R(A)
 =\sum_{(U,a)\in J}K^R_{U,a}A(K^R_{U,a})^\dagger,
 \qquad A\in M_D(\C).
\]

\subsection{Uniform geometric ergodicity in Wasserstein distance}

The two-design property controls both the averaged channel and products of Kraus operators. Inspired by the record-based comparison strategy of \cite[Sections~3.1--3.3]{Benoist2019}, we combine the Clifford two-design estimates with a direct comparison argument to obtain an explicit convergence bound uniform in $R\in U(D)$.

\begin{theorem}[Uniform geometric Wasserstein ergodicity]\label{thm:global}
For every $R\in U(D)$, the kernel $P_R$ has a unique invariant probability measure $\pi_R$. Define
\begin{equation}\label{eq:mixing-constants}
 c:=\sqrt{\frac{D-q}{D-1}},\qquad
 \lambda:=\frac{Dr-1}{D^2-1}.
\end{equation}
Then $0\le c<1$, $0<\lambda<1$, and, for every $n\ge2$ and $\mu,\nu\in\mathcal P(\X)$,
\begin{equation}\label{eq:explicit-mixing}
 W_1(\mu P_R^n,\nu P_R^n)
 \le 2\sqrt{\binom D2}\,c^{\floor{n/2}}
       +\lambda^{\ceil{n/2}}.
\end{equation}
Consequently, there exist $C_{\mathrm{mix}}<\infty$ and $\rho_{\mathrm{mix}}\in(0,1)$, depending only on $D$ and $q$, such that
\[
 \sup_{R\in U(D)}\sup_{\mu\in\mathcal P(\X)}
 W_1(\mu P_R^n,\pi_R)
 \le C_{\mathrm{mix}}\rho_{\mathrm{mix}}^n,
 \qquad n\ge0.
\]
\end{theorem}
\begin{proof}
\emph{Step 1: the averaged channel.}
The Clifford twirl of $A\mapsto\sum_a\Pi_aRA R^\dagger\Pi_a$ equals its Haar twirl. The latter acts by a scalar on the traceless matrices. Trace preservation and the trace of the superoperator therefore give
\begin{equation}\label{eq:depolarizing}
 \begin{split}
 \Phi_R(A)&=\lambda_RA+(1-\lambda_R)\frac{\Tr A}{D}I_D,\\
 1+(D^2-1)\lambda_R
 &=\sum_{U,a}|\Tr K^R_{U,a}|^2
   =\sum_a|\Tr(\Pi_aR)|^2.
 \end{split}
\end{equation}
For every $a$, Cauchy--Schwarz gives
\[
 |\Tr(\Pi_aR)|^2
 =|\Tr(\Pi_aR\Pi_a)|^2
 \le r\|\Pi_aR\Pi_a\|_{\HS}^2\le r^2.
\]
Since $qr^2=Dr$, it follows that
\[
 -\frac1{D^2-1}\le\lambda_R\le\lambda,
 \qquad |\lambda_R|\le\lambda<1.
\]
Thus, for density matrices $\rho,\sigma$ and $l\ge0$,
\begin{equation}\label{eq:barycenter-contraction}
 \frac12\|\Phi_R^l(\rho)-\Phi_R^l(\sigma)\|_1
 =\frac{|\lambda_R|^l}{2}\|\rho-\sigma\|_1
 \le\lambda^l.
\end{equation}

\emph{Step 2: contraction of the exterior square.}
Put $b_D=\binom D2$. For an operator $A$ on $\wedge^2\HH$, the two-design identity on the antisymmetric subspace yields
\[
 \frac1g\sum_U
 \|\wedge^2(\Pi_aRU)A\|_{\HS}^2
 =\frac{\binom r2}{\binom D2}\|A\|_{\HS}^2.
\]
Indeed, left multiplication by $R$ preserves the two-design property, and $\wedge^2\Pi_a$ is a projection of rank $\binom r2$. The factor $\wedge^2U^\dagger$ is unitary, while $\wedge^2(g^{-1/2}B)=g^{-1}\wedge^2B$. Hence
\begin{align*}
 \sum_{U,a}\|(\wedge^2K^R_{U,a})A\|_{\HS}
 &=\sum_a\frac1g\sum_U\|\wedge^2(\Pi_aRU)A\|_{\HS}\\
 &\le q\sqrt{\frac{r(r-1)}{D(D-1)}}\|A\|_{\HS}
 =c\|A\|_{\HS}.
\end{align*}
Iterating this inequality from $A=I_{\wedge^2\HH}$ gives, for $k\ge1$,
\begin{equation}\label{eq:word-wedge}
 \sum_{\omega\in J^k}\|\wedge^2K^R_\omega\|_{\op}
 \le\sum_{\omega\in J^k}\|\wedge^2K^R_\omega\|_{\HS}
 \le\sqrt{b_D}\,c^k.
\end{equation}
This also covers $r=1$, when every exterior square is zero.

\emph{Step 3: comparison through the branch records.}
For each $k$-word matrix $W_\omega=K^R_\omega$, choose a unit left singular vector $z_\omega$ corresponding to its largest singular value; when $W_\omega=0$, choose $z_\omega$ arbitrarily. For a unit input $x$,
\begin{align*}
 \|W_\omega x\|^2d([W_\omega x],[z_\omega])
 &=\|W_\omega x\|
   \|(I-|z_\omega\rangle\langle z_\omega|)W_\omega x\|\\
 &\le s_1(W_\omega)s_2(W_\omega)
 =\|\wedge^2W_\omega\|_{\op},
\end{align*}
with zero contribution when $W_\omega x=0$. Define
\[
 \mathcal E_k(\rho)
 :=\sum_\omega\Tr(W_\omega\rho W_\omega^\dagger)
                    \delta_{[z_\omega]},
 \qquad
 \rho_\mu:=\int_\X\rho_x\,\mu(\mathrm dx).
\]
Coupling each output with its estimator on the same word and using \eqref{eq:word-wedge}, we obtain
\[
 W_1(\mu P_R^k,\mathcal E_k(\rho_\mu))
 \le\sqrt{b_D}\,c^k.
\]
The effects $W_\omega^\dagger W_\omega$ sum to $I_D$, so
\begin{align*}
 \|\mathcal E_k(\rho)-\mathcal E_k(\sigma)\|_{\TV}
 &\le\frac12\sum_\omega
       |\Tr(W_\omega^\dagger W_\omega(\rho-\sigma))|\\
 &\le\frac12\|\rho-\sigma\|_1.
\end{align*}
Since $\operatorname{diam}(\X)=1$, $W_1\le\|\cdot\|_{\TV}$ for probability measures. Moreover, $\rho_{\mu P_R^l}=\Phi_R^l(\rho_\mu)$. Applying the preceding comparisons after $l$ preliminary steps gives
\[
 W_1(\mu P_R^{k+l},\nu P_R^{k+l})
 \le2\sqrt{b_D}\,c^k+\lambda^l.
\]
Taking $k=\floor{n/2}$ and $l=\ceil{n/2}$ proves \eqref{eq:explicit-mixing}. 

The compact Feller argument of Section~\ref{sec:framework} gives existence of an invariant probability measure. Applying \eqref{eq:explicit-mixing} to two invariant measures and letting $n\to\infty$ proves
uniqueness. Finally, set
\[
 \rho_{\mathrm{mix}}
 :=\max\{\sqrt c,\sqrt\lambda\}\in(0,1),
 \qquad
 C_{\mathrm{mix}}
 :=1+\frac{2\sqrt{\binom D2}}{\rho_{\mathrm{mix}}}.
\]
Since $W_1\le1$ and
\[
 c^{\lfloor n/2\rfloor}\le
 \rho_{\mathrm{mix}}^{n-1},
 \qquad
 \lambda^{\lceil n/2\rceil}
 \le\rho_{\mathrm{mix}}^n
 \quad(n\ge2),
\]
equation~\eqref{eq:explicit-mixing} gives the claimed uniform geometric bound for $n\ge 0$. 
\end{proof}

\subsection{Structure and stability of stationary laws}

Clifford covariance fixes the stationary moments and gives an exact optimal-transport identity relative to $S$. We also control how the invariant measure changes with $R$.

\begin{proposition}[Clifford symmetry, exact transport, and gate stability]
\label{prop:stationary-structure}
Let $u_S:=|S|^{-1}\sum_{s\in S}\delta_s$, and let $\pi_R$ be as in \cref{thm:global}. Then the following assertions hold.
\begin{enumerate}
\item \textit{Symmetry and stationary moments.}
For every $R\in U(D)$,
\[
 C_\#\pi_R=\pi_R\quad(C\in\Cl),
 \qquad \pi_I=u_S,
\]
and
\begin{equation}\label{eq:stationary-design}
 \int_\X\rho_x^{\otimes2}\,\pi_R(\mathrm dx)
 =\frac{I+\Flip}{D(D+1)}.
\end{equation}
If $m=N$, then $\pi_R=u_S$ for every $R\in U(D)$.

\item \textit{Transport to the reference law.}
For every $R\in U(D)$ and $1\le p<\infty$,
\begin{equation}\label{eq:radial-transport}
 W_p(\pi_R,u_S)^p=\int_\X r_S(x)^p\,\pi_R(\mathrm dx).
\end{equation}

\item \textit{Gate stability.}
There exists $C_{\mathrm{gate}}<\infty$, depending only on the fixed dimensions, such that
\begin{equation}\label{eq:gate-stability}
 W_1(\pi_R,\pi_{R'})
 \le C_{\mathrm{gate}}\|R-R'\|_{\op},
 \qquad R,R'\in U(D).
\end{equation}
\end{enumerate}
\end{proposition}
\begin{proof}
\emph{(i) Symmetry and moments.}
Reindexing the Clifford draw gives
\[
 P_R(Cx,CB)=P_R(x,B),\qquad C\in\Cl.
\]
Thus $C_\#\pi_R$ is invariant, and uniqueness implies $C_\#\pi_R=\pi_R$. At $R=I$, every nonzero measurement branch preserves $S$. Averaging an invariant probability of the finite restricted chain over $\Cl$ gives $u_S$, by transitivity; hence $\pi_I=u_S$. Equation~\eqref{eq:two-design} now gives
\[
 \int\rho_x^{\otimes2}\,\pi_R(\mathrm dx)
 =\int\frac1g\sum_C(C\rho_xC^\dagger)^{\otimes2}
                   \,\pi_R(\mathrm dx)
 =\frac{I+\Flip}{D(D+1)}.
\]
When $m=N$, every nonzero output is a stabilizer ray $[U^\dagger|a\rangle]$. Therefore $\pi_R$ is supported on $S$, and Clifford invariance makes it equal to $u_S$.

\emph{(ii) Optimal transport.}
Every coupling of $\pi_R$ with a probability supported on $S$ has cost at least $\pi_R(r_S^p)$. To attain this bound, set
\[
 N(x):=\{s\in S:d(x,s)=r_S(x)\},
 \qquad
 T(x,\cdot):=\frac1{|N(x)|}\sum_{s\in N(x)}\delta_s.
\]
Finiteness of $S$ makes $T$ a Borel kernel. Since Clifford transformations are isometries preserving $S$, $N(Cx)=CN(x)$. Thus $\pi_RT$ is a Clifford-invariant probability on $S$, and so $\pi_RT=u_S$. The coupling $\pi_R(\mathrm dx)T(x,\mathrm ds)$ attains the lower bound, proving \eqref{eq:radial-transport}.

\emph{(iii) A uniform Poisson bound.}
Put $b_D=\binom D2$ and
\[
 e_n(R):=\sum_{\omega\in J^n}\|\wedge^2K^R_\omega\|_{\op}
 \le\sqrt{b_D}\,c^n\quad(n\ge1),
 \qquad e_0(R)=1.
\]
For a word matrix $W$ and unit vectors $x,y$,
\[
 \min\{\|Wx\|^2,\|Wy\|^2\}
 d([Wx],[Wy])\le\|\wedge^2W\|_{\op}\,d([x],[y]),
\]
with the usual zero-output convention. The word probabilities $p_\omega(x):=\|W_\omega x\|^2$ satisfy
\[
 \sum_\omega|p_\omega(x)-p_\omega(y)|
 \le\|\rho_x-\rho_y\|_1
 =2d([x],[y]),
\]
by Kraus completeness and trace-norm contractivity under measurement. Pairing the common mass $\min\{p_\omega(x),p_\omega(y)\}$ on each word and bounding the unmatched mass by $2\|f\|_\infty$ gives
\[
 |P_R^nf(x)-P_R^nf(y)|
 \le\left[e_n(R)\Lip(f)+2\|f\|_\infty\right]
 d([x],[y]).
\]
Consequently,
\begin{equation}\label{eq:lasota}
 \Lip(P_R^nf)
 \le e_n(R)\Lip(f)+2\|f\|_\infty.
\end{equation}
For $f_R=f-\pi_R(f)$, \cref{thm:global} implies
\[
 \|P_R^nf_R\|_\infty
 \le C_{\mathrm{mix}}\rho_{\mathrm{mix}}^n\Lip(f),
 \qquad \|f_R\|_\infty\le\Lip(f).
\]
For $n=k+l$, two applications of \eqref{eq:lasota} give
\[
 \Lip(P_R^nf_R)
 \le\bigl[e_k(R)(e_l(R)+2)
       +2C_{\mathrm{mix}}\rho_{\mathrm{mix}}^l\bigr]\Lip(f).
\]
Taking $k=\floor{n/2}$ and $l=\ceil{n/2}$, and using
$e_j(R)\le\sqrt{b_D}\,c^j$ for $j\ge1$, we obtain
\[
 \Lip(P_R^nf_R)
 \le\left[
 b_Dc^n+2\sqrt{b_D}\,c^{\floor{n/2}}
 +2C_{\mathrm{mix}}\rho_{\mathrm{mix}}^{\ceil{n/2}}
 \right]\Lip(f)
\]
for $n\ge2$. Set
\[
 \rho_{\mathrm L}
 :=\max\{\sqrt c,\sqrt{\rho_{\mathrm{mix}}}\}
 \in(0,1).
\]
Combining this estimate with the uniform sup-norm decay, and enlarging the prefactor to cover $n=0,1$, yields a constant $C_{\mathrm L}<\infty$, depending only on $D$ and $q$, such that
\begin{equation}\label{eq:uniform-lip-decay}
 \|P_R^n(f-\pi_R(f))\|_{\Lip}
 \le C_{\mathrm L}\rho_{\mathrm L}^n\|f\|_{\Lip},
 \qquad
 \|f\|_{\Lip}:=\|f\|_\infty+\Lip(f).
\end{equation}
Consequently,
\[
 \begin{gathered}
 u_{R,f}:=\sum_{n\ge0}P_R^n(f-\pi_R(f)),
 \qquad (I-P_R)u_{R,f}=f-\pi_R(f),\\
 \|u_{R,f}\|_{\Lip}
 \le\frac{C_{\mathrm L}}{1-\rho_{\mathrm L}}\|f\|_{\Lip}.
 \end{gathered}
\]
For any bounded Lipschitz function $f$, the homogeneous lift $\Gamma_f$ defined in \eqref{eq:lift} satisfies
\[
 |\Gamma_f(z)-\Gamma_f(w)|
 \le\|f\|_{\Lip}(\|z\|+\|w\|)\|z-w\|.
\]
Indeed, this follows from
\[
 \begin{split}
 |\|z\|^2-\|w\|^2|
 &\le(\|z\|+\|w\|)\|z-w\|,\\
 \min\{\|z\|^2,\|w\|^2\}\,
 d([z],[w])
 &\le\min\{\|z\|,\|w\|\}\|z-w\|,
 \end{split}
\]
with the evident extension when either vector is zero. For each $U$, apply this bound to $f\circ U^\dagger$ and to $z_a=\Pi_aRUx$, $w_a=\Pi_aR'Ux$, with $\|x\|=1$. Orthogonality gives
\[
 \sum_a\|z_a\|^2=\sum_a\|w_a\|^2=1,
 \qquad
 \sum_a\|z_a-w_a\|^2\le\|R-R'\|_{\op}^2.
\]
Since Clifford conjugation preserves the Lipschitz norm, Cauchy--Schwarz and averaging over $U$ imply
\[
 \|(P_R-P_{R'})f\|_\infty
 \le 2\|f\|_{\Lip}\|R-R'\|_{\op}.
\]
The Poisson equation at $R'$ and stationarity at $R$ yield
\[
 (\pi_R-\pi_{R'})(f)
 =\pi_R\bigl((P_R-P_{R'})u_{R',f}\bigr).
\]
A $1$-Lipschitz test function, after subtraction of its value at a fixed point, has Lipschitz norm at most two. Kantorovich duality proves \eqref{eq:gate-stability}, for example with $C_{\mathrm{gate}}=4C_{\mathrm L}/(1-\rho_{\mathrm L})$.
\end{proof}

\subsection{Rank-one characterization of uniform total-variation ergodicity}

Building on the general sufficient criterion of \cref{prop:resets}, we exploit the conjugation symmetry and unitary two-design property of the Clifford group to obtain an exact algebraic characterization: uniform geometric ergodicity in total variation is equivalent to the existence of a nonzero rank-one Kraus word. We then construct explicit complete-reset sequences for three natural classes of unitary gates.

\begin{proposition}[Rank-one characterization of total-variation ergodicity]
\label{prop:clifford-TV-characterization}
Fix $R\in U(D)$, and let $\pi_R$ be the invariant
probability from \cref{thm:global}. The following
assertions are equivalent:
\begin{enumerate}
\item There exist $\ell\ge1$ and $\omega\in J^\ell$
such that $\rank K^R_\omega=1$.

\item $\delta(P_R^n)<1$ for some $n\ge1$.

\item There exist $C_R\ge1$ and $\rho_R\in(0,1)$
such that
\[
 \sup_{\mu\in\mathcal P(\X)}
 \|\mu P_R^n-\pi_R\|_{\TV}
 \le C_R\rho_R^n,\qquad n\ge0.
\]
\end{enumerate}
More precisely, if \textup{(i)} holds for
$\omega=((V_1,a_1),\ldots,(V_\ell,a_\ell))$, write
\[
 L=(V_\ell^\dagger\Pi_{a_\ell}RV_\ell)\cdots
   (V_1^\dagger\Pi_{a_1}RV_1),\qquad
 p_L=g^{1-\ell}\frac{\|L\|_{\HS}^2}{D}>0.
\]
With $a_D:=\sqrt{D/(2(D+1))}<1$, one has
\begin{equation}\label{eq:rank-one-TV-bound}
 \|\mu P_R^n-\nu P_R^n\|_{\TV}
 \le[1-p_L(1-a_D)]^{\floor{n/\ell}}
     \|\mu-\nu\|_{\TV},
 \qquad n\ge0,\quad\mu,\nu\in\mathcal P(\X).
\end{equation}
\end{proposition}

\begin{proof}
\emph{\textup{(i)} $\Rightarrow$ \textup{(ii)}.}
We use a construction that also applies to several rank-one
words with the same Clifford draw sequence. Fix
$V_1,\ldots,V_\ell\in\Cl$ and
$\mathcal A\subseteq(\F_d^m)^\ell$ such that
\[
 L_{\boldsymbol a}
 :=(V_\ell^\dagger\Pi_{a_\ell}RV_\ell)\cdots
   (V_1^\dagger\Pi_{a_1}RV_1),
 \qquad \rank L_{\boldsymbol a}\le1
 \quad(\boldsymbol a\in\mathcal A),
\]
and
$E_{\mathcal A}:=\sum_{\boldsymbol a\in\mathcal A}
 L_{\boldsymbol a}^\dagger L_{\boldsymbol a}\ne0$.
Assumption~\textup{(i)} provides such a family with
$\mathcal A$ a singleton.

For each $U\in\Cl$, select the records with Clifford draws
$(V_1U,\ldots,V_\ell U)$ and outcome tuples in
$\mathcal A$. These records are distinct, and their Kraus
products are $g^{-\ell/2}U^\dagger L_{\boldsymbol a}U$.
By Clifford one-design averaging, their total effect is
\[
 g^{-\ell}\sum_{U\in\Cl}U^\dagger E_{\mathcal A}U
 =p_{\mathcal A}I_D,
 \qquad
 p_{\mathcal A}:=g^{1-\ell}
 \frac{\Tr E_{\mathcal A}}{D}>0.
\]
Let $\mathcal A_+:=\{\boldsymbol a\in\mathcal A:
L_{\boldsymbol a}\ne0\}$.
For $\boldsymbol a\in\mathcal A_+$, write
$L_{\boldsymbol a}=\sqrt{t_{\boldsymbol a}}
 |v_{\boldsymbol a}\rangle\langle w_{\boldsymbol a}|$
with unit vectors $v_{\boldsymbol a},w_{\boldsymbol a}$,
so that
$\sum_{\boldsymbol a\in\mathcal A_+}t_{\boldsymbol a}
=\Tr E_{\mathcal A}$.
The conditional reset kernel is
\[
 B^*_{\mathcal A}([x],\cdot)
 =\frac{D}{g\Tr E_{\mathcal A}}
 \sum_{U\in\Cl}\sum_{\boldsymbol a\in\mathcal A_+}
 t_{\boldsymbol a}|\langle w_{\boldsymbol a},Ux\rangle|^2
 \delta_{[U^\dagger v_{\boldsymbol a}]}.
\]
For $\Delta=\rho_x-\rho_y$, the two-design identity gives
\[
 \frac1g\sum_U
 |\langle w,U\Delta U^\dagger w\rangle|^2
 =\frac{\Tr\Delta^2}{D(D+1)}
 =\frac{2d([x],[y])^2}{D(D+1)}
\]
for every unit vector $w$. Since passing from records
to output rays cannot increase total variation,
Cauchy--Schwarz yields
\[
 \begin{aligned}
 &\|B^*_{\mathcal A}(x,\cdot)
       -B^*_{\mathcal A}(y,\cdot)\|_{\TV}\\
 &\quad\le\frac{D}{2g\Tr E_{\mathcal A}}
 \sum_{\boldsymbol a\in\mathcal A_+}t_{\boldsymbol a}\sum_U
 |\langle w_{\boldsymbol a},U\Delta U^\dagger
      w_{\boldsymbol a}\rangle|\\
 &\quad\le\sqrt{\frac{D}{2(D+1)}}\,d([x],[y])
 \le a_D<1.
 \end{aligned}
\]
Thus Conditions~\condR\ and~\condO\ hold with
$p=p_{\mathcal A}$ and $a=a_D$.
For the singleton supplied by~\textup{(i)},
$p_{\mathcal A}=p_L>0$; hence
\cref{prop:resets} gives
\[
 \delta(P_R^\ell)\le1-p_L(1-a_D)<1,
\]
which proves~\textup{(ii)}.

\emph{\textup{(ii)} $\Rightarrow$ \textup{(iii)}.}
Choose $n_0\ge1$ with $\alpha:=\delta(P_R^{n_0})<1$,
and put
\[
 \rho_R:=\left(\frac{1+\alpha}{2}\right)^{1/n_0}
 \in(0,1),\qquad C_R:=\rho_R^{-(n_0-1)}.
\]
By Dobrushin contraction, nonexpansiveness in total
variation, and $\pi_RP_R=\pi_R$, for every $n\ge0$,
\[
 \begin{aligned}
 \|\mu P_R^n-\pi_R\|_{\TV}
 \le\rho_R^{n_0\floor{n/n_0}} \le C_R\rho_R^n,
 \end{aligned}
\]
uniformly in $\mu\in\mathcal P(\X)$.

\emph{\textup{(iii)} $\Rightarrow$ \textup{(i)}.}
By the triangle inequality,
\[
 \delta(P_R^n)\le
 2\sup_{\mu\in\mathcal P(\X)}
 \|\mu P_R^n-\pi_R\|_{\TV}\longrightarrow0.
\]
Choose $n\ge1$ with $\delta(P_R^n)<1$.
Suppose every nonzero $n$-step Kraus word has rank
at least two. Since $J^n$ is finite, there is a unit
vector $x$ outside the kernels of all nonzero word
matrices. For any two such matrices $W_\omega,W_\tau$,
\[
 W_\tau^{-1}(\C W_\omega x)
\]
is a proper complex linear subspace of $\HH$ because
$\rank W_\tau\ge2$. Choose a unit vector $y$ outside
the finite union of these subspaces. Then every output
ray from $x$ differs from every output ray from $y$.
The measures $P_R^n([x],\cdot)$ and
$P_R^n([y],\cdot)$ are mutually singular, giving
$\delta(P_R^n)=1$, a contradiction. Thus~\textup{(i)}
holds. For the word specified in the statement, the singleton
construction in the first part yields precisely
\[
 p_{\mathcal A}=p_L
 =g^{1-\ell}\frac{\|L\|_{\HS}^2}{D},
 \qquad \delta(B^*_{\mathcal A})\le a_D.
\]
By \cref{prop:resets}, iteration of the $\ell$-step
Dobrushin contraction and nonexpansiveness in total
variation give, for all $n\ge0$ and
$\mu,\nu\in\mathcal P(\X)$,
\[
 \begin{aligned}
 \|\mu P_R^n-\nu P_R^n\|_{\TV}
 &\le\delta(P_R^\ell)^{\floor{n/\ell}}
          \|\mu-\nu\|_{\TV}\\
 &\le[1-p_L(1-a_D)]^{\floor{n/\ell}}
          \|\mu-\nu\|_{\TV},
 \end{aligned}
\]
which is \eqref{eq:rank-one-TV-bound}.
\end{proof}

\begin{corollary}[Complete resets for natural unitary families]
\label{cor:clifford-reset-families}
Set $L=\ceil{N/m}$. Condition~\textup{(i)} of
\cref{prop:clifford-TV-characterization} holds for
each of the following classes of unitary gates:
\begin{enumerate}[label=\textup{(\alph*)},leftmargin=*]
\item \emph{Subsystem unitaries:}
$R=R_M\otimes I_{M^c}$, where $R_M\in U(d^m)$.

\item \emph{Product unitaries:}
$R=\bigotimes_{k=1}^N R_k$, where $R_k\in U(d)$.

\item \emph{Computational-basis diagonal unitaries:}
\[
 R=\sum_{x\in\F_d^N}z_x|x\rangle\langle x|,
 \qquad |z_x|=1.
\]
\end{enumerate}
Moreover, in each case there exist
$V_1,\ldots,V_L\in\Cl$ such that, for every
$\boldsymbol a\in(\F_d^m)^L$, the word
\begin{equation}\label{eq:relative-word}
 L_{\boldsymbol a}
 :=(V_L^\dagger\Pi_{a_L}RV_L)\cdots
   (V_1^\dagger\Pi_{a_1}RV_1)
\end{equation}
has rank at most one. Consequently,
Conditions~\condR\ and~\condO\ hold with
\begin{equation}\label{eq:reset-constants}
 p=g^{1-L},\qquad a=a_D,
\end{equation}
and the kernel $P_R$ satisfies
\[
 \|\mu P_R^n-\nu P_R^n\|_{\TV}
 \le[1-g^{1-L}(1-a_D)]^{\floor{n/L}}
   \|\mu-\nu\|_{\TV},
 \qquad n\ge0,\quad\mu,\nu\in\mathcal P(\X).
\]
\end{corollary}

\begin{proof}
Choose $m$-site sets $M_1,\ldots,M_L$ such that
\[
 M_1=M,\qquad
 \bigcup_{i=1}^{L}M_i=\{1,\ldots,N\},
\]
and qudit permutations $V_i\in\Cl$ sending
$M_i$ to $M$, with $V_1=I_D$. Put
\[
 P_{a_i}^{M_i}:=V_i^\dagger\Pi_{a_i}V_i,
 \qquad R_i:=V_i^\dagger RV_i,
 \qquad A_i:=\bigcup_{j=1}^iM_j.
\]
For cases \textup{(a)} and \textup{(b)}, let
\[
 W_i:=(P_{a_i}^{M_i}R_i)\cdots
       (P_{a_1}^{M_1}R_1).
\]
We show by induction that there are one-qudit lines
$E_k^{(i)}\subseteq\C^d$ such that
\[
 \operatorname{Ran}W_i\subseteq
 \left(\bigotimes_{k\in A_i}E_k^{(i)}\right)
       \otimes\HH_{A_i^c},
 \qquad \HH_B:=\bigotimes_{k\in B}\C^d.
\]
For $i=1$, the projection $P_{a_1}^{M_1}$ fixes
all sites in $M_1$. In case \textup{(a)}, $R_i$
acts only on $M_i$ and leaves the previously fixed
lines outside $M_i$ unchanged. In case \textup{(b)},
$R_i$ is a product of single-qudit unitaries, so it
maps previously fixed lines to one-dimensional lines
without entangling them with other sites. In both
cases, the subsequent projector fixes all sites in
$M_i$. This establishes the induction, and hence
\[
 \rank W_i\le d^{N-|A_i|},\qquad
 \rank L_{\boldsymbol a}=\rank W_L\le1.
\]

In case \textup{(c)}, all $R_i$ remain diagonal in
the computational basis and commute with all the
$P_{a_j}^{M_j}$. Therefore
\[
 L_{\boldsymbol a}
 =\left(\prod_{i=1}^L P_{a_i}^{M_i}\right)
  \left(\prod_{i=1}^L R_i\right).
\]
The first product has rank at most one because the
measurement subsets cover every site; the second factor is unitary. Thus $\rank L_{\boldsymbol a}\le1$ in all three cases.

Finally, successive outcome summation gives
\[
 \sum_{\boldsymbol a}L_{\boldsymbol a}^\dagger
 L_{\boldsymbol a}=I_D.
\]
Applying the construction in the proof of
\cref{prop:clifford-TV-characterization} with the full outcome set gives the reset probability $g^{1-L}$ and overlap bound $a_D$. Then the TV estimate follows from \cref{prop:resets} which completes the proof.
\end{proof}

\subsection{Smooth unitary perturbations}

We now replace the fixed unitary $R$ with a smooth one-parameter family $R_\theta\in U(D)$ satisfying $R_0=I_D$, and study the resulting quantum trajectories under weak perturbations. The following condition specifies the perturbation, while the subsequent proposition verifies Conditions~\condK, \condS, and~\condQ\ required for the tangent-limit and response theorems in Sections~\ref{sec:tangent}--\ref{sec:response}.

\phantomsection\label{cond:coherent}
\noindent\textbf{(H)} \textit{Smooth unitary perturbation.}
The map $\theta\mapsto R_\theta\in U(D)$ is of class $C^2$ on an open interval containing zero, with $R_0=I_D$. Write
\begin{equation}\label{eq:weak-H}
 R_\theta=I_D-i\theta H+O(\theta^2),
 \qquad H=H^\dagger.
\end{equation}
For the associated quantum trajectories, write $P_\theta=P_{R_\theta}$ and $\pi_\theta=\pi_{R_\theta}$. We denote the stationary tangent law, departure measure and occupation measure by $\pih_H$, $\Lambda_H$, and $\mu_H^{\mathrm{occ}}$.

\begin{proposition}[Reference dynamics and tangent ergodicity]
\label{prop:verification}
For the Clifford trajectory satisfying Condition~\condH, the following assertions hold.
\begin{enumerate}
\item \textit{Verification of the conditions.}
Conditions~\condK, \condS, and~\condQ\ hold with $S=\St$, $\nu=u_S$ and the following parameters:
\begin{equation}\label{eq:reference-reset}
 L=\ceil{N/m},\qquad
 p_0=g^{1-L},\qquad
 \eta=1-p_0.
\end{equation}
In particular, the reference kernel $Q=P_I|_S$
is primitive.

\item \textit{Stationary tangent limit and transport.}
The conclusions of \cref{thm:tangent} hold, and the response theorems of Section~\ref{sec:response} apply to observables satisfying their respective hypotheses. Moreover,
\[
 W_2(\pi_\theta,u_S)=O(\theta)
 \qquad(\theta\downarrow0).
\]

\item \textit{Total-variation ergodicity of the affine tangent chain.}
There exist $C_{\mathrm{tan}}<\infty$ and $\rho_{\mathrm{tan}}\in(0,1)$, depending only on the fixed reference model, such that
\[
 \sup_{\mu\in\mathcal P(\TX)}
 \|\mu\Ph_H^n-\pih_H\|_{\TV}
 \le C_{\mathrm{tan}}\rho_{\mathrm{tan}}^n,
 \qquad n\ge0.
\]
\end{enumerate}
\end{proposition}

\begin{proof}
\emph{(i) Verification of the conditions.}
By Condition~\condH\ and \eqref{eq:clifford-Kraus},
\[
 K_{U,a}(\theta)=g^{-1/2}U^\dagger\Pi_a R_\theta U,
 \qquad
 \dot K_{U,a}=-ig^{-1/2}U^\dagger\Pi_a HU.
\]
The branch set is finite, each $K_{U,a}(\theta)$ is
$C^2$, and
\[
 \sum_{U,a}K_{U,a}(\theta)^\dagger K_{U,a}(\theta)
 =\frac1g\sum_U U^\dagger R_\theta^\dagger
       \Bigl(\sum_a\Pi_a\Bigr)R_\theta U=I_D.
\]
Thus Condition~\condK\ holds. At $R_0=I_D$, every nonzero branch preserves $S$.
The measurement-cover construction of
\cref{cor:clifford-reset-families} yields a set
$\mathcal E\subseteq J^L$ of rank-at-most-one Kraus
words whose nonzero output rays belong to $S$, with
\[
 \sum_{\omega\in\mathcal E}
 K_{\omega,0}^\dagger K_{\omega,0}=p_0I_D,
 \qquad L=\ceil{N/m},\quad p_0=g^{1-L}.
\]
For each $s\in S$, choose $\tau(\omega,s)\in S$
as follows. For a selected nonzero word, use its
input-independent range ray; for a selected zero
word, choose any element of $S$. For an unselected
word, use $[K_{\omega,0}s]$ if $K_{\omega,0}s\ne0$,
and otherwise choose any element of $S$.

Then $E_{\tau(\omega,s)}K_{\omega,0}s=0$ for
every word, and
$E_{\tau(\omega,s)}K_{\omega,0}=0$ for all
selected words. For unselected words,
\[
 E_{\tau(\omega,s)}K_{\omega,0}
 =E_{\tau(\omega,s)}K_{\omega,0}E_s.
\]
Consequently,
\[
 \begin{aligned}
 \sum_{\omega\in J^L}K_{\omega,0}^\dagger
     E_{\tau(\omega,s)}K_{\omega,0}
 &=\sum_{\omega\notin\mathcal E}
 E_sK_{\omega,0}^\dagger
     E_{\tau(\omega,s)}K_{\omega,0}E_s\\
 &\preceq E_s\Bigl(\sum_{\omega\notin\mathcal E}
 K_{\omega,0}^\dagger K_{\omega,0}\Bigr)E_s
 =(1-p_0)E_s.
 \end{aligned}
\]
This verifies Condition~\condQ\ with
$\eta=1-p_0$.

To establish Condition~\condS, let $B_0$ be the
conditional reset kernel on $S$. At $R_0=I_D$,
the measurement-cover projectors commute, and
their nonzero products are rank-one computational
basis projectors, indexed by $z\in\F_d^N$.
Thus, for $s\in S$,
\[
 B_0(s,\cdot)=\frac1g\sum_{U\in\Cl}
  \sum_{z\in\F_d^N}|\langle z,Us\rangle|^2
     \delta_{[U^\dagger z]}.
\]
Clifford transitivity on $S$ gives
\[
 B_0(s,t)=\frac D{|S|}|\langle t,s\rangle|^2,
 \qquad s,t\in S.
\]
The projective two-design identity for the
uniform stabilizer ensemble then yields
\[
 \begin{aligned}
 B_0^2(s,t)
 &=\frac{D^2}{|S|^2}\sum_{u\in S}
      |\langle t,u\rangle|^2
      |\langle u,s\rangle|^2\\
 &=\frac D{|S|(D+1)}
       \bigl(1+|\langle s,t\rangle|^2\bigr)>0.
 \end{aligned}
\]
Since $Q^L\ge p_0B_0$, it follows that
$Q^{2L}\ge p_0^2B_0^2>0$. Hence $Q$ is
primitive, and \cref{prop:stationary-structure}
identifies its invariant probability as $u_S$.
This proves Condition~\condS.

\emph{(ii) Stationary perturbation limits.}
The verified conditions allow us to apply
\cref{thm:tangent} and the response results for
observables satisfying their respective hypotheses.
Moreover, \eqref{eq:radial-transport} and
\eqref{eq:stationary-concentration} give
\[
 W_2(\pi_\theta,u_S)^2
 =\pi_\theta(r_S^2)\le C\theta^2,
\]
which proves the stated $W_2$ estimate.

\emph{(iii) Coupling the affine tangent chains.}
The lower bound on $Q^{2L}$ obtained in
part~\textup{(i)} gives the uniform minorization
\[
 Q^{2L}(s,\cdot)\ge\varepsilon_b u_S,
 \qquad
 \varepsilon_b:=p_0^2\frac D{D+1}\in(0,1).
\]
At successive $2L$-step endpoints, couple two
copies of the reference chain by using their
common $\varepsilon_bu_S$ component.
Conditional on each pair of endpoints, sample
the intermediate marked paths according to
their original conditional laws. Since the
base-chain branch probabilities do not depend
on the tangent vectors, this yields a valid
coupling of the two affine tangent chains.
The base chains can be kept together after
they meet, and their meeting time $\tau_b$
satisfies
\[
 \PP(\tau_b>2Lk)\le(1-\varepsilon_b)^k,
 \qquad k\ge0.
\]

After the bases meet, use identical branch
records. For every current reference state,
the selected reset event in the next $L$ steps
has probability $p_0$, since its Kraus effects
sum to $p_0I_D$. For each selected word with
positive reference probability, the homogeneous
$L$-step tangent map vanishes: its Kraus product
has rank at most one, so its projection
transverse to the output ray is zero.
The two tangent vectors therefore coincide
after such a word, because they have the same
affine forcing.

The additional waiting time $\tau_r$ satisfies
\[
 \PP(\tau_r>kL\mid\text{history at }\tau_b)
 \le(1-p_0)^k,
 \qquad k\ge0.
\]
Thus the complete coupling time
$\tau\le\tau_b+\tau_r$ obeys
\[
 \PP(\tau>n)
 \le(1-\varepsilon_b)^{\floor{n/(4L)}}
      +(1-p_0)^{\floor{n/(2L)}},
\]
where $0^0=1$. Since $\varepsilon_b\le p_0$ and $0<\varepsilon_b<1$, we may take
\[
 \rho_{\mathrm{tan}}
 :=(1-\varepsilon_b)^{1/(4L)}\in(0,1),
 \qquad
 C_{\mathrm{tan}}
 :=\frac2{1-\varepsilon_b},
\]
so that
$\PP(\tau>n)\le
C_{\mathrm{tan}}\rho_{\mathrm{tan}}^n$.
Coupling one copy with the invariant law
$\pih_H$ and using the coupling inequality
proves the required total-variation bound,
uniformly in the initial probability measure
and in $H$.
\end{proof}

\section{Clifford-driven tangent dynamics and stationary response}\label{sec:energy}

The preceding section established the ergodic structure of Clifford-driven quantum trajectories and verified the hypotheses of the general perturbative framework. Moreover, the Clifford structure allows us to explore a more precise characterization of stationary behavior, beginning with the stationary tangent law and its dependence on the perturbing Hamiltonian. We then pass from these local dynamics to the stationary measure on projective space.

\subsection{Stabilizer measurement statistics}

The Clifford structure makes the measurement statistics of uniform stabilizer states explicitly accessible. We collect the outcome counts and conditional second moments needed to derive the exact tangent-energy formula in the next subsection.

\begin{lemma}[Outcome counts and conditional stabilizer moments]
\label{lem:stabilizer-count}
Let $d$ be prime, $N\ge1$, and $1\le m\le N$. Set
$D=d^N$, $q=d^m$, and $r=d^{N-m}$, and let $\psi$ be
uniform on $\mathcal S_{d,N}$. For a unit representative of its
ray, write
\[
 p_a=\|\Pi_a\psi\|^2,\qquad
 k_M(\psi)=|\{a\in\F_d^m:p_a>0\}|.
\]
Then the following statements hold.
\begin{enumerate}
\item \textit{Number of nonzero outcomes.}
\begin{equation}\label{eq:support-count}
 \E k_M(\psi)=\frac{dD+1}{dr+1}.
\end{equation}

\item \textit{Born-weighted postselection.}
For a computational-basis measurement on any fixed $j$ sites,
$0\le j\le N$, and $b\in\F_d^j$, order the measured sites first.
In this item, let
$\xi_b=(\langle b|\otimes I)\psi$ and
$p_b=\|\xi_b\|^2$; when $p_b>0$, put
$\chi_b=\xi_b/\sqrt{p_b}$.
Then, for every function $f$ on $\mathcal S_{d,N-j}$,
\begin{equation}\label{eq:weighted-postselection}
 \E[p_b f([\chi_b])]
 =\frac{d^{-j}}{|\mathcal S_{d,N-j}|}
   \sum_{s\in\mathcal S_{d,N-j}}f(s),
\end{equation}
where the summand on the left is defined to be zero when
$p_b=0$, and $\mathcal S_{d,0}$ consists of the unique ray
of $\C$.

\item \textit{One measured qudit.}
Suppose $m=1$ and $N\ge2$, so that $r=d^{N-1}$.
Write $\HH=\HH_A\otimes\HH_B$, where
$\HH_A=\C^d$ is the measured qudit and
$\HH_B=\C^r$ is the remaining subsystem. Define
\[
 \rho_A:=\Tr_{\HH_B}(|\psi\rangle\langle\psi|),
 \qquad
 \chi_a:=\frac{(\langle a|\otimes I_r)\psi}{\sqrt{p_a}}
 \quad(p_a>0).
\]
Then exactly one of the following three events occurs:
\[
\begin{aligned}
 \mathsf E_{\mathrm{det}}
  &:=\{\rank\rho_A=1,\ k_M(\psi)=1\},\\
 \mathsf E_{\mathrm{prod}}
  &:=\{\rank\rho_A=1,\ k_M(\psi)=d\},\\
 \mathsf E_{\mathrm{ent}}
  &:=\{\rho_A=I_d/d\}.
\end{aligned}
\]
Their probabilities are
\begin{equation}\label{eq:one-qudit-cases}
 \begin{aligned}
 \PP(\mathsf E_{\mathrm{det}})
   &=\frac1{dr+1},\\
 \PP(\mathsf E_{\mathrm{prod}})
   &=\frac d{dr+1},\\
 \PP(\mathsf E_{\mathrm{ent}})
   &=\frac{d(r-1)}{dr+1}.
 \end{aligned}
\end{equation}

Define the Clifford orbit
\[
 \mathcal O:=
 \left\{
 \bigl([C|0,\ldots,0\rangle],
       [C|1,0,\ldots,0\rangle]\bigr):
 C\in\mathcal C_{d,N-1}
 \right\},
\]
and let $u_{\mathcal O}$ and $u_{\mathcal S_{d,N-1}}$ denote the uniform probability measures on $\mathcal O$ and $\mathcal S_{d,N-1}$, respectively. On
$\mathsf E_{\mathrm{prod}}\cup\mathsf E_{\mathrm{ent}}$,
\[
 p_a=\frac1d,\qquad a\in\F_d.
\]
Moreover, on the respective events,
\[
\begin{aligned}
 \mathsf E_{\mathrm{prod}}:\quad
 &[\chi_a]=[\chi_b] &&(a,b\in\F_d),\\
 &\mathcal L([\chi_a]\mid\mathsf E_{\mathrm{prod}})
   =u_{\mathcal S_{d,N-1}} &&(a\in\F_d);\\
 \mathsf E_{\mathrm{ent}}:\quad
 &\langle\chi_a,\chi_b\rangle=\delta_{ab}
   &&(a,b\in\F_d),\\
 &\mathcal L(([\chi_a],[\chi_b])
    \mid\mathsf E_{\mathrm{ent}})
   =u_{\mathcal O} &&(a\ne b).
\end{aligned}
\]

\item \textit{Conditional second moments.}
Under the assumptions of \textup{(iii)}, let $A\in M_r(\C)$.
For $a,b\in\F_d$ with $a\ne b$,
\begin{align}
 \E\!\left[|\langle\chi_a,A\chi_a\rangle|^2
       \mid\mathsf E_{\mathrm{prod}}\right]
  &=\frac{\|A\|_{\HS}^2+|\Tr A|^2}{r(r+1)},
     \label{eq:same-ray-moment}\\
 \E\!\left[|\langle\chi_a,A\chi_b\rangle|^2
       \mid\mathsf E_{\mathrm{ent}}\right]
  &=\frac{\|A\|_{\HS}^2-|\Tr A|^2/r}{r^2-1}.
     \label{eq:orthogonal-ray-moment}
\end{align}
Moreover, for every $a\in\F_d$,
\[
 \E\bigl[|\chi_a\rangle\langle\chi_a|
     \mid\mathsf E_{\mathrm{prod}}\bigr]
 =
 \E\bigl[|\chi_a\rangle\langle\chi_a|
     \mid\mathsf E_{\mathrm{ent}}\bigr]
 =\frac{I_r}{r}.
\]
\end{enumerate}
\end{lemma}

\begin{proof}
\emph{(i) Number of nonzero outcomes.}
We first recall the stabilizer-state count. On $n$ prime-dimensional
qudits, the computational support of a stabilizer ray is an affine
subspace of $\F_d^n$, and its nonzero amplitudes have equal modulus.
On an affine support of dimension $k$, the Pauli eigenvalue equations
restrict the phases, modulo a global phase, to a quadratic function. For odd $d$, these phases are parameterized by
\[
 \omega_d^{\frac12 x^TAx+b\cdot x},
 \qquad A=A^T\in M_k(\F_d),\quad b\in\F_d^k.
\]
For $d=2$, they are parameterized by
\[
 i^{\sum_{h=1}^k c_hx_h}
 (-1)^{\sum_{h<l}b_{hl}x_hx_l},
 \qquad c_h\in\mathbb Z_4,\quad b_{hl}\in\F_2.
\]
In either case the number of phase choices is
$d^{k(k+3)/2}$. If $\genfrac{[}{]}{0pt}{}{n}{k}_d$
denotes the Gaussian binomial coefficient, there are
$d^{n-k}\genfrac{[}{]}{0pt}{}{n}{k}_d$ affine supports of
dimension $k$. The Gaussian-binomial identity
\begin{equation}\label{eq:q-binomial}
 \sum_{k=0}^n\genfrac{[}{]}{0pt}{}{n}{k}_d
 d^{k(k-1)/2}x^k
 =\prod_{h=0}^{n-1}(1+xd^h)
\end{equation}
therefore gives
\begin{equation}\label{eq:stabilizer-number}
 |\mathcal S_{d,n}|
 =\sum_{k=0}^n d^{n-k}\genfrac{[}{]}{0pt}{}{n}{k}_d
 d^{k(k+3)/2}
 =d^n\prod_{h=1}^n(1+d^h),
\end{equation}
including $|\mathcal S_{d,0}|=1$.

Fix the measured outcome $\boldsymbol 0\in\F_d^m$.
Conditional on its having positive probability, the remaining ray is a
stabilizer ray. By Clifford transitivity on the remaining subsystem,
each possible remaining ray has the same number of stabilizer preimages.
We count those whose remaining ray is
$[|0\rangle^{\otimes(N-m)}]$. Their affine computational supports meet
$\{\boldsymbol0\}\times\F_d^{N-m}$ exactly at the origin, and hence
are linear subspaces. A support of dimension $k$ is precisely the
graph of a linear map from a $k$-dimensional subspace of $\F_d^m$
into $\F_d^{N-m}$. There are
$\genfrac{[}{]}{0pt}{}{m}{k}_d r^k$ such supports;
every allowed quadratic phase yields the required conditional ray.
Consequently, the number of preimages of any fixed remaining ray is
\[
 \sum_{k=0}^m\genfrac{[}{]}{0pt}{}{m}{k}_d
       r^k d^{k(k+3)/2}
 =\prod_{h=0}^{m-1}(1+rd^{h+2}),
\]
by \eqref{eq:q-binomial}. Using \eqref{eq:stabilizer-number} and
$D=qr$, we obtain
\[
 \begin{aligned}
 \PP(p_{\boldsymbol0}>0)
 &=\frac{|\mathcal S_{d,N-m}|}{|\mathcal S_{d,N}|}
      \prod_{h=0}^{m-1}(1+rd^{h+2})\\
 &=\frac{dD+1}{q(dr+1)}.
 \end{aligned}
\]
Measured-subsystem Pauli translations permute the $q$ outcomes while
preserving the uniform stabilizer ensemble. Thus
\[
 \E k_M(\psi)
 =\sum_{a\in\F_d^m}\PP(p_a>0)
 =q\,\PP(p_{\boldsymbol0}>0)
 =\frac{dD+1}{dr+1},
\]
as claimed.

\emph{(ii) Born-weighted postselection.}
For any $j$ and $b$ as in the statement, nonzero computational-basis
postselection of a stabilizer ray gives a stabilizer ray on the
remaining $N-j$ qudits. Consider the finite measure
\[
 \mu_b(f):=\E\bigl[p_bf([\chi_b])\bigr],
\]
with zero contribution when $p_b=0$. Since uniform stabilizer rays
form a projective one-design,
\[
 \mu_b(1)=\E p_b
 =\Tr\!\left[(|b\rangle\langle b|\otimes I_{d^{N-j}})
                 \frac{I_D}{D}\right]=d^{-j}.
\]
For every Clifford unitary $C$ on the remaining subsystem, the map
$\psi\mapsto(I_{d^j}\otimes C)\psi$ preserves the uniform stabilizer
ensemble and $p_b$, while sending $[\chi_b]$ to $[C\chi_b]$.
Thus $\mu_b$ is Clifford-invariant. Transitivity on
$\mathcal S_{d,N-j}$ and its total mass give precisely
\eqref{eq:weighted-postselection}. This also covers $j=0,N$.

\emph{(iii) One measured qudit.}
Write $\HH=\HH_A\otimes\HH_B$ as in the statement.
If $\mathcal G_\psi$ is the stabilizer group of $\psi$,
its Pauli expansion gives
\[
 |\psi\rangle\langle\psi|
 =\frac1D\sum_{g\in\mathcal G_\psi}g,
 \qquad
 \rho_A=\frac1d\sum_{h\in\mathcal G_{\psi,A}}h,
\]
where $\mathcal G_{\psi,A}$ consists of the single-qudit Pauli
operators $h$ for which $h\otimes I_r\in\mathcal G_\psi$.
Since $d$ is prime, this abelian subgroup has order either $1$
or $d$. Accordingly, $\rho_A=I_d/d$ or $\rho_A$ is a rank-one
stabilizer projector. These cases are disjoint and exhaustive.

Let $t:=\PP(\rank\rho_A=1)$. The stabilizer two-design identity
and the swap trick yield
\[
 \E\Tr\rho_A^2=\frac{d+r}{dr+1}.
\]
Since $\Tr\rho_A^2$ equals $1$ in the pure case and $1/d$ in
the maximally mixed case,
\[
 t+\frac{1-t}{d}=\frac{d+r}{dr+1},
 \qquad
 t=\frac{d+1}{dr+1}.
\]
If $\rho_A$ is pure, then $\psi=\phi\otimes\eta$ for
single-qudit and $(N-1)$-qudit stabilizer rays $[\phi]$ and
$[\eta]$. Conditional on this event, these factors are independent
and uniform: the correspondence between pairs of local stabilizer
rays and product stabilizer rays is bijective.
By \eqref{eq:stabilizer-number}, there are $d(d+1)$ single-qudit
stabilizer rays, of which exactly $d$ are computational-basis rays.
Every other one is an eigenray of a Pauli $X^uZ^v$ with $u\ne0$;
as $d$ is prime, its computational amplitudes all have modulus
$d^{-1/2}$. It follows that the pure-marginal case splits into
$k_M(\psi)=1$ and $k_M(\psi)=d$, with probabilities
\[
 \PP(\mathsf E_{\mathrm{det}})
 =t\frac1{d+1}=\frac1{dr+1},
 \qquad
 \PP(\mathsf E_{\mathrm{prod}})
 =t\frac d{d+1}=\frac d{dr+1}.
\]
For $\mathsf E_{\mathrm{prod}}$, the factors satisfy
$\chi_a=e^{i\alpha_a}\eta$ for suitable phases, so all their
rays coincide; their common ray is uniform on
$\mathcal S_{d,N-1}$.

On $\mathsf E_{\mathrm{ent}}$ we have $\rho_A=I_d/d$, and hence
\[
 \PP(\mathsf E_{\mathrm{ent}})=1-t
 =\frac{d(r-1)}{dr+1},
 \qquad
 p_a=\frac1d,
 \qquad
 \langle\chi_a,\chi_b\rangle=\delta_{ab}.
\]
It remains to identify the joint distribution of the conditional
rays. Put $n=N-1$ and let
$L_\psi\subset\F_d^2\oplus\F_d^{2n}$ be the Lagrangian Pauli
label space of the global stabilizer. The condition
$\rho_A=I_d/d$ means that
$L_\psi\cap(\F_d^2\oplus\{0\})=\{0\}$.
With $\perp_A$ denoting symplectic orthogonality on the first
summand and $L_\psi=L_\psi^\perp$, we have
\[
 (\operatorname{pr}_A L_\psi)^{\perp_A}
 =\{u\in\F_d^2:(u,0)\in L_\psi^\perp\}
 =\{0\}.
\]
Thus $\operatorname{pr}_A L_\psi=\F_d^2$, and its
kernel has dimension $n-1$.
Choose $n-1$ independent stabilizer labels in this kernel,
and one whose first-qudit component is the label of $Z$.
Their remaining-system components are $n$ independent,
commuting Pauli labels. After postselection on the first qudit,
the conditional vectors $\chi_a$ are joint eigenvectors of the
corresponding remaining-system Paulis. The eigenvalue of the
last generator varies with $a$, so the $d$ conditional rays
are distinct members of a common stabilizer basis.

A Clifford on $\HH_B$ maps this basis to the computational basis;
an affine invertible computational relabeling sends any ordered
pair of distinct basis labels to $(0,e_1)$. Thus, for fixed
$a\ne b$, the pair $([\chi_a],[\chi_b])$ belongs to the orbit
$\mathcal O$ defined in the statement. Finally, the event
$\mathsf E_{\mathrm{ent}}$ and the uniform initial ensemble are
invariant under $I_d\otimes\mathcal C_{d,N-1}$. As this group
acts transitively on $\mathcal O$, the conditional law of the
ordered pair is $u_{\mathcal O}$, completing~\textup{(iii)}.

\emph{(iv) Conditional second moments.}
Let $Q_a:=|\chi_a\rangle\langle\chi_a|$ and let $\Flip$ be the
swap operator on $\C^r\otimes\C^r$.
By the conditional uniformity established in~\textup{(iii)}
and the stabilizer two-design identity,
\[
 \E[Q_a\otimes Q_a\mid\mathsf E_{\mathrm{prod}}]
 =\frac{I_{r^2}+\Flip}{r(r+1)}.
\]
For $a\ne b$, the ordered pair on $\mathsf E_{\mathrm{ent}}$
is uniform on the Clifford orbit of an orthogonal pair.
Since the Clifford group is a unitary two-design, its tensor-square
average lies in $\operatorname{span}\{I_{r^2},\Flip\}$.
Moreover,
\[
 \Tr\E[Q_a\otimes Q_b\mid\mathsf E_{\mathrm{ent}}]=1,
 \qquad
 \Tr\!\left[\Flip\,
   \E[Q_a\otimes Q_b\mid\mathsf E_{\mathrm{ent}}]\right]=0,
\]
the second identity following from
$\langle\chi_a,\chi_b\rangle=0$.
Solving for the two coefficients gives
\[
 \E[Q_a\otimes Q_b\mid\mathsf E_{\mathrm{ent}}]
 =\frac{I_{r^2}-\Flip/r}{r^2-1}.
\]
For unit vectors $u,v\in\C^r$ and $A\in M_r(\C)$,
\[
 |\langle u,Av\rangle|^2
 =\Tr\!\left[(A\otimes A^\dagger)\Flip
    (|u\rangle\langle u|\otimes|v\rangle\langle v|)\right],
\]
and
\[
 \Tr[(A\otimes A^\dagger)\Flip]=\|A\|_{\HS}^2,
 \qquad
 \Tr(A\otimes A^\dagger)=|\Tr A|^2.
\]
Substitution proves \eqref{eq:same-ray-moment} and
\eqref{eq:orthogonal-ray-moment}. Taking either partial trace
of the two conditional tensor moments shows, for every $a$,
\[
 \E[Q_a\mid\mathsf E_{\mathrm{prod}}]
 =\E[Q_a\mid\mathsf E_{\mathrm{ent}}]=\frac{I_r}{r},
\]
which completes the proof.
\end{proof}

\subsection{Exact tangent energy and effective Hamiltonians}

For Clifford-driven quantum trajectories, the symmetry of the Clifford group and the algebraic structure of stabilizer states allow for a precise quantitative characterization of stationary tangent fluctuations. In particular, the Clifford two-design property, together with \cref{lem:stabilizer-count}, enables an exact evaluation of the stationary tangent energy and a characterization of the Hamiltonian directions that produce nontrivial fluctuations.

The measurement projectors distinguish three orthogonal components of a Hermitian generator $H$:
\begin{equation}\label{eq:H-decomposition}
 \begin{split}
 H_0&=\sum_a\frac{\Tr(\Pi_aH)}r\Pi_a,\\
 H_{\mathrm c}&=\sum_a\Pi_aH\Pi_a-H_0,\\
 H_{\mathrm{nc}}&=H-\sum_a\Pi_aH\Pi_a.
 \end{split}
\end{equation}
Then $H=H_0+H_{\mathrm c}+H_{\mathrm{nc}}$. The component $H_0$ is scalar on each measurement sector, $H_{\mathrm c}$ is traceless within every diagonal block, and $H_{\mathrm{nc}}$ consists of the off-diagonal blocks. These components are Hilbert--Schmidt orthogonal. Define
\begin{equation}\label{eq:energies}
 E_{\mathrm c}(H)=\frac{\|H_{\mathrm c}\|_{\HS}^2}{D},
 \qquad
 E_{\mathrm{nc}}(H)=\frac{\|H_{\mathrm{nc}}\|_{\HS}^2}{D},
\end{equation}
and define the real subspace of measurement-sector phases by
\[
 \mathcal A_M=\left\{\sum_a h_a\Pi_a:h_a\in\R\right\}.
\]
Thus $H_0$ is the Hilbert--Schmidt projection of $H$ onto $\mathcal A_M$. The next theorem gives an exact recursion for tangent energy and determines its stationary value.

\begin{theorem}[Exact tangent energy and stationary susceptibility]
\label{thm:energy}
Suppose Condition~\condH\ holds with perturbing Hamiltonian $H\in\Herm(D)$ for the Clifford-driven model on $N$ qudits of prime local dimension $d$, with $1\le m\le N$ measured qudits. Set
\[
 D=d^N,\qquad q=d^m,\qquad r=d^{N-m}.
\]
Write $V(s,v)=\|v\|$, and let $\pih_H$ denote the unique stationary law of the affine tangent kernel $\Ph_H$. Set
\begin{align}
 \sigma&=\frac{(r-1)(dD+1)}{(D-1)(dr+1)},
 &B(H)&=\frac r{r+1}E_{\mathrm c}(H)
             +\frac{d(r-1)}{dr+1}E_{\mathrm{nc}}(H),
             \label{eq:sigma-B}\\
 C_{\mathrm c}&=\frac{(D-1)(dr+1)}{(d+1)(q-1)(r+1)},
 &C_{\mathrm{nc}}&=\frac{d(r-1)(D-1)}{(d+1)r(q-1)}.
             \label{eq:C-coefficients}
\end{align}
Then the following statements hold.

\begin{enumerate}
\item \textit{One-step tangent energy.}
For every $s\in S$ and $v\in s^\perp$,
\begin{equation}\label{eq:exact-energy}
 \Pt V^2(s,v)=\sigma\|v\|^2,
 \qquad
 \Ph_HV^2(s,v)=\sigma\|v\|^2+B(H),
\end{equation}
where $0\le\sigma<1$.

\item \textit{Energy evolution and stationary susceptibility.}
For any affine tangent chain $(S_n,V_n)_{n\ge0}$ whose
initial law has finite second moment,
\begin{equation}\label{eq:energy-recurrence}
 \E\|V_n\|^2=\sigma^n\E\|V_0\|^2
                   +\frac{1-\sigma^n}{1-\sigma}B(H),
 \qquad n\ge0,
\end{equation}
with $\sigma^0:=1$. In particular,
\begin{equation}\label{eq:M2-closed}
 M_2(H):=\pih_H(V^2)
       =\frac{B(H)}{1-\sigma}
       =C_{\mathrm c}E_{\mathrm c}(H)+C_{\mathrm{nc}}E_{\mathrm{nc}}(H).
\end{equation}

\item \textit{Effective Hamiltonian directions.}
If $m<N$, then $0<C_{\mathrm{nc}}<C_{\mathrm c}$, and
\begin{equation}\label{eq:effective-directions}
 \begin{aligned}
 H\in\mathcal A_M
 &\Longleftrightarrow b_{j,s}(H)=0
       \quad(s\in S,\ j\in J_+(s))\\
 &\Longleftrightarrow M_2(H)=0.
 \end{aligned}
\end{equation}
Moreover,
\begin{equation}\label{eq:sharp-energy}
 C_{\mathrm{nc}}\frac{\|H-H_0\|_{\HS}^2}{D}
       \le M_2(H)\le
 C_{\mathrm c}\frac{\|H-H_0\|_{\HS}^2}{D},
\end{equation}
with equality precisely when
\[
\begin{aligned}
 M_2(H)=C_{\mathrm{nc}}\frac{\|H-H_0\|_{\HS}^2}{D}
   &\Longleftrightarrow H_{\mathrm c}=0,\\
 M_2(H)=C_{\mathrm c}\frac{\|H-H_0\|_{\HS}^2}{D}
   &\Longleftrightarrow H_{\mathrm{nc}}=0.
\end{aligned}
\]
Thus $[H]\mapsto M_2(H)$ is a positive-definite real
quadratic form on $\Herm(D)/\mathcal A_M$. For the exact
exponential path,
\[
 H\in\mathcal A_M\quad\Longrightarrow\quad
 P_{e^{-i\theta H}}=P_I\qquad(\theta\in\R).
\]
\end{enumerate}
\end{theorem}

\begin{proof}
\emph{(i) One-step tangent energy.}
Fix $s\in S$ and $v\in s^\perp$. Let $G_s$ be the
stabilizer of $s$ in the projective Clifford group, with
unitary representatives chosen to fix $s$. Clifford
symmetry gives
\begin{equation}\label{eq:input-stabilizer-average}
 \frac1{|G_s|}\sum_{C\in G_s}Cv=0,
 \qquad
 \frac1{|G_s|}\sum_{C\in G_s}
       C|v\rangle\langle v|C^\dagger
 =\frac{\|v\|^2}{D-1}E_s.
\end{equation}
Indeed, for $|0\rangle^{\otimes N}$, averaging over diagonal Pauli operators gives a zero vector mean and a diagonal second-moment matrix. The latter has zero entry at the computational label $0$, and the computational permutations induced by $\operatorname{GL}(N,d)$ act transitively on the nonzero labels. The resulting averages are $0$ and $\|v\|^2E_s/(D-1)$, respectively. These are invariant under $G_s$, so further averaging does not change them. Clifford transitivity extends the identities to every $s\in S$. For a uniformly drawn $U\in\Cl$, put $\psi=Us$ and after ordering the measured subsystem first, write
\[
 p_a=\|\Pi_a\psi\|^2,
 \qquad
 \Pi_a\psi=\sqrt{p_a}\,|a\rangle\otimes\chi_a
 \quad(p_a>0),
\]
where $\chi_a\in\C^r$ is a unit vector. Define
\[
 T_a:=\Pi_a-|a\rangle\langle a|\otimes
                    |\chi_a\rangle\langle\chi_a|
 \quad(p_a>0).
\]
Then $T_a$ is an orthogonal projector satisfying
\[
 T_a\psi=0,\qquad \Tr T_a=r-1.
\]
Conditional on $\psi$, choose the global phases of the
Clifford representatives so that $Us=\psi$. The remaining
Clifford draws form a coset of $G_s$; hence
\eqref{eq:input-stabilizer-average} implies
\[
 \E(Uv\mid\psi)=0,
 \qquad
 \E\bigl(|Uv\rangle\langle Uv|\mid\psi\bigr)
 =\frac{\|v\|^2}{D-1}(I_D-\rho_\psi).
\]
For $j=(U,a)$ with $p_a>0$, the definitions
\eqref{eq:clifford-Kraus} and \eqref{eq:A-b}, together with
$\dot R_0=-iH$, give
\[
 \begin{aligned}
 p_j(s)\|A_{j,s}v\|^2
   &=g^{-1}\|T_aUv\|^2,\\
 p_j(s)\|A_{j,s}v+b_{j,s}(H)\|^2
   &=g^{-1}\|T_a(Uv-iH\psi)\|^2.
 \end{aligned}
\]
In the first identity, conditioning on $\psi$ and using
$T_a\rho_\psi=0$ yields
\[
 \begin{aligned}
 \Pt V^2(s,v)
 &=\frac{\|v\|^2}{D-1}
   \E_\psi\sum_{a:p_a>0}\Tr T_a\\
 &=\frac{r-1}{D-1}\E_\psi k_M(\psi)\,\|v\|^2
 =\sigma\|v\|^2,
 \end{aligned}
\]
where the last equality follows from \eqref{eq:support-count}.
For the affine kernel, the conditional mean of $Uv$ is zero,
so the cross term vanishes upon expanding the second identity.
Consequently,
\begin{equation}\label{eq:B-raw}
 \Ph_HV^2(s,v)=\sigma\|v\|^2+B(H),
 \qquad
 B(H)=\E_\psi\sum_{a:p_a>0}\|T_aH\psi\|^2.
\end{equation}
It remains to evaluate $B(H)$. If $r=1$, then $T_a=0$ on every positive branch, so
$B(H)=0$. This agrees with \eqref{eq:sigma-B}, because $H_{\mathrm c}=0$ and the coefficient of $E_{\mathrm{nc}}(H)$ vanishes. Henceforth assume $r>1$. Identify $H_{ab}:=\Pi_aH\Pi_b$ with an $r\times r$
block. Averaging over the diagonal Pauli operators on the
measured subsystem removes cross terms between distinct
input blocks: these operators preserve $p_a$ and $T_a$,
while multiplying the vector in block $b$ by its character.
It follows that
\begin{equation}\label{eq:block-forcing}
 \begin{split}
 B(H)&=\sum_{a,b}B_{ab},\\
 B_{ab}&:=\E_\psi\!\left[
  \one_{\{p_a>0\}}p_b
  \left(\|H_{ab}\chi_b\|^2
   -|\langle\chi_a,H_{ab}\chi_b\rangle|^2\right)
  \right].
 \end{split}
\end{equation}
A summand with $p_ap_b=0$ is defined as zero.

For $a=b$, \eqref{eq:weighted-postselection} with $j=m$ identifies the Born-weighted conditional-ray measure as $q^{-1}$ times the uniform stabilizer law on the remaining $N-m$ qudits. Applying the stabilizer projective two-design identity \eqref{eq:two-design} on this $r$-dimensional subsystem gives
\[
 \begin{aligned}
 B_{aa}
 &=\frac1q\left[
   \frac{\Tr(H_{aa}^2)}r
   -\frac{\Tr(H_{aa}^2)+(\Tr H_{aa})^2}{r(r+1)}
   \right]\\
 &=\frac1{q(r+1)}
   \left[\Tr(H_{aa}^2)-\frac{(\Tr H_{aa})^2}{r}\right].
 \end{aligned}
\]
Since $H_{aa}$ is Hermitian, summing over $a$ yields
\[
 \sum_aB_{aa}
 =\frac{\|H_{\mathrm c}\|_{\HS}^2}{q(r+1)}
 =\frac r{r+1}E_{\mathrm c}(H).
\]

Now fix $a\ne b$ and put
$t=\|H_{ab}\|_{\HS}^2$ and $u=|\Tr H_{ab}|^2$.
A measured-subsystem affine Clifford permutation sends the
ordered pair $(a,b)$ to $(0,e_1)$. By Born-weighted
postselection on the other $m-1$ measured qudits,
\eqref{eq:weighted-postselection} reduces the expectation
to one measured qudit with $r$ remaining dimensions and
contributes a factor $d^{-(m-1)}$. The event
$\mathsf E_{\mathrm{det}}$ contributes zero, whereas on
$\mathsf E_{\mathrm{prod}}$ and $\mathsf E_{\mathrm{ent}}$
the selected outcome has probability $1/d$. Using
\eqref{eq:one-qudit-cases} and the conditional moments
\eqref{eq:same-ray-moment}--\eqref{eq:orthogonal-ray-moment},
we obtain
\[
 \begin{aligned}
 B_{ab}
 &=\frac{d^{-(m-1)}}d\Bigg[
   \frac d{dr+1}
    \left(\frac tr-\frac{t+u}{r(r+1)}\right)\\
 &\hspace{60pt}+
   \frac{d(r-1)}{dr+1}
    \left(\frac tr-\frac{t-u/r}{r^2-1}\right)
   \Bigg]\\
 &=\frac{d^{-(m-1)}(r-1)}{r(dr+1)}\,t
  =\frac{d(r-1)}{D(dr+1)}\|H_{ab}\|_{\HS}^2.
 \end{aligned}
\]
In particular, the $|\Tr H_{ab}|^2$ terms cancel exactly.
Therefore
\[
 \sum_{a\ne b}B_{ab}
 =\frac{d(r-1)}{dr+1}E_{\mathrm{nc}}(H).
\]
Together with the diagonal contribution, this proves \eqref{eq:exact-energy}.

\emph{(ii) Energy evolution and stationary susceptibility.}
Using $D=qr$, we have
\[
 1-\sigma
 =\frac{(d+1)r(q-1)}{(D-1)(dr+1)}>0.
\]
For the affine tangent chain, set
$m_n:=\E\|V_n\|^2$. The one-step identity yields
\[
 m_{n+1}=\sigma m_n+B(H),
\]
so iteration gives \eqref{eq:energy-recurrence}, with
$\sigma^0=1$ also when $\sigma=0$.
By \cref{prop:verification,prop:affine}, the invariant law
$\pih_H$ has finite second moment. Integrating
\eqref{eq:exact-energy} against this law and using
stationarity yields
\[
 M_2(H)=\sigma M_2(H)+B(H),
 \qquad M_2(H)=\frac{B(H)}{1-\sigma}.
\]
Substituting the two coefficients of $B(H)$ and the
expression for $1-\sigma$ gives exactly the constants
$C_{\mathrm c}$ and $C_{\mathrm{nc}}$ in
\eqref{eq:C-coefficients}, proving \eqref{eq:M2-closed}.

\emph{(iii) Effective Hamiltonian directions.}
Suppose $m<N$, so that $r>1$. The coefficients in
\eqref{eq:C-coefficients} satisfy $C_{\mathrm{nc}}>0$ and
\[
 C_{\mathrm c}-C_{\mathrm{nc}}
 =\frac{(D-1)(r+d)}
       {(d+1)(q-1)r(r+1)}>0.
\]
By orthogonality of the decomposition
\eqref{eq:H-decomposition},
\[
 \|H-H_0\|_{\HS}^2
 =\|H_{\mathrm c}\|_{\HS}^2
  +\|H_{\mathrm{nc}}\|_{\HS}^2.
\]
The closed formula \eqref{eq:M2-closed} therefore gives
\eqref{eq:sharp-energy}. Because
$C_{\mathrm c}>C_{\mathrm{nc}}$, equality in the lower
bound holds exactly when $H_{\mathrm c}=0$, and equality
in the upper bound exactly when $H_{\mathrm{nc}}=0$.

For every $s\in S$, evaluating the affine identity
\eqref{eq:exact-energy} at $v=0$ yields
\[
 B(H)=\sum_{j\in J_+(s)}p_j(s)\|b_{j,s}(H)\|^2.
\]
All the weights in this sum are strictly positive. Since
$1-\sigma>0$ and both coefficients in
\eqref{eq:M2-closed} are positive, it follows that
\[
 \begin{aligned}
 M_2(H)=0
 &\Longleftrightarrow B(H)=0\\
 &\Longleftrightarrow
    b_{j,s}(H)=0\quad(s\in S,\ j\in J_+(s))\\
 &\Longleftrightarrow
    H_{\mathrm c}=H_{\mathrm{nc}}=0
 \Longleftrightarrow H\in\mathcal A_M.
 \end{aligned}
\]
Moreover, $H\mapsto(H_{\mathrm c},H_{\mathrm{nc}})$ is
real-linear, and its kernel is $\mathcal A_M$. The
positive-definite quadratic-form assertion on
$\Herm(D)/\mathcal A_M$ follows directly from
\eqref{eq:M2-closed}. Finally, for $H\in\mathcal A_M$ write
$H=\sum_a h_a\Pi_a$ with $h_a\in\R$. For the exact
exponential path $R_\theta=e^{-i\theta H}$,
\[
 \Pi_aR_\theta=e^{-i\theta h_a}\Pi_a,
 \qquad
 K^{R_\theta}_{U,a}
 =e^{-i\theta h_a}K^{I_D}_{U,a}.
\]
These branchwise phases change neither the Born probabilities nor the output rays. Hence
$P_{e^{-i\theta H}}=P_I$ for every $\theta\in\R$, as asserted.
\end{proof}

\subsection{Finite-memory tangent dynamics and exponential integrability}
\label{subsec:tangent-moments}

The preceding subsection gives an exact formula for the stationary tangent energy and identifies the Hamiltonian directions responsible for nontrivial tangent fluctuations. The Clifford--stabilizer structure allows us to go further: every nonzero reference Kraus word has equal nonzero singular values, yielding a uniform bound on its projective derivative along stabilizer rays, independent of the word length. Combined with the exact quadratic contraction, this property leads to an almost surely finite-memory representation of the stationary tangent law, together with moment bounds at every finite order and exponential integrability.

A \emph{stabilizer-code projector} on $\HH=\C^D$, $D=d^N$, is the orthogonal projector onto a joint eigenspace of commuting Pauli operators. After choosing their phases, it has the form
\[
 P_{\mathcal G}=\frac1{|\mathcal G|}\sum_{h\in\mathcal G}h,
 \qquad \rank P_{\mathcal G}=\frac D{|\mathcal G|},
\]
where $\mathcal G$ is an abelian Pauli subgroup with no nonidentity scalar. Its order and the projector's rank are powers of $d$. We use fixed unit representatives of the rays in $S$, as in \eqref{eq:A-b}.

\begin{lemma}[Reference Kraus words and uniform tangent bounds]
\label{lem:flat-reference-words}
Let $W=K_{\omega,0}\ne0$ be a reference Kraus word,
and set $k_W:=\rank W$. Then there exist
$\gamma_W>0$ and a stabilizer-code projector $P_W$
such that
\begin{equation}\label{eq:flat-word-effect}
 W^\dagger W=\gamma_WP_W,
 \qquad
 \rank P_W=k_W\in\{d^j:0\le j\le N\}.
\end{equation}
In particular, all nonzero singular values of $W$ equal $\sqrt{\gamma_W}$. If $k_W=1$, the range of $W$ is a stabilizer ray in $S$. For $s\in S$ with $Ws\ne0$, let $t=[Ws]\in S$.
Using the fixed unit representatives of $s$ and $t$, write $Ws=\lambda_{W,s}t$, and define
\[
 \alpha_{W,s}:=\langle s,P_Ws\rangle>0,
 \qquad
 u_{W,s}:=\frac{P_Ws}{\sqrt{\alpha_{W,s}}},
 \qquad
 \mathsf A_{W,s}:=\lambda_{W,s}^{-1}
                    E_tW|_{s^\perp}.
\]
Then
\begin{equation}\label{eq:flat-word-tangent}
 \begin{aligned}
 \alpha_{W,s}&\ge\frac{k_W}{D},\\
 \mathsf A_{W,s}^\dagger\mathsf A_{W,s}
 &=\left.
   \frac{P_W-|u_{W,s}\rangle\langle u_{W,s}|}
        {\alpha_{W,s}}
   \right|_{s^\perp}.
 \end{aligned}
\end{equation}
Consequently,
$\mathsf A_{W,s}=0$ if and only if $k_W=1$.
If $k_W>1$, then
\begin{equation}\label{eq:uniform-word-derivative}
 \|\mathsf A_{W,s}\|_{\op}\le\sqrt{\frac Dd},
 \qquad
 \|\mathsf A_{W,s}\|_{\HS}^2
 =\frac{k_W-1}{\alpha_{W,s}}\ge d-1.
\end{equation}
The Hilbert--Schmidt norm is taken on the complex
tangent space $s^\perp$.
\end{lemma}

\begin{proof}
\emph{Flat spectrum and stabilizer output.}
Write $W=g^{-\ell/2}Q_\ell\cdots Q_1$, where each
$Q_i=U_i^\dagger\Pi_{a_i}U_i$ is a projector onto a joint
Pauli eigenspace. The maximally entangled vector
\[
 |\Omega_D\rangle=D^{-1/2}
       \sum_{x\in\F_d^N}|x,x\rangle
\]
is a stabilizer vector on $2N$ qudits. Since a nonzero
Pauli postselection preserves stabilizer states,
\[
 |\Psi_W\rangle
 :=\frac{(W\otimes I_D)|\Omega_D\rangle}
         {\|(W\otimes I_D)|\Omega_D\rangle\|}
\]
is a stabilizer vector as well.

Recall that the reduced density matrix of a pure stabilizer
state is a normalized stabilizer-code projector. Indeed, if $\mathcal G$
is its stabilizer group on $2N$ qudits, then
\[
 |\Psi_W\rangle\langle\Psi_W|
   =\frac1{D^2}\sum_{h\in\mathcal G}h,
 \qquad
 \Tr_{\mathrm{out}}|\Psi_W\rangle\langle\Psi_W|
   =\frac1D\sum_{h\in\mathcal G_{\mathrm{in}}}h
   =\frac{P_{\mathrm{in}}}{\rank P_{\mathrm{in}}},
\]
where $\mathcal G_{\mathrm{in}}$ consists of the Pauli
operators $h$ with $I_D\otimes h\in\mathcal G$, and
$P_{\mathrm{in}}$ is the corresponding stabilizer-code
projector. In particular, $\rank P_{\mathrm{in}}$ is a
power of $d$. On the other hand, the Choi identity gives
\[
 \Tr_{\mathrm{out}}|\Psi_W\rangle\langle\Psi_W|
       =\frac{(W^\dagger W)^T}{\Tr(W^\dagger W)}.
\]
Transposition preserves stabilizer-code projectors, since
complex conjugation maps Pauli operators to Pauli operators
up to phases. Thus $P_W:=P_{\mathrm{in}}^T$ is a
stabilizer-code projector, and
\[
 W^\dagger W=\gamma_WP_W,
 \qquad
 \gamma_W:=\frac{\Tr(W^\dagger W)}{k_W}>0,
 \qquad
 \rank P_W=k_W\in\{d^j:0\le j\le N\},
\]
proving \eqref{eq:flat-word-effect}. Hence every nonzero
singular value of $W$ equals $\sqrt{\gamma_W}$.
If $k_W=1$, the output marginal
$WW^\dagger/\Tr(W^\dagger W)$ is a
rank-one stabilizer-code projector, so the range ray of
$W$ belongs to $S$. More generally, for $s\in S$ with
$Ws\ne0$, successive nonzero Pauli postselections of $s$
show that $[Ws]\in S$.

\emph{Overlap with a reference ray.}
Fix $s\in S$ with $Ws\ne0$. Write
\[
 P_W=\frac1{|\mathcal G_W|}\sum_{h\in\mathcal G_W}h,
 \qquad |\mathcal G_W|=\frac{D}{k_W}.
\]
For a Pauli operator $h$, the stabilizer property of $s$
implies that $\langle s,hs\rangle$ is either zero or a
phase; it is nonzero exactly when $h$ acts as a scalar on
$s$. Hence
\[
 \mathcal G_{W,s}
 :=\{h\in\mathcal G_W:\langle s,hs\rangle\ne0\}
\]
is a subgroup, and
$\chi_s(h):=\langle s,hs\rangle$ defines a character
of $\mathcal G_{W,s}$. Character orthogonality yields
\[
 \alpha_{W,s}
 =\frac1{|\mathcal G_W|}
      \sum_{h\in\mathcal G_{W,s}}\chi_s(h)
 \in\left\{0,\frac{|\mathcal G_{W,s}|}
                         {|\mathcal G_W|}\right\}.
\]
But $\|Ws\|^2=\gamma_W\alpha_{W,s}>0$, so
\[
 \alpha_{W,s}
 =\frac{|\mathcal G_{W,s}|}{|\mathcal G_W|}
 \ge\frac1{|\mathcal G_W|}=\frac{k_W}{D}.
\]

\emph{Tangent derivative and uniform bounds.}
Along the word $W=K_{j_\ell,0}\cdots K_{j_1,0}$,
let $t_0=s$ and let $t_i\in S$ be the reference output
ray after $i$ steps, with fixed unit representatives and
$K_{j_i,0}t_{i-1}=\lambda_i t_i$.
Every $\lambda_i$ is nonzero, and
$\lambda_{W,s}=\prod_{i=1}^\ell\lambda_i$.
Since
\[
 E_{t_i}K_{j_i,0}(I_D-E_{t_{i-1}})=0,
\]
the intermediate projections in the product of the
single-step derivatives \eqref{eq:A-b} can be removed.
The chain rule therefore gives
\[
 A_{j_\ell,t_{\ell-1}}\cdots A_{j_1,s}
  =\lambda_{W,s}^{-1}E_tW|_{s^\perp}
  =\mathsf A_{W,s},
 \qquad t=t_\ell.
\]

From \eqref{eq:flat-word-effect} and
$Ws=\lambda_{W,s}t$ we obtain
\[
 |\lambda_{W,s}|^2=\gamma_W\alpha_{W,s},
 \qquad
 W^\dagger E_tW
 =\gamma_W\left(P_W-
       \frac{P_W|s\rangle\langle s|P_W}
            {\alpha_{W,s}}\right).
\]
As $u_{W,s}=P_Ws/\sqrt{\alpha_{W,s}}$ is a unit
vector in $\operatorname{Ran}P_W$, this implies
\[
 \mathsf A_{W,s}^\dagger\mathsf A_{W,s}
 =\left.\frac{P_W-|u_{W,s}\rangle\langle u_{W,s}|}
                  {\alpha_{W,s}}\right|_{s^\perp}.
\]
The numerator is an orthogonal projector of rank
$k_W-1$ and maps $s$ to zero. Hence
$\mathsf A_{W,s}=0$ exactly when $k_W=1$. If $k_W>1$,
then $k_W\ge d$, and
\[
 \|\mathsf A_{W,s}\|_{\op}^2
 =\frac1{\alpha_{W,s}}
 \le\frac{D}{k_W}\le\frac Dd,
 \qquad
 \|\mathsf A_{W,s}\|_{\HS}^2
 =\frac{k_W-1}{\alpha_{W,s}}
 \ge k_W-1\ge d-1,
\]
using $\alpha_{W,s}\le1$. These are
\eqref{eq:flat-word-tangent} and
\eqref{eq:uniform-word-derivative}.
\end{proof}

For $m<N$, let $\sigma\in(0,1)$ be given by \eqref{eq:sigma-B}, and put
\begin{equation}\label{eq:moment-tail-constants}
 C_A:=\sqrt{D/d},\qquad
 C_{\mathrm{surv}}
 :=\frac{(D-1)(d^2+1)}{(d-1)(dD+1)},\qquad
 \kappa:=-\ln\sigma.
\end{equation}
For $p>0$, define
\begin{equation}\label{eq:moment-order-constant}
 a_p:=
 \begin{cases}
 C_{\mathrm{surv}}^{\,1-p/2},&0<p\le2,\\
 C_A^{\,p-2},&p\ge2.
 \end{cases}
\end{equation}
For a Hermitian generator $H$, write
\[
 \mathfrak b_H:=\max_{s\in S,\,j\in J_+(s)}\|b_{j,s}(H)\|,
 \qquad
 M_p(H):=\pih_H(V^p)\quad(p>0).
\]
The following theorem provides a stronger characterization of the stationary tangent law: an almost surely finite-memory representation with matching geometric tail bounds and, when $r=d$, an exact shifted geometric law, together with moment bounds at every finite order and exponential integrability.

\begin{theorem}[Finite memory, moment bounds, and exponential integrability]
\label{thm:tangent-exponential-moments}
Suppose Condition~\condH\ holds and $m<N$.
Let $(S_n,J_n)_{n\in\mathbb Z}$ be the two-sided
stationary marked reference chain constructed in the
proof of \cref{prop:affine}, with
\[
 \PP(S_n=s)=u_S(s),\qquad
 \PP(J_n=j\mid S_n=s)=p_j(s).
\]
Set
\[
 A_n:=A_{J_n,S_n},\qquad
 \mathsf A_n:=A_{n-1}\cdots A_0\quad(n\ge1),
 \qquad
 \mathsf A_0:=I_{S_0^\perp},
\]
and write $\PP_s(\cdot):=\PP(\cdot\mid S_0=s)$.
Then the following statements hold.

\begin{enumerate}
\item \textit{Contraction at every moment order.}
For every $s\in S$ and $n\ge1$,
\begin{equation}\label{eq:tangent-survival}
 \sigma^{n-1}\le\PP_s(\mathsf A_n\ne0)
 \le\min\{1,C_{\mathrm{surv}}\sigma^n\}.
\end{equation}
At $n=0$, this probability equals $1$ and the stated
upper bound remains valid. If $r=d$, then
$\PP_s(\mathsf A_n\ne0)=\sigma^{n-1}$ for every $n\ge1$.
Moreover, for every $s\in S$, $n\ge0$, $p>0$, and $v\in s^\perp$,
\begin{equation}\label{eq:all-order-tangent-contraction}
 \Pt^nV^p(s,v)\le a_p\sigma^n\|v\|^p.
\end{equation}

\item \textit{Finite stationary memory.}
Define
\begin{equation}\label{eq:tangent-memory-time}
 T:=\inf\{n\ge1:A_{-1}\cdots A_{-n}=0\}.
\end{equation}
Then $T<\infty$ almost surely, and for every $n\ge1$,
\begin{equation}\label{eq:tangent-memory-tail}
 \sigma^{n-1}\le\PP(T>n)
 \le\min\{1,C_{\mathrm{surv}}\sigma^n\}.
\end{equation}
Also $\PP(T>0)=1$. The exponential decay rate is exact:
\[
\lim_{n\to\infty}\PP(T>n)^{1/n}=\sigma.
\]
For $t\ge0$,
\[
 \E e^{tT}<\infty\quad\Longleftrightarrow\quad t<\kappa.
\]
In the special case $r=d$, the memory time has the exact
shifted geometric law
\[
 \PP(T=k)=(1-\sigma)\sigma^{k-2},\qquad k\ge2.
\]
Moreover, the stationary backward series of
\cref{cor:jets} terminates almost surely:
\begin{equation}\label{eq:tangent-finite-series}
 \begin{aligned}
 V_H
 &=\sum_{k=1}^{\infty}
   A_{-1}\cdots A_{-k+1}
   b_{J_{-k},S_{-k}}(H)\\
 &=\sum_{k=1}^{T}
   A_{-1}\cdots A_{-k+1}
   b_{J_{-k},S_{-k}}(H),
 \end{aligned}
\end{equation}
where the product for $k=1$ is the identity and
every summand with $k>T$ vanishes almost surely.
In particular,
\[
 \mathcal L(S_0,V_H)=\pih_H.
\]
Furthermore,
\begin{equation}\label{eq:tangent-forcing-bound}
 \mathfrak b_H\le\sqrt q\,\|H-H_0\|_{\op},
 \qquad
 \|V_H\|\le C_A\mathfrak b_HT
 \quad\text{almost surely}.
\end{equation}

\item \textit{Stationary moments and exponential integrability.}
For every $p>0$, one has $M_p(H)<\infty$.
In particular, for $1\le p<\infty$,
\begin{equation}\label{eq:stationary-all-moments}
 M_p(H)^{1/p}
 \le\frac{a_p^{1/p}\mathfrak b_H}
          {1-\sigma^{1/p}}.
\end{equation}
If $\mathfrak b_H>0$, then, for every
$0\le t<\kappa/(C_A\mathfrak b_H)$,
\begin{equation}\label{eq:stationary-exponential-moment}
 \pih_H(e^{tV})
 \le1+C_{\mathrm{surv}}
       \frac{e^{tC_A\mathfrak b_H}-1}
            {1-\sigma e^{tC_A\mathfrak b_H}}
 <\infty.
\end{equation}
If $\mathfrak b_H=0$, then
\[
 \pih_H=\frac1{|S|}
          \sum_{s\in S}\delta_{(s,0)}.
\]
\end{enumerate}
\end{theorem}

\begin{proof}
\emph{(i) Contraction at every moment order.}
Fix $s\in S$. By the homogeneous identity in
\eqref{eq:exact-energy} and the Markov property, for
$n\ge0$ and $v\in s^\perp$,
\[
 \E_s\|\mathsf A_n v\|^2=\sigma^n\|v\|^2.
\]
Summing over an orthonormal basis of $s^\perp$ gives
\[
 \E_s\|\mathsf A_n\|_{\HS}^2=(D-1)\sigma^n.
\]
By \cref{lem:flat-reference-words}, for every $n\ge0$
the product $\mathsf A_n$ has operator norm at most
$C_A$ almost surely (the case $n=0$ is immediate).
Fix $n\ge1$, and write $W_\omega:=K_{\omega,0}$ for
$\omega\in J^n$. For each power $k$ of $d$ with
$d\le k\le r$, let
\[
 z_{n,k}:=\sum_{\substack{\omega\in J^n\\
                              \rank W_\omega=k}}
               \|W_\omega s\|^2.
\]
This quantity does not depend on $s\in S$: replacing
all Clifford draws $U_i$ in a word by $U_iC$ sends
$W_\omega$ to $C^\dagger W_\omega C$ and preserves its
rank, and $\Cl$ acts transitively on $S$.

For a nonzero word of rank $k$, write
$W_\omega^\dagger W_\omega=\gamma_\omega P_\omega$
as in \eqref{eq:flat-word-effect}, and put
$\alpha_{\omega,s}=\langle s,P_\omega s\rangle$.
Every rank-$k$ stabilizer-code projector is Clifford
conjugate to a computational-basis postselection projector
on $N-\log_d k$ sites, by symplectic reduction of its
commuting Pauli generators and correction of their phases.
Therefore
\eqref{eq:support-count}, applied with $r=k$ and
$q=D/k$, gives
\[
 \PP_{s\sim u_S}(\alpha_{\omega,s}>0)
 =\frac{k(dD+1)}{D(dk+1)}.
\]
The stabilizer one-design identity also yields
$\E_{s\sim u_S}\alpha_{\omega,s}=k/D$, whence
\[
 z_{n,k}=\frac{k}{D}
       \sum_{\substack{\omega\in J^n\\
                        \rank W_\omega=k}}\gamma_\omega.
\]
For a record with $W_\omega s\ne0$,
\cref{lem:flat-reference-words} gives
$\|\mathsf A_{W_\omega,s}\|_{\HS}^2
=(k-1)/\alpha_{\omega,s}$. A record with
$W_\omega s=0$ has zero Born weight. Consequently,
averaging over the uniform stabilizer ray and then over
the Born record gives the rank-resolved identity
\[
 \begin{aligned}
 (D-1)\sigma^n
 &=\E_{s\sim u_S}\E_s\|\mathsf A_n\|_{\HS}^2\\
 &=\sum_{\substack{k=d^j\\ d\le k\le r}}
   \frac{(k-1)(dD+1)}{dk+1}\,z_{n,k}.
 \end{aligned}
\]
The function $k\mapsto(k-1)/(dk+1)$ is strictly
increasing. Set
$z_n:=\sum_{d\le k\le r}z_{n,k}$, where the sum runs
over powers of $d$. By \cref{lem:flat-reference-words},
$z_n=\PP_s(\mathsf A_n\ne0)$, and hence
\[
 \frac{(d-1)(dD+1)}{d^2+1}\,z_n
 \le(D-1)\sigma^n
 \le\frac{(r-1)(dD+1)}{dr+1}\,z_n
 =(D-1)\sigma z_n.
\]
Rearranging and using the definition of
$C_{\mathrm{surv}}$ proves
\eqref{eq:tangent-survival}. For $n=0$,
$\PP_s(\mathsf A_0\ne0)=1\le C_{\mathrm{surv}}$,
since $D\ge d^2$. If $r=d$, the only Kraus rank
contributing to $z_n$ is $d$, so both inequalities
above become equalities and $z_n=\sigma^{n-1}$. The case $n=0$ of \eqref{eq:all-order-tangent-contraction}
is immediate since $a_p\ge1$. For the remaining moment
bounds, assume $n\ge1$.

If $p\ge2$ and $v\ne0$, the uniform operator bound and
the second-moment identity yield
\[
 \begin{aligned}
 (\Pt^nV^p)(s,v)
 &=\E_s\|\mathsf A_nv\|^p\\
 &\le C_A^{p-2}\|v\|^{p-2}
          \E_s\|\mathsf A_nv\|^2
 =C_A^{p-2}\sigma^n\|v\|^p.
 \end{aligned}
\]
For $0<p<2$, the vector $\mathsf A_nv$ vanishes on
$\{\mathsf A_n=0\}$. H\"older's inequality, followed by
the second-moment identity and
\eqref{eq:tangent-survival}, gives
\[
 \begin{aligned}
 (\Pt^nV^p)(s,v)
 &\le\bigl(\E_s\|\mathsf A_nv\|^2\bigr)^{p/2}
       \PP_s(\mathsf A_n\ne0)^{1-p/2}\\
 &\le C_{\mathrm{surv}}^{\,1-p/2}
       \sigma^n\|v\|^p.
 \end{aligned}
\]
The case $v=0$ is immediate. The definition of $a_p$ in
\eqref{eq:moment-order-constant} completes
\eqref{eq:all-order-tangent-contraction}.

\emph{(ii) Finite stationary memory.}
By stationarity of the two-sided reference chain,
\[
 \PP(T>n)
 =\PP(A_{-1}\cdots A_{-n}\ne0)
 =\sum_{s\in S}u_S(s)
       \PP_s(\mathsf A_n\ne0),\qquad n\ge1.
\]
Thus \eqref{eq:tangent-survival} proves the two-sided
bound \eqref{eq:tangent-memory-tail}. Also $\PP(T>0)=1$;
since $\sigma<1$, the upper bound implies $T<\infty$
almost surely. If $r=d$, the equality case in
\eqref{eq:tangent-survival} gives
$\PP(T>n)=\sigma^{n-1}$ for $n\ge1$. Taking successive
differences proves
$\PP(T=k)=(1-\sigma)\sigma^{k-2}$ for $k\ge2$.

For $z>1$, the tail-sum identity, valid also with an
infinite expectation, reads
\[
 \E z^T=1+(z-1)\sum_{n=0}^\infty
                     z^n\PP(T>n).
\]
The matching upper and lower bounds show that this is
finite precisely when $z\sigma<1$. The case $z=1$
is immediate. With $z=e^t$, this proves the stated
exponential-moment threshold $t<\kappa$ for $t\ge0$. If $T<\infty$, then
$A_{-1}\cdots A_{-T}=0$. Every longer backward product
contains this zero product on the left; in particular,
\[
 A_{-1}\cdots A_{-k+1}=0\qquad(k>T).
\]
Thus the backward series of \cref{cor:jets}, constructed
using the common stationary reference chain of
\cref{prop:affine}, has only finitely many nonzero terms:
\[
 \begin{aligned}
 V_H
 &=\sum_{k=1}^{\infty}A_{-1}\cdots A_{-k+1}
                      b_{J_{-k},S_{-k}}(H)\\
 &=\sum_{k=1}^{T}A_{-1}\cdots A_{-k+1}
                      b_{J_{-k},S_{-k}}(H).
 \end{aligned}
\]
The empty product for $k=1$ is the identity, and the
stationary recursion in \cref{prop:affine,cor:jets}
gives $\mathcal L(S_0,V_H)=\pih_H$. This proves
\eqref{eq:tangent-finite-series}.

It remains to control the affine forcing. Since
$H_0=\sum_a h_a\Pi_a$, we have
$\Pi_aH_0=h_a\Pi_a$ and hence, on each positive
reference branch $j=(U,a)$,
\[
 \dot K_{U,a}(H_0)=-ih_aK_{U,a}(0),
 \qquad b_{j,s}(H_0)=0.
\]
The map $H\mapsto b_{j,s}(H)$ is real-linear, so
$b_{j,s}(H)=b_{j,s}(H-H_0)$. Put
$p_a=\|\Pi_aUs\|^2>0$. As $Us$ is a stabilizer state
and $\Pi_a$ is a rank-$r$ stabilizer-code projector,
the overlap estimate in \cref{lem:flat-reference-words}
gives $p_a\ge r/D=1/q$. From \eqref{eq:A-b}
and $|\lambda_{j,s}|^2=g^{-1}p_a$,
\[
 \begin{aligned}
 \|b_{j,s}(H)\|
 &\le\frac{\|\Pi_a(H-H_0)Us\|}{\sqrt{p_a}}\\
 &\le\sqrt q\,\|H-H_0\|_{\op}.
 \end{aligned}
\]
Taking the maximum gives the first inequality in
\eqref{eq:tangent-forcing-bound}. Every product in the
finite series, including the empty product, has
operator norm at most $C_A$ by
\cref{lem:flat-reference-words}. Therefore
\[
 \|V_H\|
 \le\sum_{k=1}^{T}C_A\mathfrak b_H
 =C_A\mathfrak b_HT
 \quad\text{almost surely},
\]
which proves the second inequality.

\emph{(iii) Stationary moments and exponential integrability.}
For $k\ge1$, set
\[
 Y_k(H):=A_{-1}\cdots A_{-k+1}
                    b_{J_{-k},S_{-k}}(H).
\]
Condition on the marked history through the transition
into $S_{-k+1}$. The vector
$b_{J_{-k},S_{-k}}(H)\in S_{-k+1}^\perp$ is then fixed,
and the following $k-1$ transitions have the usual
reference-chain law. By
\eqref{eq:all-order-tangent-contraction}, for $p\ge1$,
\[
 \E\bigl[\|Y_k(H)\|^p\mid\mathcal F_{-k+1}^{-}\bigr]
 \le a_p\sigma^{k-1}
       \|b_{J_{-k},S_{-k}}(H)\|^p
 \le a_p\sigma^{k-1}\mathfrak b_H^p,
\]
where $\mathcal F_{-k+1}^{-}$ denotes this marked history,
as in \cref{prop:affine}. Thus
\[
 \|Y_k(H)\|_{L^p}
 \le a_p^{1/p}\sigma^{(k-1)/p}\mathfrak b_H.
\]
Minkowski's inequality and the finite-memory
representation now give
\[
 \begin{aligned}
 M_p(H)^{1/p}
 &=\|V_H\|_{L^p}
 \le\sum_{k=1}^{\infty}\|Y_k(H)\|_{L^p}\\
 &\le a_p^{1/p}\mathfrak b_H
          \sum_{k=1}^{\infty}\sigma^{(k-1)/p}
 =\frac{a_p^{1/p}\mathfrak b_H}{1-\sigma^{1/p}},
 \end{aligned}
\]
for every $p\ge1$. In particular, $M_1(H)<\infty$;
for $0<p<1$, Jensen's inequality gives
$M_p(H)\le M_1(H)^p<\infty$. 

Finally, suppose $\mathfrak b_H>0$, and put
$z=e^{tC_A\mathfrak b_H}$ for
$0\le t<\kappa/(C_A\mathfrak b_H)$. Then
$1\le z<\sigma^{-1}$. Since $T$ is integer-valued
and $T\ge1$, the tail-sum identity and
\eqref{eq:tangent-memory-tail} yield
\[
 \begin{aligned}
 \pih_H(e^{tV})
 &=\E e^{t\|V_H\|}
 \le\E z^T\\
 &=1+(z-1)\sum_{n=0}^{\infty}z^n\PP(T>n)\\
 &\le1+C_{\mathrm{surv}}(z-1)
                  \sum_{n=0}^{\infty}(z\sigma)^n\\
 &=1+C_{\mathrm{surv}}
           \frac{e^{tC_A\mathfrak b_H}-1}
                {1-\sigma e^{tC_A\mathfrak b_H}}<\infty.
 \end{aligned}
\]
This is \eqref{eq:stationary-exponential-moment}.
If $\mathfrak b_H=0$, every affine forcing vanishes,
so $V_H=0$ almost surely. Since the stationary base
marginal is $u_S$, we obtain
$\pih_H=|S|^{-1}\sum_{s\in S}\delta_{(s,0)}$.
\end{proof}

The constants in \eqref{eq:moment-tail-constants} depend only on $d,N,m$. For $m<N$, set $h(H):=\|H-H_0\|_{\op}$. When $h(H)>0$, \cref{thm:tangent-exponential-moments} yields
\begin{equation}\label{eq:normalized-tangent-exponential}
 \pih_H\!\left(
 \exp\frac{\kappa V}{2C_A\sqrt q\,h(H)}
 \right)
 \le 1+\frac{C_{\mathrm{surv}}}{\sqrt{\sigma}}.
\end{equation}
Thus the normalized tangent magnitude $V/h(H)$ has a uniform exponential-moment bound, depending only on $d,N,m$. In particular, the stationary tangent laws have uniformly controlled exponential tails on bounded sets of effective Hamiltonians. When $m=N$, one has $\sigma=0$ and $A_{j,s}=b_{j,s}(H)=0$ on every positive reference branch. Hence
\[
 \pih_H=\frac1{|S|}\sum_{s\in S}\delta_{(s,0)},
 \qquad
 \pih_H(e^{tV})=1\quad(t\ge0),
\]
so exponential integrability holds directly in this case.

The linear coupling of stationary tangent laws established in \cref{cor:jets}, together with the all-order moment bounds above, yields quantitative Wasserstein stability with respect to the perturbing Hamiltonian. The characterization of effective Hamiltonian directions in \cref{thm:energy} further identifies the associated norm structure on $\Herm(D)/\mathcal A_M$.

\begin{corollary}[Hamiltonian stability at every finite moment order]
\label{cor:tangent-Hamiltonian-stability}
Assume $m<N$. For each $H\in\Herm(D)$, let $\pih_H$
denote the stationary affine tangent law associated with the
perturbing Hamiltonian $H$ in Condition~\condH.
Then, for every $H,G\in\Herm(D)$ and $1\le p<\infty$,
\begin{equation}\label{eq:all-order-Hamiltonian-stability}
 \begin{aligned}
 W_p^{\widehat d}(\pih_H,\pih_G)
 \le M_p(H-G)^{1/p}\le\frac{a_p^{1/p}\sqrt q}{1-\sigma^{1/p}}
       \|(H-G)-(H-G)_0\|_{\op}.
 \end{aligned}
\end{equation}
Here $(H-G)_0$ is the measurement-sector component in \eqref{eq:H-decomposition}, and $W_p^{\widehat d}$ is the Wasserstein distance. Moreover, for every $1\le p<\infty$,
\[
 \|[H]\|_{p,\mathrm{tan}}:=M_p(H)^{1/p},
 \qquad [H]\in\Herm(D)/\mathcal A_M,
\]
defines a norm on the real quotient $\Herm(D)/\mathcal A_M$.
\end{corollary}

\begin{proof}
Fix the common two-sided stationary reference chain
$(S_n,J_n)_{n\in\mathbb Z}$ of \cref{prop:affine}.
Every Hermitian $H$ is realized by the normalized path
$R_\theta=e^{-i\theta H}$; by \cref{cor:jets}, its
stationary tangent law depends only on the first-order
generator. In the finite-memory representation
\eqref{eq:tangent-finite-series}, the homogeneous maps and
the memory time $T$ are independent of $H$, whereas
$b_{j,s}(H)$ is real-linear in $H$. Consequently, on
this common probability space,
\[
 V_{\alpha H+\beta G}=\alpha V_H+\beta V_G
 \qquad(H,G\in\Herm(D),\ \alpha,\beta\in\R)
\]
almost surely, with $\mathcal L(S_0,V_H)=\pih_H$. The joint law of $(S_0,V_H)$ and $(S_0,V_G)$ is a
coupling of $\pih_H$ and $\pih_G$ supported on pairs with the same base ray. Hence
\[
 \begin{aligned}
 W_p^{\widehat d}(\pih_H,\pih_G)
 &\le \bigl(\E\|V_H-V_G\|^p\bigr)^{1/p}\\
 &=\bigl(\E\|V_{H-G}\|^p\bigr)^{1/p}
 =M_p(H-G)^{1/p}\\
 &\le\frac{a_p^{1/p}\mathfrak b_{H-G}}
              {1-\sigma^{1/p}}\\
 &\le\frac{a_p^{1/p}\sqrt q}{1-\sigma^{1/p}}
       \|(H-G)-(H-G)_0\|_{\op},
 \end{aligned}
\]
where the last two inequalities follow from
\eqref{eq:stationary-all-moments} and
\eqref{eq:tangent-forcing-bound}.

We next verify the quotient-norm assertion.
For $K\in\mathcal A_M$, \cref{thm:energy} gives
$b_{j,s}(K)=0$ on every positive reference branch.
The common finite-memory representation then implies
$V_K=0$ and $V_{H+K}=V_H$ almost surely. Thus
$M_p(H+K)=M_p(H)$, so $\|[H]\|_{p,\mathrm{tan}}$
is well defined on $\Herm(D)/\mathcal A_M$. Its absolute homogeneity follows from
$V_{tH}=tV_H$:
\[
 \|t[H]\|_{p,\mathrm{tan}}
 =|t|\,\|[H]\|_{p,\mathrm{tan}}
 \qquad(t\in\R).
\]
Similarly, Minkowski's inequality and
$V_{H+G}=V_H+V_G$ yield
\[
 \|[H]+[G]\|_{p,\mathrm{tan}}
 \le\|[H]\|_{p,\mathrm{tan}}+\|[G]\|_{p,\mathrm{tan}}.
\]
Finally, since $M_p(H)<\infty$ and $M_2(H)<\infty$,
\[
 \begin{aligned}
 \|[H]\|_{p,\mathrm{tan}}=0
 &\Longleftrightarrow V_H=0\quad\text{almost surely}\\
 &\Longleftrightarrow M_2(H)=0\\
 &\Longleftrightarrow H\in\mathcal A_M,
 \end{aligned}
\]
where the last equivalence is \eqref{eq:effective-directions}
under $m<N$. Thus the induced seminorm is positive
definite on the quotient, completing the proof.
\end{proof}

\subsection{Excursion occupation and Wasserstein response}
\label{subsec:excursion-transport}

The preceding analysis characterizes stationary tangent fluctuations near the stabilizer set $S$. The stationary response on projective space also involves rare departures generated by branches of zero reference probability. Following such a departure, the evolution until reaching $S$ forms an \emph{excursion}, whose cumulative contribution is described by the occupation measure $\mu_H^{\mathrm{occ}}$. We now control these excursions and combine them with the tangent dynamics to characterize the stationary Wasserstein response.

For the reference Markov chain $(X_n)_{n\ge0}$ with kernel $P_0$, define the first hitting time of $S$ by
\[
 \tau_S:=\inf\{n\ge0:X_n\in S\},
\]
with $\inf\varnothing=\infty$. This stopping time marks the end of an excursion starting outside $S$. The departure measure $\Lambda_H$, the killed kernel $P_0^\circ$, and the occupation measure $\mu_H^{\mathrm{occ}}$ are defined in \eqref{eq:Lambda}--\eqref{eq:Xi}. The following proposition establishes a uniform exponential tail bound for $\tau_S$, thereby strengthening the weighted occupation estimate of \cref{thm:quadratic} to finite total occupation mass and providing an exponential truncation bound for the occupation series.

\begin{proposition}
\label{prop:finite-Clifford-occupation}
Suppose Condition~\condH\ holds and $m<N$.
Let $\PP_x$ denote the law of the reference Markov
chain with kernel $P_0$ started at $X_0=x$.
With $\sigma\in(0,1)$ and $C_{\mathrm{surv}}$
as in \eqref{eq:sigma-B} and
\eqref{eq:moment-tail-constants}, respectively,
the following statements hold.

For every $x\in\X$ and $n\ge0$,
\begin{equation}\label{eq:Clifford-return-tail}
 \PP_x(\tau_S>n)
 \le\min\{1,C_{\mathrm{surv}}\sigma^n\}.
\end{equation}
Moreover, the departure intensity satisfies
\begin{equation}\label{eq:Clifford-departure-mass}
 \Lambda_H(\X\setminus S)
 \le 2M_2(H)+\frac2D\|H-H_0\|_{\HS}^2
 <\infty.
\end{equation}
Consequently, $\mu_H^{\mathrm{occ}}$ is a finite measure, with
\begin{equation}\label{eq:Clifford-occupation-mass}
 \mu_H^{\mathrm{occ}}(\X\setminus S)
 \le\frac{C_{\mathrm{surv}}}{1-\sigma}
       \Lambda_H(\X\setminus S)<\infty.
\end{equation}
Furthermore, for every bounded Borel function
$F:\X\setminus S\to\R$ and integer $n_0\ge0$,
\begin{equation}\label{eq:Clifford-occupation-truncation}
 \left|
 \mu_H^{\mathrm{occ}}(F)
 -\sum_{n=0}^{n_0-1}
       \Lambda_H\bigl((P_0^\circ)^nF\bigr)
 \right|
 \le
 \frac{C_{\mathrm{surv}}\sigma^{n_0}}{1-\sigma}
 \Lambda_H(\X\setminus S)\|F\|_\infty,
\end{equation}
where the sum is zero when $n_0=0$.
\end{proposition}

\begin{proof}
\emph{Uniform exponential return.}
For $n\ge1$, let
\[
 \mathsf F_n:=
 \sum_{\substack{\omega\in J^n\\\rank K_{\omega,0}\ge2}}
 K_{\omega,0}^\dagger K_{\omega,0}.
\]
For any $C\in\Cl$, replacing each Clifford draw $U_i$
in a record by $U_iC$ gives a bijection of $J^n$ and
replaces its Kraus word $K_{\omega,0}$ by
$C^\dagger K_{\omega,0}C$. Since rank is preserved,
$C^\dagger\mathsf F_n C=\mathsf F_n$. Clifford one-design
averaging therefore gives
\[
 \mathsf F_n=\frac{\Tr\mathsf F_n}{D}I_D=:z_nI_D,
 \qquad 0\le z_n\le1,
\]
where the last inequality follows from Kraus completeness.
For every unit representative $x$ of a ray in $\X$,
\[
 \sum_{\substack{\omega\in J^n\\\rank K_{\omega,0}\ge2}}
 \|K_{\omega,0}x\|^2=z_n.
\]

Now fix $s\in S$. On each record with
$K_{\omega,0}s\ne0$, \cref{lem:flat-reference-words}
identifies the homogeneous tangent product with the
projective derivative of $K_{\omega,0}$ and gives
\[
 \mathsf A_n\ne0
 \quad\Longleftrightarrow\quad
 \rank K_{\omega,0}\ge2.
\]
Records for which $K_{\omega,0}s=0$ have zero Born
probability. Thus the preceding identity and
\eqref{eq:tangent-survival} imply
\[
 z_n=\PP_s(\mathsf A_n\ne0)
 \le\min\{1,C_{\mathrm{surv}}\sigma^n\}.
\]
Every nonzero rank-one reference Kraus word has its
range in $S$, by \cref{lem:flat-reference-words}. Since
$S$ is invariant under $P_0$, for every $x\in\X$,
\[
 \PP_x(\tau_S>n)
 =\PP_x(X_n\notin S)
 \le z_n,
 \qquad n\ge1.
\]
For $n=0$, the bound follows from
$\PP_x(\tau_S>0)\le1\le C_{\mathrm{surv}}$.
This proves \eqref{eq:Clifford-return-tail}.

\emph{Departure intensity.}
Write $B:=H-H_0$. For $j=(U,a)\in J_0(s)$,
we have $K_js=0$. Since $H_0=\sum_a h_a\Pi_a$,
\[
 \dot K_j(H_0)s=-ih_aK_js=0,
 \qquad
 z_j^+(s,v)=K_jv+\dot K_j(B)s.
\]
By the definition \eqref{eq:Lambda}, dropping the
restriction that the limiting output ray lies outside
$S$ and using $\|u+w\|^2\le2\|u\|^2+2\|w\|^2$
give
\[
 \begin{aligned}
 \Lambda_H(\X\setminus S)
 &\le2\int_{\TX}\sum_{j\in J}\|K_jv\|^2
                 \,\pih_H(\mathrm d(s,v))\\
 &\quad+2\int_{\TX}\sum_{j\in J}
        \|\dot K_j(B)s\|^2\,\pih_H(\mathrm d(s,v)).
 \end{aligned}
\]
By Kraus completeness, the first integral equals
$M_2(H)$. For every $s\in S$, Clifford one-design
averaging and $\sum_a\Pi_a=I_D$ yield
\[
 \begin{aligned}
 \sum_{j\in J}\|\dot K_j(B)s\|^2
 &=\frac1g\sum_{U\in\Cl}\sum_a\|\Pi_aBUs\|^2\\
 &=\frac1g\sum_{U\in\Cl}\|BUs\|^2
 =\frac{\Tr(B^2)}{D}
 =\frac{\|B\|_{\HS}^2}{D}.
 \end{aligned}
\]
As $M_2(H)<\infty$ by \cref{thm:energy}, this proves
\eqref{eq:Clifford-departure-mass}.

\emph{Finite total occupation.}
For $x\in\X\setminus S$, the definition of the killed
kernel gives
\[
 (P_0^\circ)^n\one(x)=\PP_x(\tau_S>n)
 \le C_{\mathrm{surv}}\sigma^n,
 \qquad n\ge0.
\]
Using the occupation series \eqref{eq:Xi} and Tonelli's
theorem, we obtain
\[
 \begin{aligned}
 \mu_H^{\mathrm{occ}}(\X\setminus S)
 &=\sum_{n=0}^\infty
       \Lambda_H\bigl((P_0^\circ)^n\one\bigr)\\
 &\le C_{\mathrm{surv}}\Lambda_H(\X\setminus S)
       \sum_{n=0}^\infty\sigma^n\\
 &=\frac{C_{\mathrm{surv}}}{1-\sigma}
       \Lambda_H(\X\setminus S)<\infty,
 \end{aligned}
\]
which proves \eqref{eq:Clifford-occupation-mass}.

\emph{Exponential truncation.}
Let $F:\X\setminus S\to\R$ be bounded and Borel,
and fix an integer $n_0\ge0$. Since $P_0^\circ$ is
sub-Markov,
\[
 |(P_0^\circ)^nF|
 \le\|F\|_\infty(P_0^\circ)^n\one.
\]
The preceding estimate ensures absolute convergence of
the occupation series. Consequently,
\[
 \begin{aligned}
 \left|\mu_H^{\mathrm{occ}}(F)-\sum_{n=0}^{n_0-1}
             \Lambda_H\bigl((P_0^\circ)^nF\bigr)\right|
 &\le\|F\|_\infty\sum_{n=n_0}^\infty
             \Lambda_H\bigl((P_0^\circ)^n\one\bigr)\\
 &\le C_{\mathrm{surv}}\|F\|_\infty
       \Lambda_H(\X\setminus S)
       \sum_{n=n_0}^\infty\sigma^n\\
 &=\frac{C_{\mathrm{surv}}\sigma^{n_0}}{1-\sigma}
       \Lambda_H(\X\setminus S)\|F\|_\infty.
 \end{aligned}
\]
This includes $n_0=0$, with the partial sum interpreted as zero, and proves \eqref{eq:Clifford-occupation-truncation}.
\end{proof}

The exact transport identity \eqref{eq:radial-transport} makes the distance to the uniform stabilizer law an observable response. The following corollary shows its leading term changes at $p=2$: lower orders retain only tangent fluctuations, the second order combines tangent and excursion contributions, and higher orders detect the excursion term whenever it is nonzero.

\begin{corollary}[Stationary Wasserstein response at every finite order]
\label{cor:Clifford-Wasserstein-response}
Suppose Condition~\condH\ holds. For every fixed
$1\le p<\infty$, as $\theta\downarrow0$,
\begin{equation}\label{eq:Clifford-Wasserstein-response}
 W_p(\pi_\theta,u_S)^p=
 \begin{cases}
 \theta^pM_p(H)+o(\theta^p),
   &1\le p<2,\\[2pt]
 \theta^2\bigl[M_2(H)+\mu_H^{\mathrm{occ}}(r_S^2)\bigr]
   +o(\theta^2),
   &p=2,\\[2pt]
 \theta^2\mu_H^{\mathrm{occ}}(r_S^p)+o(\theta^2),
   &2<p<\infty.
 \end{cases}
\end{equation}
If $m<N$ and $H\notin\mathcal A_M$, the coefficients
for $1\le p\le2$ are strictly positive.
If $m=N$, then $\pi_\theta=u_S$ for every $\theta$.
\end{corollary}

\begin{proof}
For $p\ge1$, set $F_p:=r_S^p$. Since $r_S$ is
$1$-Lipschitz and takes values in $[0,1]$, the function
$F_p$ is bounded and globally Lipschitz. By the exact
transport identity \eqref{eq:radial-transport},
\[
 W_p(\pi_\theta,u_S)^p=\pi_\theta(F_p).
\]
For each $s\in S$, the ray $s$ is the unique nearest
point of $S$ in a sufficiently small neighborhood of $s$.
Hence, in the chart \eqref{eq:charts},
\begin{equation}\label{eq:radial-chart-response}
 F_p([s+v])
 =\frac{\|v\|^p}{(1+\|v\|^2)^{p/2}}
\end{equation}
for all sufficiently small $v\in s^\perp$.
We apply the stationary response results of
\cref{cor:subquadratic,thm:quadratic}, whose hypotheses
are verified for the Clifford model by
\cref{prop:verification}.

\emph{Case $1\le p<2$.}
By \eqref{eq:radial-chart-response},
\[
 \theta^{-p}F_p([s+\theta v])
 =\frac{\|v\|^p}{(1+\theta^2\|v\|^2)^{p/2}}
 \longrightarrow\|v\|^p
\]
uniformly on bounded subsets of $\TX$. Since
$F_p=r_S^p$, it satisfies the global bound in
\cref{def:subquadratic}. Thus $F_p$ is a subquadratic
observable of order $p$, with leading local term $V^p$.
By \cref{cor:subquadratic},
\[
 \pi_\theta(F_p)
 =\theta^p\pih_H(V^p)+o(\theta^p)
 =\theta^pM_p(H)+o(\theta^p).
\]

\emph{Case $p=2$.}
Equation~\eqref{eq:radial-chart-response} gives
\[
 F_2([s+v])=\|v\|^2+O(\|v\|^4),
\]
so $F_2$ is an ordinary quadratic observable with
$q_{F_2}(s,v)=\|v\|^2$. By \cref{thm:quadratic},
\[
 \pi_\theta(F_2)
 =\theta^2\bigl[M_2(H)+\mu_H^{\mathrm{occ}}(r_S^2)\bigr]
   +o(\theta^2).
\]

\emph{Case $2<p<\infty$.}
In this case, \eqref{eq:radial-chart-response} implies
\[
 F_p(s)=0,\qquad
 \mathrm dF_p|_s=0,\qquad
 D^2(F_p\circ\kappa_s^{-1})(0)=0
 \quad(s\in S).
\]
Moreover, $F_p$ is $C^2$ in a neighborhood of $S$.
It is therefore an ordinary quadratic observable
with $q_{F_p}=0$. By \cref{thm:quadratic},
\[
 \pi_\theta(F_p)
 =\theta^2\mu_H^{\mathrm{occ}}(r_S^p)+o(\theta^2).
\]
The occupation integral is finite, since
$0\le r_S^p\le r_S^2$ and
$\mu_H^{\mathrm{occ}}(r_S^2)<\infty$ by \cref{thm:quadratic}.
Together with the transport identity, these three cases
prove \eqref{eq:Clifford-Wasserstein-response}.

Suppose $m<N$ and $H\notin\mathcal A_M$.
By \cref{thm:energy}, $M_2(H)>0$, so the stationary
tangent law assigns positive probability to
$\{(s,v):v\ne0\}$. Consequently,
\[
 M_p(H)>0\qquad(p>0).
\]
Since $\mu_H^{\mathrm{occ}}$ is a nonnegative measure, the leading
coefficients in \eqref{eq:Clifford-Wasserstein-response}
are strictly positive for $1\le p\le2$. Finally, suppose $m=N$. By
\cref{prop:stationary-structure},
$\pi_\theta=u_S$ for every admissible $\theta$.
\end{proof}

\subsection{Response positivity and smooth observables}

Having characterized stationary tangent fluctuations and excursion occupation, we now turn to the response of general observables. The exact tangent-energy formula provides a criterion for strict positivity of the leading response coefficient of nonnegative observables with positive local terms. Moreover, Clifford symmetry allows us to transform any $f\in C^2(\X;\R)$, by averaging over the Clifford group and subtracting its reference mean, into an ordinary quadratic observable in the sense of \cref{def:quadratic}, without changing its stationary response. This yields a second-order response formula for arbitrary smooth observables.

\begin{corollary}[Positive response and smooth stationary averages]
\label{cor:clifford-response}
Suppose Condition~\condH\ holds. Then the following
statements hold.

\begin{enumerate}
\item \textit{Positive leading response.}
Assume $m<N$. Let $F\ge0$ belong to one of the
classes in
\cref{def:subquadratic,def:quadratic,def:logarithmic},
and denote its leading local term by
\[
 g_F=
 \begin{cases}
 f_k,&F\text{ is subquadratic of order }k,\\
 q_F,&F\text{ is ordinary quadratic},\\
 q_{\log},&F\text{ is logarithmic quadratic}.
 \end{cases}
\]
If
\[
 g_F(s,v)>0
 \qquad(s\in S,\ v\in s^\perp\setminus\{0\}),
\]
then the leading coefficient in the corresponding response formula \eqref{eq:subquadratic-response}, \eqref{eq:log-response} or \eqref{eq:quadratic-response} is strictly positive if and only if $H\notin\mathcal A_M$.

\item \textit{Smooth observables.}
For every $f\in C^2(\X;\R)$, define
\begin{equation}\label{eq:Clifford-average}
 f^{\Cl}(x)=\frac1g\sum_{C\in\Cl}f(Cx),
 \qquad
 F_f=f^{\Cl}-u_S(f).
\end{equation}
Then $F_f$ is an ordinary quadratic observable
in the sense of \cref{def:quadratic}; in particular,
\[
 F_f|_S=0,\qquad \mathrm dF_f|_S=0.
\]
Moreover, as $\theta\downarrow0$,
\begin{equation}\label{eq:smooth-response}
 \begin{split}
 \pi_\theta(f)
 &=u_S(f)+\theta^2\mathcal L_H(f)+o(\theta^2),\\
 \mathcal L_H(f)
 &=\pih_H(q_{F_f})+\mu_H^{\mathrm{occ}}(F_f).
 \end{split}
\end{equation}
The map $f\mapsto\mathcal L_H(f)$ is real-linear.
In particular,
\[
 \lim_{\theta\downarrow0}
 \frac{\pi_\theta(f)-u_S(f)}{\theta}=0.
\]
\end{enumerate}
\end{corollary}

\begin{proof}
Suppose first that $H\notin\mathcal A_M$. By
\cref{thm:energy}, $M_2(H)>0$, and hence
\[
 \pih_H(\{(s,v):v\ne0\})>0.
\]
Since $F\ge0$, its leading local term $g_F$ is
nonnegative. The assumed strict positivity away
from the zero section therefore implies
$\pih_H(g_F)>0$. This is the coefficient in
\eqref{eq:subquadratic-response} and
\eqref{eq:log-response}. For ordinary quadratic
observables, the coefficient in
\eqref{eq:quadratic-response} satisfies
\[
 \pih_H(q_F)+\mu_H^{\mathrm{occ}}(F)
 \ge\pih_H(q_F)>0,
\]
since $\mu_H^{\mathrm{occ}}$ is a nonnegative measure.

Conversely, suppose $H\in\mathcal A_M$. By
\cref{thm:energy}, $M_2(H)=0$, so $\pih_H$ is
supported on the zero section, where every
leading local term vanishes. Moreover, writing
$H=\sum_a h_a\Pi_a$, we have
\[
 \dot K_{U,a}(H)=-ih_aK_{U,a}(0).
\]
Thus, for every $j=(U,a)\in J_0(s)$,
\[
 z_j^+(s,0)=K_j0+\dot K_j(H)s=0.
\]
It follows from \eqref{eq:Lambda}--\eqref{eq:Xi}
that $\Lambda_H=\mu_H^{\mathrm{occ}}=0$. Hence all three
response coefficients vanish, proving (i). By Clifford invariance of $\pi_\theta$ and
transitivity of the Clifford action on $S$,
\[
 \pi_\theta(f)=\pi_\theta(f^{\Cl}),
 \qquad
 f^{\Cl}(s)=u_S(f)\quad(s\in S).
\]
Consequently, $F_f|_S=0$.

Fix $s\in S$, and let $G_s$ be its stabilizer
in the projective Clifford group. Choose unitary
representatives fixing $s$. Since $f^{\Cl}$
is Clifford invariant, its differential satisfies
\[
 \begin{aligned}
 \mathrm d(f^{\Cl})_s(v)
 &=\frac1{|G_s|}\sum_{C\in G_s}
       \mathrm d(f^{\Cl})_s(Cv)\\
 &=\mathrm d(f^{\Cl})_s
       \left(\frac1{|G_s|}\sum_{C\in G_s}Cv\right)
 =0,
 \end{aligned}
\]
where the last equality follows from
\eqref{eq:input-stabilizer-average}.
Thus $\mathrm dF_f|_S=0$. Since $F_f\in C^2(\X;\R)$ and $\X$ is compact,
$F_f$ is globally Lipschitz. Hence $F_f$ is
an ordinary quadratic observable in the sense
of \cref{def:quadratic}. Applying
\cref{thm:quadratic} and using
\[
 \pi_\theta(f)-u_S(f)=\pi_\theta(F_f)
\]
gives \eqref{eq:smooth-response}. Finally, the maps $f\mapsto F_f$ and
$F_f\mapsto q_{F_f}$ are real-linear.
Since $\pih_H$ and $\mu_H^{\mathrm{occ}}$ are fixed measures,
$f\mapsto\mathcal L_H(f)$ is real-linear.
\end{proof}

\section{Stationary magic response}\label{sec:resources}

We now apply the stationary response framework to Gross--Wigner mana and stabilizer R\'enyi entropies. We first identify their local expansions near the stabilizer set and verify the hypotheses of the corresponding observable classes. Combining these resource-specific inputs with the general response theorems and the Clifford-specific tangent-energy and excursion-occupation results, we derive their weak-injection stationary asymptotics. These yield a linear mana response, whose coefficient defines a norm on effective Hamiltonian directions when $m<N$, and three order-dependent regimes for stabilizer R\'enyi entropies.

Throughout this section, $S=\St$, $R_\theta$ satisfies Condition~\condH, and $\pi_\theta=\pi_{R_\theta}$. The quantity
\[
 \pi_\theta(F)=\int_\X F(x)\,\pi_\theta(\mathrm dx)
\]
is the stationary average of a pure-state observable. All limits are taken as $\theta\downarrow0$. Resource logarithms have base two; $\ln$ denotes the natural logarithm.

\subsection{Resource observables and local asymptotics}

We first consider Gross--Wigner mana for odd prime local dimension $d$. For $u=(p,b)\in\F_d^{2N}$, the Gross phase-point operators \cite{Gross2006}
are defined by
\[
 A_{p,b}=\sum_{x\in\F_d^N}
          \omega_d^{2p\cdot x}|b+x\rangle\langle b-x|,
 \qquad \omega_d=e^{2\pi i/d}.
\]
They satisfy
\[
 A_u^\dagger=A_u,\quad A_u^2=I,\quad
 \Tr A_u=1,\quad \Tr(A_uA_w)=D\one_{\{u=w\}}.
\]
The Wigner function and mana are
\begin{equation}\label{eq:mana-definition}
 W_x(u)=D^{-1}\Tr(A_u\rho_x),
 \qquad
 \mathcal M(x)=\log_2\sum_u|W_x(u)|.
\end{equation}
For a stabilizer $s$, $W_s=D^{-1}\one_{L_s}$, where $L_s\subset\F_d^{2N}$ is an affine Lagrangian set of size $D$. For $v\perp s$, set
\begin{equation}\label{eq:mana-germ}
 \begin{split}
 \ell^W_{s,u}(v)
   &=D^{-1}\Tr\bigl[A_u(|v\rangle\langle s|+|s\rangle\langle v|)\bigr],\\
 \mathfrak m_s(v)&=\frac1{\ln2}\sum_{u\notin L_s}|\ell^W_{s,u}(v)|.
 \end{split}
\end{equation}
We next consider stabilizer R\'enyi entropies,
defined for any prime local dimension $d$.
Choose unitary representatives
$(W_a)_{a\in\F_d^{2N}}$ of the projective Pauli classes, with $W_0=I_D$ and $\Tr(W_a^\dagger W_b)=D\one_{\{a=b\}}$. Define
\begin{equation}\label{eq:SRE-data}
 x_a(x)=|\Tr(W_a\rho_x)|^2,
 \qquad T_\alpha(x)=D^{-1}\sum_a x_a(x)^\alpha,
\end{equation}
and
\begin{equation}\label{eq:SRE-definition}
 \begin{split}
 S_\alpha(x)&=\frac{\log_2T_\alpha(x)}{1-\alpha},
       \qquad 0<\alpha<\infty,\quad\alpha\ne1,\\
 S_1(x)&=-D^{-1}\sum_a x_a(x)\log_2x_a(x),
 \qquad 0\log0=0.
 \end{split}
\end{equation}
Pauli Parseval gives $\sum_a x_a(x)=D$. Thus $S_\alpha$ is the order-$\alpha$ R\'enyi entropy of the probability vector $(x_a(x)/D)_a$ minus $\log_2D$, with the Shannon convention at $\alpha=1$. For $d=2$, this agrees with the original stabilizer R\'enyi entropy of \cite{Leone2022}; for prime $d>2$, it is the generalized-Pauli extension studied in \cite{WangLi2023}. For $s\in S$, put
\[
 \mathcal T_s=\{a:|\langle s,W_as\rangle|=1\}.
\]
The first variation of a Pauli expectation and the local term for $0<\alpha<1$ are
\begin{equation}\label{eq:Pauli-differential}
 \begin{split}
 \ell^P_{s,a}(v)&=\langle v,W_as\rangle+\langle s,W_av\rangle,\\
 f_\alpha(s,v)&=\frac1{D(1-\alpha)\ln2}
       \sum_{a\notin \mathcal T_s}|\ell^P_{s,a}(v)|^{2\alpha},
       \qquad 0<\alpha<1.
 \end{split}
\end{equation}

\begin{lemma}[Local resource terms and observable classes]
\label{lem:resource-germs}
All local asymptotics below hold uniformly on each bounded tangent set
$\mathcal K_R=\{(s,v)\in\TX:\|v\|\le R\}$,
for every $R<\infty$.
\begin{enumerate}
\item \textit{Gross--Wigner mana.}
For odd prime $d$, mana is nonnegative, Clifford
invariant, and globally Lipschitz, with
$\mathcal M^{-1}(0)=S$. Its local expansion is
\[
 \mathcal M([s+\theta v])
 =\theta\mathfrak m_s(v)+O_R(\theta^2).
\]
For each $s\in S$, the function $\mathfrak m_s$
defines a norm on the real tangent space $s^\perp$.
Moreover, there exists $C_{\mathcal M}<\infty$ such
that
\[
 \mathcal M(x)\le C_{\mathcal M}r_S(x),
 \qquad x\in\X.
\]
Hence mana is a subquadratic observable of order one.

\item \textit{Basic properties of stabilizer R\'enyi entropies.}
For every finite $\alpha>0$, the function
$S_\alpha$ is nonnegative, continuous, and Clifford
invariant, with $S_\alpha^{-1}(0)=S$.

\item \textit{Local expansions and observable classes.}
For each fixed $\alpha>0$, the stabilizer R\'enyi
entropies satisfy
\begin{align}
 S_\alpha([s+\theta v])
 &=\theta^{2\alpha}f_\alpha(s,v)
   +o_R(\theta^{2\alpha}),
   &&0<\alpha<1,
   \label{eq:SRE-low-germ}\\
 S_1([s+\theta v])
 &=\theta^2\ln(1/\theta)
   \frac4{\ln2}\|v\|^2
   +o_R\bigl(\theta^2\ln(1/\theta)\bigr),
   &&\alpha=1,
   \label{eq:SRE-one-germ}\\
 S_\alpha([s+\theta v])
 &=\theta^2\frac{2\alpha}{(\alpha-1)\ln2}
   \|v\|^2+o_R(\theta^2),
   &&\alpha>1.
   \label{eq:SRE-high-germ}
\end{align}
For $0<\alpha<1$, there exists $C_\alpha<\infty$
such that
\[
 S_\alpha(x)\le C_\alpha r_S(x)^{2\alpha},
 \qquad x\in\X.
\]
Consequently, for $0<\alpha<1$, $S_\alpha$ is a subquadratic observable of order $2\alpha$. At $\alpha=1$, $S_1$ is a logarithmic quadratic observable of exponent $\zeta=1$, with $q_{\log}=4V^2/\ln2$. For $\alpha>1$, $S_\alpha$ is an ordinary quadratic observable, with local quadratic term
\[
 q_{S_\alpha}
 =\frac{2\alpha}{(\alpha-1)\ln2}V^2.
\]
All the leading local terms are strictly
positive for $v\ne0$. 
\end{enumerate}
\end{lemma}

\begin{proof}
\emph{(i) Gross--Wigner mana.}
The Wigner components are real and sum to one, so
\[
 \sum_u|W_x(u)|\ge1,\qquad \mathcal M(x)\ge0.
\]
Since $\|A_u\|_{\op}=1$,
\[
 \sum_u|W_x(u)-W_y(u)|
 \le D\|\rho_x-\rho_y\|_1
 =2D\,d(x,y).
\]
As $\log_2 t$ is $(\ln2)^{-1}$-Lipschitz on
$[1,\infty)$, mana is globally Lipschitz.
Clifford covariance of the Wigner function gives Clifford invariance, while Gross's discrete Hudson theorem identifies its zero set with $S$. Fix $s\in S$. For $u\in L_s$,
$\langle s,A_us\rangle=DW_s(u)=1$. Since $A_u$
is a Hermitian involution, $A_us=s$, and hence
$\ell^W_{s,u}(v)=0$ for every $v\perp s$.
The normalized projector satisfies
\begin{equation}\label{eq:projector-expansion}
 \rho_{[s+\theta v]}
 =\frac{\rho_s+\theta\dot\rho_{s,v}
             +\theta^2|v\rangle\langle v|}
        {1+\theta^2\|v\|^2},
 \qquad
 \dot\rho_{s,v}
 =|v\rangle\langle s|+|s\rangle\langle v|.
\end{equation}
Consequently, as $\theta\downarrow0$,
\[
 W_{[s+\theta v]}(u)=
 \begin{cases}
 D^{-1}+O_R(\theta^2),&u\in L_s,\\
 \theta\ell^W_{s,u}(v)+O_R(\theta^2),
      &u\notin L_s.
 \end{cases}
\]
Using $\bigl||a+b|-|a|\bigr|\le|b|$ and
expanding the logarithm, we obtain
\[
 \mathcal M([s+\theta v])
 =\frac{\theta}{\ln2}
    \sum_{u\notin L_s}|\ell^W_{s,u}(v)|
    +O_R(\theta^2)
 =\theta\mathfrak m_s(v)+O_R(\theta^2).
\]

The maps $\ell^W_{s,u}$ are real-linear, so
$\mathfrak m_s$ is a seminorm on the real tangent
space $s^\perp$. Orthogonality of the phase-point
operators gives
\begin{equation}\label{eq:Wigner-Parseval-tangent}
 \sum_{u\notin L_s}|\ell^W_{s,u}(v)|^2
 =\frac1D\Tr(\dot\rho_{s,v}^2)
 =\frac2D\|v\|^2.
\end{equation}
Thus $\mathfrak m_s(v)=0$ only when $v=0$,
proving that $\mathfrak m_s$ is a norm.
Finally, global Lipschitz continuity and
$\mathcal M|_S=0$ imply
\[
 \mathcal M(x)\le C_{\mathcal M}r_S(x),
 \qquad x\in\X.
\]
This verifies the subquadratic observable
conditions with order one.

\emph{(ii) Basic properties of stabilizer R\'enyi entropies.}
Pauli Parseval and unitarity give
\begin{equation}\label{eq:Pauli-normalization}
 \sum_a x_a(x)=D,\qquad 0\le x_a(x)\le1.
\end{equation}
These identities imply $S_\alpha\ge0$ for every finite $\alpha>0$, with equality precisely when $x_a(x)\in\{0,1\}$ for all $a$. By the standard characterization of stabilizer states in terms of maximal commuting Pauli subgroups \cite{Gottesman1997,Gross2006}, this is equivalent to $x\in S$. Hence $S_\alpha^{-1}(0)=S$. Continuity follows from \eqref{eq:SRE-definition}, while Clifford invariance follows from the permutation of projective Pauli labels under Clifford conjugation.

\emph{(iii) Local expansions and observable classes.}
Fix $s\in S$ and $v\perp s$.
For $a\in\mathcal T_s$, both $W_as$ and
$W_a^\dagger s$ are scalar multiples of $s$,
so the first variation of the Pauli expectation
vanishes. For $a\notin\mathcal T_s$, its value
at $s$ is zero and its first variation is
$\ell^P_{s,a}(v)$. By Pauli Parseval,
\begin{equation}\label{eq:Pauli-Parseval-tangent}
 \frac1D\sum_{a\notin\mathcal T_s}
       |\ell^P_{s,a}(v)|^2
 =\frac1D\sum_a
       |\Tr(W_a\dot\rho_{s,v})|^2
 =\Tr(\dot\rho_{s,v}^2)
 =2\|v\|^2.
\end{equation}
Using \eqref{eq:projector-expansion}, we obtain
\[
 x_a([s+\theta v])=
 \begin{cases}
 1+O_R(\theta^2),&a\in\mathcal T_s,\\
 \theta^2|\ell^P_{s,a}(v)+O_R(\theta)|^2,
       &a\notin\mathcal T_s.
 \end{cases}
\]
Since $\sum_a x_a=D$ and
$|\mathcal T_s|=D$, it follows that
\begin{equation}\label{eq:on-support-decrease}
 \frac1D\sum_{a\in\mathcal T_s}
       \bigl(x_a([s+\theta v])-1\bigr)
 =-2\theta^2\|v\|^2+o_R(\theta^2).
\end{equation}

For $0<\alpha<1$, continuity of
$z\mapsto|z|^{2\alpha}$ on bounded sets gives
\[
 T_\alpha([s+\theta v])
 =1+\frac{\theta^{2\alpha}}D
       \sum_{a\notin\mathcal T_s}
          |\ell^P_{s,a}(v)|^{2\alpha}
       +o_R(\theta^{2\alpha}),
\]
where the $O_R(\theta^2)$ changes on
$\mathcal T_s$ are of smaller order.
Expanding the logarithm proves
\eqref{eq:SRE-low-germ}.

The same pointwise estimates, with $v$ now
taken sufficiently small, yield
\[
 0\le T_\alpha([s+v])-1
       \le C\|v\|^{2\alpha}.
\]
Since $r_S([s+v])$ is comparable to $\|v\|$
near $s$, this gives
$S_\alpha\le C r_S^{2\alpha}$ near $S$.
Compactness of $\X$ and the positive lower
bound for $r_S$ away from a neighborhood of
$S$ extend the estimate to all of $\X$.
Thus $S_\alpha$ is a subquadratic observable
of order $2\alpha$.

At $\alpha=1$, for $a\notin\mathcal T_s$ write
\[
 x_a([s+\theta v])=\theta^2y_{a,\theta}(s,v),
 \qquad
 y_{a,\theta}(s,v)\longrightarrow
       |\ell^P_{s,a}(v)|^2
\]
uniformly on $\mathcal K_R$. Then
\[
 \begin{aligned}
 -x_a\log_2x_a
 &=\frac{2\theta^2\ln(1/\theta)}{\ln2}
       y_{a,\theta}
   -\theta^2y_{a,\theta}\log_2y_{a,\theta}.
 \end{aligned}
\]
The function $y\mapsto y\log y$, continuously
extended by zero at $y=0$, is bounded on
compact intervals. The terms with
$a\in\mathcal T_s$ contribute $O_R(\theta^2)$.
Consequently, \eqref{eq:Pauli-Parseval-tangent}
gives
\[
 S_1([s+\theta v])
 =\theta^2\ln(1/\theta)
       \frac4{\ln2}\|v\|^2
   +o_R\bigl(\theta^2\ln(1/\theta)\bigr),
\]
proving \eqref{eq:SRE-one-germ}. It remains to verify the regularity condition
for the logarithmic quadratic class.
The function
\[
 z\longmapsto-|z|^2\ln|z|^2,
 \qquad 0\ln0:=0,
\]
is $C^1$ on $\C\simeq\R^2$, so
$S_1\in C^1(\X)$.
Write $G_s(v)=S_1([s+v])$. For
$a\notin\mathcal T_s$, the corresponding
Pauli expectation is $O(\|v\|)$ and has
bounded first derivatives. For
$a\in\mathcal T_s$, its squared modulus
equals $1+O(\|v\|^2)$ and has derivative
$O(\|v\|)$. Hence, uniformly in $s\in S$,
\[
 |G_s(v)|
 \le C\|v\|^2\ln\frac e{\|v\|},
 \qquad
 \|DG_s(v)\|
 \le C\|v\|\ln\frac e{\|v\|}
\]
for all sufficiently small nonzero $v$.
Using
\[
 r_S([s+v])
 =\frac{\|v\|}{\sqrt{1+\|v\|^2}}
\]
near $s$, together with
$\delta_S(z)=\|z\|r_S([z])$, differentiation
of $\Gamma_{S_1}(z)=\|z\|^2S_1([z])$ yields
\[
 \|\nabla\Gamma_{S_1}(z)\|
 \le C\delta_S(z)
       \ln\frac{e\|z\|}{\delta_S(z)}
\]
near the nonzero stabilizer lines, with the
right-hand side interpreted as zero on
those lines. On their complement, the bound
follows from compactness of the unit sphere,
$C^1$ regularity, and a positive lower
bound for $\delta_S(z)/\|z\|$. Homogeneity
extends it to all $z\ne0$. Thus
\eqref{eq:log-gradient} holds with $\zeta=1$,
and $S_1$ is a logarithmic quadratic
observable with $q_{\log}=4V^2/\ln2$.

Finally, let $\alpha>1$. The contribution
to $T_\alpha$ from $a\notin\mathcal T_s$
is $o_R(\theta^2)$. Taylor expansion at
$x_a=1$, together with
\eqref{eq:on-support-decrease}, gives
\[
 T_\alpha([s+\theta v])
 =1-2\alpha\theta^2\|v\|^2
       +o_R(\theta^2).
\]
Expanding $\log_2T_\alpha$ proves
\eqref{eq:SRE-high-germ}.
Moreover, $z\mapsto|z|^{2\alpha}$ is $C^2$
on $\C$ for $\alpha>1$, and the identity
Pauli gives $T_\alpha\ge D^{-1}$.
Thus $S_\alpha\in C^2(\X)$ and is globally
Lipschitz. Since it is nonnegative and
vanishes on $S$, its differential vanishes
there. Hence $S_\alpha$ is an ordinary
quadratic observable, with
\[
 q_{S_\alpha}
 =\frac{2\alpha}{(\alpha-1)\ln2}V^2.
\]

Strict positivity of the mana local term
follows from \eqref{eq:Wigner-Parseval-tangent}.
For $0<\alpha<1$, the identity
\eqref{eq:Pauli-Parseval-tangent} implies
$f_\alpha(s,v)>0$ whenever $v\ne0$.
The same positivity is immediate for the
local terms at $\alpha\ge1$.
All expansions are uniform on each
$\mathcal K_R$, since $S$ and the phase-space
and Pauli label sets are finite.
\end{proof}

\subsection{Mana response and its induced Hamiltonian norm}

Mana belongs to the subquadratic class, so its stationary coefficient is obtained directly from the tangent law. The linear coupling in \cref{cor:jets} and the nondegeneracy in \cref{thm:energy} give the additional norm structure below.

\begin{theorem}[Stationary mana response and its induced Hamiltonian norm]
\label{thm:mana}
Let $d$ be an odd prime and suppose
Condition~\condH\ holds. Then the following
statements hold.

\begin{enumerate}
\item \textit{First-order stationary response.}
As $\theta\downarrow0$,
\begin{equation}\label{eq:mana-response}
 \begin{split}
 \overline{\mathcal M}_H(\theta)
 :=\pi_\theta(\mathcal M)
 &=\theta\kappa_{\mathcal M}(H)+o(\theta),\\
 \kappa_{\mathcal M}(H)
 &=\int_{\TX}\mathfrak m_s(v)\,
       \pih_H(\mathrm d(s,v)).
 \end{split}
\end{equation}

\item \textit{Induced Hamiltonian norm.}
If $m<N$, the map
\[
 [H]\longmapsto\kappa_{\mathcal M}(H)
\]
defines a norm on the real quotient space
$\Herm(D)/\mathcal A_M$. In particular,
\[
 \kappa_{\mathcal M}(H)>0
 \quad\Longleftrightarrow\quad
 H\notin\mathcal A_M.
\]

\item \textit{Full measurement.}
If $m=N$, then $\pi_R(\mathcal M)=0$
for every $R\in U(D)$.
\end{enumerate}
\end{theorem}

\begin{proof}
\emph{(i) First-order stationary response.}
By \cref{prop:verification}, the Clifford model satisfies
the reference conditions. By \cref{lem:resource-germs},
mana is a subquadratic observable of order one, with
leading local term $\mathfrak m$. Applying
\cref{cor:subquadratic} gives
\[
 \pi_\theta(\mathcal M)
 =\theta\pih_H(\mathfrak m)+o(\theta),
\]
which is \eqref{eq:mana-response}.

\emph{(ii) Induced Hamiltonian norm.}
The first-order Kraus direction
\[
 h(H)_{U,a}:=-ig^{-1/2}U^\dagger\Pi_a HU
\]
is real-linear in $H$. Every such direction is
realized by the normalized unitary path
$R_\theta=e^{-i\theta H}$. Thus \cref{cor:jets}
provides a common coupling $(S_0,V_H)$ with
$\mathcal L(S_0,V_H)=\pih_H$ and
\[
 V_{H+G}=V_H+V_G,\qquad
 V_{tH}=tV_H\quad(t\in\R)
\]
almost surely. Since each $\mathfrak m_s$ is a norm and $S$ is
finite, there exists $C<\infty$ such that
$\mathfrak m_s(v)\le C\|v\|$ for all $(s,v)\in\TX$.
Consequently,
\[
 0\le\kappa_{\mathcal M}(H)
 =\E\mathfrak m_{S_0}(V_H)
 \le C M_2(H)^{1/2}<\infty.
\]
The norm properties of $\mathfrak m_s$ and the
linear coupling give
\[
 \begin{aligned}
 \kappa_{\mathcal M}(tH)
 &=|t|\kappa_{\mathcal M}(H),\\
 \kappa_{\mathcal M}(H+G)
 &\le\kappa_{\mathcal M}(H)
       +\kappa_{\mathcal M}(G).
 \end{aligned}
\]
If $G\in\mathcal A_M$, then $M_2(G)=0$ by
\cref{thm:energy}, so $V_G=0$ almost surely.
Therefore
\[
 \kappa_{\mathcal M}(H+G)
 =\kappa_{\mathcal M}(H),
\]
and $\kappa_{\mathcal M}$ is well-defined on
$\Herm(D)/\mathcal A_M$.

Finally, since $\mathfrak m_s(v)=0$ if and only
if $v=0$,
\[
 \kappa_{\mathcal M}(H)=0
 \quad\Longleftrightarrow\quad
 V_H=0\ \text{almost surely}
 \quad\Longleftrightarrow\quad
 M_2(H)=0.
\]
For $m<N$, \cref{thm:energy} identifies the last
condition with $H\in\mathcal A_M$. Hence the
induced function on $\Herm(D)/\mathcal A_M$ is
a norm.

\emph{(iii) Full measurement.}
If $m=N$, \cref{prop:stationary-structure}
gives $\pi_R=u_S$ for every $R\in U(D)$.
Since mana vanishes on $S$, we obtain
$\pi_R(\mathcal M)=0$.
\end{proof}

\subsection{Three regimes of stabilizer R\'enyi response}

By \cref{lem:resource-germs}, stabilizer R\'enyi entropies belong to the subquadratic, logarithmic quadratic, and ordinary quadratic observable classes for $0<\alpha<1$, $\alpha=1$, and $\alpha>1$, respectively. We now combine the corresponding stationary response theorems with the Clifford-specific tangent-energy and excursion-occupation results to characterize their weak-injection stationary response at every finite positive order.

\begin{theorem}[Stationary stabilizer R\'enyi response]
\label{thm:sre}
Suppose Condition~\condH\ holds. For each fixed
$0<\alpha<\infty$, define
$\overline S_{\alpha,H}(\theta):=\pi_\theta(S_\alpha)$.
Then, as $\theta\downarrow0$,
\begin{equation}\label{eq:all-order-response}
 \overline S_{\alpha,H}(\theta)=
 \begin{cases}
 \theta^{2\alpha}\kappa_\alpha(H)
   +o(\theta^{2\alpha}),
   &0<\alpha<1,\\[2pt]
 \theta^2\ln(1/\theta)\kappa_{1,\log}(H)
   +o\bigl(\theta^2\ln(1/\theta)\bigr),
   &\alpha=1,\\[2pt]
 \theta^2\kappa_\alpha^{(2)}(H)
   +o(\theta^2),
   &1<\alpha<\infty.
 \end{cases}
\end{equation}
The coefficients are finite and nonnegative,
with the following expressions and properties.

\begin{enumerate}
\item \textit{Orders below one.}
For $0<\alpha<1$,
\begin{equation}\label{eq:kappa-low}
 \kappa_\alpha(H)=\pih_H(f_\alpha).
\end{equation}

\item \textit{Order one.}
The logarithmic response coefficient is
\begin{equation}\label{eq:kappa-log}
 \begin{split}
 \kappa_{1,\log}(H)
 &=\frac4{\ln2}M_2(H)\\
 &=\frac4{\ln2}
   \bigl[C_{\mathrm c}E_{\mathrm c}(H)
        +C_{\mathrm{nc}}E_{\mathrm{nc}}(H)\bigr].
 \end{split}
\end{equation}

\item \textit{Orders above one.}
For $1<\alpha<\infty$,
\begin{equation}\label{eq:kappa-high}
 \kappa_\alpha^{(2)}(H)
 =\frac{2\alpha}{(\alpha-1)\ln2}M_2(H)
   +\mu_H^{\mathrm{occ}}(S_\alpha).
\end{equation}

\item \textit{Nondegeneracy and full measurement.}
If $m<N$, the corresponding response coefficient
in each of the three regimes is strictly positive
if and only if $H\notin\mathcal A_M$. Moreover,
\[
 [H]\longmapsto\kappa_{1,\log}(H)
\]
defines a positive definite real quadratic form
on $\Herm(D)/\mathcal A_M$. If $m=N$, then $\pi_R=u_S$ for every $R\in U(D)$,
and hence $\pi_R(S_\alpha)=0$ for every finite
$\alpha>0$.
\end{enumerate}
\end{theorem}

\begin{proof}
By \cref{prop:verification}, the Clifford model
satisfies the reference conditions required by
the stationary response theorems.

\emph{(i) Orders below one.}
For $0<\alpha<1$, \cref{lem:resource-germs}
shows that $S_\alpha$ is a subquadratic observable
of order $2\alpha$, with leading local term
$f_\alpha$. Applying \cref{cor:subquadratic} gives
\[
 \theta^{-2\alpha}\overline S_{\alpha,H}(\theta)
 \longrightarrow \pih_H(f_\alpha)
 =\kappa_\alpha(H),
\]
proving \eqref{eq:kappa-low} and the first line
of \eqref{eq:all-order-response}.

\emph{(ii) Order one.}
By \cref{lem:resource-germs}, $S_1$ is a
logarithmic quadratic observable with
$\zeta=1$ and $q_{\log}=4V^2/\ln2$.
Thus \cref{thm:log} yields
\[
 \frac{\overline S_{1,H}(\theta)}
      {\theta^2\ln(1/\theta)}
 \longrightarrow
 \frac4{\ln2}\pih_H(V^2)
 =\frac4{\ln2}M_2(H).
\]
The explicit expression in \eqref{eq:kappa-log}
then follows from \eqref{eq:M2-closed}.

\emph{(iii) Orders above one.}
For $\alpha>1$, \cref{lem:resource-germs}
shows that $S_\alpha$ is an ordinary quadratic
observable with
\[
 q_{S_\alpha}
 =\frac{2\alpha}{(\alpha-1)\ln2}V^2.
\]
Applying \cref{thm:quadratic}, we obtain
\[
 \begin{aligned}
 \theta^{-2}\overline S_{\alpha,H}(\theta)
 &\longrightarrow
   \pih_H(q_{S_\alpha})+\mu_H^{\mathrm{occ}}(S_\alpha)\\
 &=\frac{2\alpha}{(\alpha-1)\ln2}M_2(H)
   +\mu_H^{\mathrm{occ}}(S_\alpha),
 \end{aligned}
\]
which proves \eqref{eq:kappa-high} and the
third line of \eqref{eq:all-order-response}.
All coefficients are nonnegative. Their
finiteness follows from the tangent moment
bounds and, for the occupation term, from
$S_\alpha\le C_\alpha r_S^2$ and
$\mu_H^{\mathrm{occ}}(r_S^2)<\infty$ in \cref{thm:quadratic}.

\emph{(iv) Nondegeneracy and full measurement.}
Suppose $m<N$. By \cref{lem:resource-germs},
all the corresponding leading local terms
are strictly positive away from the zero
section. The nondegeneracy criterion in
\cref{cor:clifford-response} therefore gives
strict positivity of each response coefficient
if and only if $H\notin\mathcal A_M$. Moreover, \eqref{eq:kappa-log} and
\cref{thm:energy} show that
\[
 [H]\longmapsto\kappa_{1,\log}(H)
 =\frac4{\ln2}M_2(H)
\]
is a positive definite real quadratic form
on $\Herm(D)/\mathcal A_M$.

Finally, if $m=N$,
\cref{prop:stationary-structure} gives
$\pi_R=u_S$ for every $R\in U(D)$.
Since $S_\alpha$ vanishes on $S$,
$\pi_R(S_\alpha)=0$ for every finite
$\alpha>0$.
\end{proof}

\section{Applications and connections to prior work}
\label{sec:applications}

The Kraus-product analysis and stationary-response theory developed above provide tools for examining structural assumptions and extracting quantitative information in monitored quantum dynamics. We illustrate two applications. The first establishes an exact singular-value criterion for Born-conditioned Haar preservation, allowing a logical-state ansatz to be verified record by record and identifying when it fails. The second embeds a previously studied magic-injection protocol into the general framework, yielding explicit Hamiltonian-energy and rare-excursion contributions to its stationary response.

\subsection{Conditional Haar preservation in quantum trajectories}

For Haar-distributed initial states, Scocco \emph{et al.}~ \cite{Scocco2026RiseFall} used the following Haar-code ansatz at $\theta_M=0$ in their Eq.~(25):
\[
 |\psi_\nu\rangle\overset{\mathrm d}{=}
 C\bigl(|\psi_{\mathrm{Haar}}^{(\nu)}\rangle\otimes
 |0\rangle^{\otimes(N-\nu)}\bigr),
\]
where $C$ is a Clifford unitary and $|\psi_{\mathrm{Haar}}^{(\nu)}\rangle$ is Haar distributed on $\nu$ logical qubits. The stabilizer-code decomposition alone does not establish the Haar law of the logical component after conditioning on measurement outcomes. We give a criterion that makes this distinction precise.

For every nonzero finite-dimensional complex Hilbert space $L$, let $\mathsf h_L$ denote normalized Haar measure on $\mathbb P(L)$. Fix $W\in M_D(\C)\setminus\{0\}$, and define
\begin{equation}\label{eq:haar-born-disintegration}
 \begin{aligned}
 \mu_W(\dd[x])
   &:=\frac{D\|Wx\|^2}{\Tr(W^\dagger W)}\,
      \mathsf h_\HH(\dd[x]),\\
 T_W([x])&:=[Wx]\quad(Wx\ne0),\qquad
 \eta_W:=(T_W)_\#\mu_W.
 \end{aligned}
\end{equation}
Here $x$ is any unit representative of $[x]$. The measure $\mu_W$ is a probability measure, and $T_W$ is defined $\mu_W$-almost surely. If $W$ is the Kraus product of a finite record, then $\Tr(W^\dagger W)/D$ is the probability of that record for a Haar initial state, and $\eta_W$ is its conditional output law. Finally, let $Q_W$ be the orthogonal projector onto $\operatorname{Ran}W$ and set $k_W:=\rank W$.

\begin{proposition}[Born-conditioned Haar preservation]
\label{prop:haar-code-disintegration}
The Born-conditioned output law $\eta_W$ satisfies
\begin{equation}\label{eq:haar-word-criterion}
 \begin{aligned}
 \int\rho_y\,\eta_W(\dd[y])
  &=\frac{WW^\dagger}{\Tr(W^\dagger W)},\\
 \eta_W=\mathsf h_{\operatorname{Ran}W}
  &\quad\Longleftrightarrow\quad
  WW^\dagger=\frac{\Tr(W^\dagger W)}{k_W}\,Q_W.
 \end{aligned}
\end{equation}
Hence $\eta_W$ is Haar on its support if and only if all nonzero singular values of $W$ are equal.
\end{proposition}

\begin{proof}
The first identity follows from
$\int\rho_x\,\mathsf h_\HH(\dd[x])=I_D/D$ and
\eqref{eq:haar-born-disintegration}. If
$\eta_W=\mathsf h_{\operatorname{Ran}W}$, its first moment
is $Q_W/k_W$, proving necessity. Conversely, suppose
$WW^\dagger=\gamma Q_W$ with $\gamma>0$.
Then $W=\sqrt\gamma\,VP$ for an orthogonal projector
$P$ and a partial isometry $V$ with $V^\dagger V=P$
and $VV^\dagger=Q_W$.
For Haar $[x]$, the projected direction $[Px]$ is Haar
on $\mathbb P(\operatorname{Ran}P)$ and independent of
$\|Px\|^2$. Since the density in
\eqref{eq:haar-born-disintegration} is then proportional
to $\|Px\|^2$, Born weighting does not change this
direction law. The isometry $V$ pushes it forward to
$\mathsf h_{\operatorname{Ran}W}$.
\end{proof}

In the model of \cite{Scocco2026RiseFall}, the
$\theta_M=0$ cycle consists of a uniform global Clifford
$U$, a $Z_1$ measurement, and the inverse Clifford
$U^\dagger$. This is exactly \eqref{eq:clifford-Kraus}
with $d=2$, $M=\{1\}$, and $R=I_D$.
For every finite record $\omega\in J^t$ with
$W_\omega:=K_{\omega,0}\ne0$,
\cref{lem:flat-reference-words} gives
\[
 W_\omega^\dagger W_\omega=\gamma_\omega P_\omega,
 \qquad
 W_\omega W_\omega^\dagger=\gamma_\omega Q_\omega,
\]
where $P_\omega$ is a stabilizer-code projector and
$Q_\omega$ projects onto $\operatorname{Ran}W_\omega$.
The stabilizer--Choi argument in the proof of that lemma,
now applied to the output marginal, shows that $Q_\omega$
is also a stabilizer-code projector.
Thus $\rank Q_\omega=2^{\nu_\omega}$ for some
$\nu_\omega\in\{0,\ldots,N\}$, and there exists a
Clifford unitary $C_\omega$ with
\[
 \operatorname{Ran}W_\omega
 =C_\omega\bigl(\C^{2^{\nu_\omega}}
       \otimes |0\rangle^{\otimes(N-\nu_\omega)}\bigr).
\]
By \cref{prop:haar-code-disintegration}, for a Haar
initial state,
\begin{equation}\label{eq:haar-code-representation}
 \mathcal L(X_t^0\mid\omega)
 =\mathcal L\Bigl(
 \bigl[C_\omega\bigl(|\zeta_{\nu_\omega}\rangle
      \otimes|0\rangle^{\otimes(N-\nu_\omega)}\bigr)\bigr]
 \Bigr),\qquad
 [\zeta_{\nu_\omega}]
 \sim\mathsf h_{\C^{2^{\nu_\omega}}}.
\end{equation}
This proves the ansatz of
\cite[Eq.~(25)]{Scocco2026RiseFall}
exactly at every finite time, even conditional on each nonzero record. For $\theta_M>0$, the same exact recordwise Haar
property need not persist. Each single-step branch still has flat nonzero singular values, since for any $R\in U(D)$,
\[
 K^R_{U,a}(K^R_{U,a})^\dagger
 =g^{-1}U^\dagger\Pi_aU,
\]
but a product of branches may have unequal nonzero
singular values. For example, take $N=2$, $M=\{1\}$,
$R_\varphi=e^{-i\varphi X_1}$ with
$0<\varphi\le\pi/8$, and the two-step record with
$U_1=I_4$, $U_2=\operatorname{CNOT}_{2\to1}$,
and $a_1=a_2=0$. Writing
$|v_\varphi\rangle=e^{i\varphi X}|0\rangle$,
one finds
\begin{equation}\label{eq:haar-breakdown-word}
 \begin{aligned}
 W_\varphi
  &:=K^{R_\varphi}_{U_2,0}K^{R_\varphi}_{U_1,0}\\
  &=g^{-1}\bigl(
       \cos\varphi\,|00\rangle\langle v_\varphi,0|
       -i\sin\varphi\,|11\rangle\langle v_\varphi,1|
     \bigr),\\
 \frac{W_\varphi W_\varphi^\dagger}
      {\Tr(W_\varphi W_\varphi^\dagger)}
  &=\cos^2\varphi\,|00\rangle\langle00|
     +\sin^2\varphi\,|11\rangle\langle11|.
 \end{aligned}
\end{equation}
The record has probability $1/(4g^2)>0$ for a Haar initial state, and its two nonzero singular values are $g^{-1}\cos\varphi\ne g^{-1}\sin\varphi$. Thus its conditional output is not Haar on the two-dimensional code $\operatorname{span}\{|00\rangle,|11\rangle\}$. For the rotation path of \cite{Scocco2026RiseFall}, $\varphi=\pi\theta_M/8$ up to a global phase, so the exact recordwise Haar-code property fails for a positive-probability record at every $\theta_M\in(0,1]$. This does not by itself exclude an approximate Haar description after averaging over records.

\subsection{From a specific model to a general framework}

The monitored circuit analyzed in \cite{ZhenEtAl2026CMPv2} is a direct specialization of \eqref{eq:clifford-Kraus}. Take $M=\{1\}$, so that $m=1$, $q=d$, and $r=d^{N-1}$, and choose
\begin{equation}\label{eq:previous-T-specialization}
 R_\theta=\bigl(F_d^\dagger D_a^{(d)}(\theta)F_d\bigr)_1,
 \qquad
 D_a^{(d)}(\theta)=\diag(e^{-i\theta h_k})_{k\in\F_d},
\end{equation}
where $F_d$ is the discrete Fourier transform and $(h_k)$ is the fixed real phase lift of the $T$-type path in that work (with $a$ needed only for $d>3$). The same Clifford draw, Born rule, projection, and inverse Clifford therefore give exactly the same transition kernel.

Writing $\bar h=d^{-1}\sum_kh_k$, the perturbing Hamiltonian of \eqref{eq:weak-H} is $H=(F_d^\dagger\diag(h_k)F_d)_1$. Since the Fourier basis is mutually unbiased with the measured computational basis,
\begin{equation}\label{eq:previous-H-energy}
 H_0=\bar h I_D,\qquad H_{\mathrm c}=0,\qquad
 E_{\mathrm{nc}}(H)=\frac1d\sum_{k\in\F_d}(h_k-\bar h)^2>0.
\end{equation}
Thus $H\notin\mathcal A_M$. For $N\ge2$, the nondegeneracy statements in \cref{thm:energy,thm:mana,thm:sre} recover the strictly positive linear stationary mana response for odd prime $d$ and the strictly positive quadratic qubit $2$-SRE response proved in \cite{ZhenEtAl2026CMPv2}. The present formulation also gives their exact tangent-energy contribution, since \eqref{eq:M2-closed} reduces to
\[
 M_2(H)=C_{\mathrm{nc}}E_{\mathrm{nc}}(H),\qquad
 C_{\mathrm{nc}}
 =\frac{d(r-1)(dr-1)}{(d+1)r(d-1)}.
\]
In particular, for the qubit path $D^{(2)}(\theta)=\diag(1,e^{i\pi\theta/4})$, $E_{\mathrm{nc}}(H)=\pi^2/64$, and \cref{thm:sre} gives the explicit bound
\begin{equation}\label{eq:previous-S2-lower}
 \kappa_2^{(2)}(H)
 =\frac4{\ln2}M_2(H)+\mu_H^{\mathrm{occ}}(S_2)
 \ge\frac{\pi^2(r-1)(2r-1)}{24r\ln2}>0,
 \qquad r=2^{N-1}>1.
\end{equation}
The additional occupation term $\mu_H^{\mathrm{occ}}(S_2)$ is nonnegative. This specialization illustrates the additional information supplied by the general framework. The same first-order Hamiltonian data determine the leading stationary responses at all fixed finite positive R\'enyi orders, including the logarithmically enhanced regime at order one. The quadratic coefficient separates an explicit tangent-energy contribution from a nonnegative rare-excursion term, with \eqref{eq:previous-S2-lower} providing a quantitative lower bound. More generally, the measurement-projector span characterizes precisely the Hamiltonian directions invisible to the leading resource response under partial measurement. Thus, the framework both recovers the earlier model-specific asymptotics and supplies new structural and quantitative information.

\section*{Statements and Declarations}

\subsection*{Funding}

This work was partially supported by the National Key R\&D Program of China (Grant No.~\allowbreak 2024YFB4504004), the National Natural Science Foundation of China (Grant No.~92576114), the Guangdong Provincial Quantum Science Strategic Initiative (Grant Nos.~GDZX2403008 and GDZX2503001), the CCF-Tencent Rhino-Bird Open Research Fund, the Guangdong Provincial Key Lab of Integrated Communication, Sensing and Computation for Ubiquitous Internet of Things (Grant No.~2023B1212010007), the Quantum Science Center of Guangdong-Hong Kong-Macao Greater Bay Area, and the Education Bureau of Guangzhou Municipality.

\subsection*{Author Contributions}

G.Z. developed the main theoretical framework, carried out the mathematical analysis and wrote the original draft. X.Y. assisted in reviewing the technical approach and surveying the relevant literature. C.Z. contributed to shaping the research direction and designing the overall structure of the manuscript. R.C. and X.W. provided overarching scientific guidance on the choice of topic, the research direction, and the development of the core content. All authors discussed the results, reviewed and edited the manuscript, and approved the final version.

\subsection*{Use of Artificial Intelligence}

Building on the earlier arXiv version of this work \cite{ZhenEtAl2026CMPv2}, the authors used OpenAI's ChatGPT (GPT-6 series) as an interactive research and writing aid. The tool assisted in identifying generalizable mechanisms underlying the earlier proofs, developing and examining broader mathematical formulations and technical arguments, and exploring the scope and limitations of the resulting theoretical framework. It was also used to help organize and refine the manuscript. All mathematical arguments, interpretations, references, and final editorial decisions were independently checked and approved by the authors, who take full responsibility for the scientific content of the paper.

\subsection*{Competing Interests}

The authors have no relevant financial or non-financial interests to disclose.

\subsection*{Data and Code Availability}

This work is entirely theoretical. No datasets and code are required to establish
the results.

\end{document}